\documentclass[accepted=2026-01-17,a4paper,twocolumn,11pt]{quantumarticle}
\pdfoutput=1

\usepackage[utf8]{inputenc}
\usepackage[english]{babel}
\usepackage[T1]{fontenc}
\usepackage{tikz}
\usetikzlibrary{patterns}
\usetikzlibrary{calc}
\usepackage{lipsum}

\usepackage{subcaption}
\usepackage[ruled,vlined]{algorithm2e}
\usepackage{booktabs}
\usepackage{array}
\usepackage{graphicx}
\usepackage{physics}

\usepackage[numbers,sort&compress]{natbib}
\usepackage{amsmath,amssymb,amsfonts}

\usepackage{textcomp}
\usepackage{xcolor}
\usepackage{pgfplots,pgfplotstable}
\usepackage{hyperref}
\usepackage[nameinlink,capitalize]{cleveref} 
\pgfplotsset{compat=newest} 
\definecolor{one}{HTML}{c23b23}
\definecolor{two}{HTML}{E77700}
\definecolor{three}{HTML}{FFCA33}
\definecolor{four}{HTML}{03c03c}
\definecolor{orange}{HTML}{C49500}

\begin{document}

\title{A Multilevel Framework for Partitioning Quantum Circuits}

\author{Felix Burt}
\affiliation{Electrical and Electronic Engineering, Imperial College London, South Kensington Campus, London SW7 2AZ, UK}
\email{f.burt23@imperial.ac.uk}
\affiliation{Imperial Quantum Engineering, Science and Technology Centre (QuEST)}
\author{Kuan-Cheng Chen}
\affiliation{Electrical and Electronic Engineering, Imperial College London, South Kensington Campus, London SW7 2AZ, UK}
\email{kuan-cheng.chen17@imperial.ac.uk}
\affiliation{Imperial Quantum Engineering, Science and Technology Centre (QuEST)}
\author{Kin K. Leung}
\affiliation{Electrical and Electronic Engineering, Imperial College London, South Kensington Campus, London SW7 2AZ, UK}
\email{kin.leung@imperial.ac.uk}

\maketitle

\begin{abstract}
Executing quantum algorithms over distributed quantum systems requires quantum circuits to be divided into sub-circuits which communicate via entanglement-based teleportation. Naively mapping circuits to qubits over multiple quantum processing units (QPUs) results in large communication overhead, increasing both execution time and noise. This can be minimised by optimising the assignment of qubits to QPUs and the methods used for covering non-local operations. Formulations that are general enough to capture the spectrum of teleportation possibilities lead to complex problem instances which can be difficult to solve effectively. This highlights a need to exploit the wide range of heuristic techniques used in the graph partitioning literature. This paper formalises and extends existing constructions for graphical quantum circuit partitioning and designs a new objective function that captures further possibilities for non-local operations via \textit{nested state teleportation}. We adapt the well-known Fiduccia-Mattheyses heuristic to the constraints and problem objective and explore multilevel techniques that coarsen hypergraphs and partition at multiple levels of granularity. We find that this reduces runtime and improves solution quality of standard partitioning. We place these techniques within a larger framework, through which we can extract full distributed quantum circuits including teleportation instructions. We compare the entanglement requirements and runtimes with state-of-the-art methods, finding that we achieve the lowest entanglement costs in most cases. Averaging over a wide range of circuits, we reduce the entanglement requirements by 35\% compared with the next best-performing method. We also find that our techniques can scale to much larger circuit sizes than competing methods, provided the number of partitions is not too large.
\end{abstract}

\section{Introduction}\label{sec:intro}
Distributed quantum computing (DQC) is becoming an increasingly popular paradigm for building scalable quantum computers \cite{caleffiDistributedQuantumComputing2024}. The ability to share entanglement among nodes in a network of quantum processing units (QPUs) grants access to the full quantum computational power across the network, at the expense of the additional time and noise overheads incurred by entanglement distribution \cite{isailovicInterconnectionNetworksScalable2006}. Key milestones for facilitating DQC have recently been demonstrated \cite{akhtarHighfidelityQuantumMatterlink2023,krutyanskiyEntanglementTrappedIonQubits2023,oreillyFastPhotonMediatedEntanglement2024,incDistributedQuantumComputing2024,almanaklyDeterministicRemoteEntanglement2025,sahaHighfidelityRemoteEntanglement2025}, while early-stage experimental demonstrations have given credibility to the approach \cite{chouDeterministicTeleportationQuantum2018a, wanQuantumGateTeleportation2019,mainDistributedQuantumComputing2025a}.

Despite these important achievements, DQC systems face a number of challenges that are not present in traditional, monolithic systems, which must be overcome to achieve the desired scaling. Most notably, the overhead introduced by entanglement distribution is significant. This can lead to deep, slow circuits that require frequent inter-QPU entanglement sharing. Estimates suggest operations requiring shared entanglement are at least an order of magnitude slower than those executed on a single QPU \cite{angARQUINArchitecturesMultinode2024}. Coupling high demands for entanglement with high noise and latency can quickly render circuits infeasible to execute.

A growing body of literature is concerned with minimising this additional overhead, using optimised routines for assigning logical qubits to QPUs and covering resulting non-local operations \cite{barralReviewDistributedQuantum2025, zomorodi-moghadamOptimizingTeleportationCost2018,andres-martinezAutomatedDistributionQuantum2019, daeiOptimizedQuantumCircuit2020,bakerTimeslicedQuantumCircuit2020, dadkhahNewApproachOptimization2021,ferrariCompilerDesignDistributed2021,gsundaramEfficientDistributionQuantum2021,nikahdAutomatedWindowbasedPartitioning2021,wuEntanglementefficientBipartitedistributedQuantum2023a,wuAutoCommFrameworkEnabling2022,wuQuCommOptimizingCollective2023,cuomoOptimizedCompilerDistributed2023,ferrariModularQuantumCompilation2023a,escofetHungarianQubitAssignment2023a,sunkelApplyingEvolutionaryAlgorithm2024,chenCircuitPartitioningTransmission2024,cramptonGeneticApproachMinimising2025,cambiucciHypergraphicPartitioningFramework2025,promponasCompilerDistributedQuantum2025,kaurOptimizedQuantumCircuit2025,escofetRevisitingMappingQuantum2025,russoTeleSABRELayoutSynthesis2025a}. The resulting optimisation problems promise reductions in entanglement requirements but are typically NP-hard \cite{maoQubitAllocationDistributed2023}, with no polynomial-time solutions for obtaining global optima. However, similar problems have been faced and extensively explored in networking and VLSI circuit design, in which very mature, effective and efficient algorithms for partitioning graphs and circuits have been developed \cite{kernighanEfficientHeuristicProcedure1970,fiducciaLinearTimeHeuristicImproving1982,sanchisMultipleWayNetworkPartitioning1989a,congParallelBottomupClustering1993,johannesPartitioningVLSICircuits1996,huangPartitioningbasedStandardcellGlobal1997,karypisMultilevelHypergraphPartitioning1997,karypisFastHighQuality1998,karypisMultilevelWayHypergraph1999,meyerhenkePartitioningComplexNetworks2014,schlagHighQualityHypergraphPartitioning2023}. 

This article aims to combine a number of methods and techniques employed for large-scale hypergraph partitioning and apply them to the unique case of quantum circuit partitioning. The result is a multilevel partitioning framework, inspired by state-of-the-art graph partitioners such as METIS \cite{karypisMETISSoftwarePackage1997}, hMETIS \cite{karypisFastHighQuality1998} and KaHyPar \cite{schlagHighQualityHypergraphPartitioning2023}, that exploits the particular structure of quantum circuits. The problem is broken down into a number of stages, each of which can be adapted to the particular instance. The stages are \textit{transpilation}, \textit{graph conversion}, \textit{gate grouping}, \textit{coarsening}, \textit{partitioning}, \textit{refinement} and \textit{circuit extraction}. 

We use existing transpilation techniques to convert circuits into an appropriate gate set, then convert the resulting circuit to a \textit{temporally-extended} graph representing interaction of qubits over time. We then generalise this graph to a hypergraph by grouping gates together based on their ``compatibility'' for multi-gate teleportation, i.e., whether or not they can be teleported using the same entanglement resource. We use this hypergraph structure to define a cost function which captures the entanglement cost of the resulting partitioning, considering the possibility of state teleportation, gate teleportation, multi-gate teleportation and a new protocol referred to as \textit{nested state teleportation}, in which we partially collapse gate teleportation procedures into state teleportation. We then design a variant of the Fiduccia-Mattheyses (FM) algorithm \cite{fiducciaLinearTimeHeuristicImproving1982} that is compatible with the unique constraints of the problem and tweaked to be resistant to local minima. Following this, we introduce a multilevel framework that applies partitioning at varying levels of temporal granularity. We investigate three different strategies for coarsening hypergraphs along the time axis, using the FM algorithm to refine the partitions at each level of granularity. We show that this both improves the time efficiency of the algorithm and improves the partitioning quality in terms of entanglement costs. The final stage produces a distributed quantum circuit, including all necessary instructions for teleportation of qubits and gates between QPUs. 

We demonstrate the effectiveness of the framework by comparing the entanglement costs achieved by the best of our multilevel techniques with those obtained by state-of-the-art methods in the literature. Our techniques are shown to be the most effective at reducing entanglement requirements across a wide range of circuit types, including those with high connectivity and depth, while existing methods tend to be effective in a restricted number of cases. On average, we achieve a 35\% improvement in the average ratio of entangled pairs of qubits, or \textit{e-bits}, to two-qubit gates compared with the best benchmark method. In most cases, these results are achieved with a lower amount of computation time. Furthermore, the multilevel framework allows the complexity of the problem to be tailored to the available computational resources and can thus be applied to large-scale circuits with hundreds of qubits and gates with proper use of coarsening. 

We note that, while the methods investigated here achieve state-of-the-art performance, they operate under an assumption of all-to-all connectivity between QPUs, which may not be feasible in large-scale systems. Additionally, we restrict our investigations to networks with fewer than $20$ QPUs, since the partitioning problem scales with the size of the network. However, in follow-up work we tackle both of these issues, adapting the framework to arbitrary network topologies and larger numbers of QPUs \cite{burtEntanglementEfficientDistributionQuantum2025}. Both the current and follow-up work form the open-source repository \texttt{disqco}, which is available on GitHub \cite{FelixburtDISQCO}, and can be used to reproduce the results presented here.
\section{Background}\label{sec:background}

\subsection{Distributed Quantum Computing}\label{sec:DQC}

Distributed quantum computing (DQC) is concerned with the execution of quantum algorithms across multiple linked quantum processing units (QPUs). While this can feasibly be achieved using classical communication via circuit cutting and circuit knitting \cite{pengSimulatingLargeQuantum2020,tangCutQCUsingSmall2021,tangCuttingQuantumCircuits2022,basuFragQCEfficientQuantum2024,tomeshDivideConquerCombinatorial2023a,bechtoldInvestigatingEffectCircuit2023,chatterjeeQurzonPrototypeDivide2022,brandhoferOptimalPartitioningQuantum2024}, these represent a near-term form of DQC that is limited by the exponential scaling of the sampling overhead required to recover the correct results. The alternative is entanglement-based DQC, which requires links capable of sharing purely quantum information among QPUs. This is typically implemented by optical fibres transmitting photonic qubits \cite{divincenzoPhysicalImplementationQuantum2000}. Qubit transmission over fibres is used to distribute pairs of entangled qubits between geographically separated locations. Shared entangled pairs are known as \textit{e-bits} or \textit{EPR pairs} and represent a consumable resource required for distributed quantum computation. 

Two QPUs holding a shared e-bit can use local operations and classical communication (LOCC) to teleport the state of a qubit that is actively involved in a computation from one QPU to the other \cite{bennettTeleportingUnknownQuantum1993}. This is referred to as \textit{state teleportation}. Alternatively, the e-bit can directly moderate a non-local controlled-unitary operation \cite{gottesmanDemonstratingViabilityUniversal1999,eisertOptimalLocalImplementation2000}, analogously referred to as \textit{gate teleportation}. Gate teleportation has been shown to be a flexible procedure, that can cover multiple non-local operations using the same shared e-bit \cite{yimsiriwattanaGeneralizedGHZStates2004,eisertOptimalLocalImplementation2000,wuEntanglementefficientBipartitedistributedQuantum2023a}, creating a large design space for optimising the entanglement requirements of distributed quantum circuits.

\subsection{Non-local gate coverage}\label{sec:NLGC}

Non-local gate coverage is a key target for optimisation in DQC. Any method of dividing or partitioning a quantum circuit results in some number of non-local operations. We can formalise this by defining a set of QPUs as $Q$, and a set of logical qubits $q$. Partitioning a quantum circuit requires making a \textit{partition assignment} $\phi : q \to Q$ which maps logical qubits to QPUs. For any two-qubit gate, defined partly by the qubits on which it operates as $g(q_{i}, q_{j})$, the gate is non-local under $\phi$ if $\phi(q_{i}) \neq \phi(q_{j})$. The cost of a partitioning is thus closely related to the non-local operations present in the circuit. However, since there are many ways to cover non-local operations, the cost of a partitioning does not directly correspond to the number of non-local operations. Rather, it is a quantity that depends on the method chosen for covering non-local operations. 

In general, a given non-local operation can be covered in three different ways. First, state teleportation can be used to teleport the state of a qubit from one QPU to another, modifying the assignment function $\phi$, allowing the gate to be executed locally. Second, if the gate is a controlled-unitary operation, it can be executed remotely using gate teleportation, a procedure that entangles the control qubit with a communication qubit in the target QPU, then uses the linked communication qubit to control the application of the unitary. Both methods require a single e-bit, though state teleportation may modify the locality of future gates in the circuit. Third, when certain compatibility conditions are met, separate gate teleportation procedures can be merged into a single gate teleportation, which consumes just one e-bit \cite{wuEntanglementefficientBipartitedistributedQuantum2023a}. This has been referred to as the EJPP protocol \cite{wuEntanglementefficientBipartitedistributedQuantum2023a}, burst communication \cite{wuAutoCommFrameworkEnabling2022}, extended gate teleportation \cite{burtGeneralisedCircuitPartitioning2024b}, or simply teleportation \cite{yimsiriwattanaDistributedQuantumComputing2004}. For simplicity, we refer to the full process of teleporting multiple gates within the same process as \textit{multi-gate teleportation}. 

\subsubsection{Multi-gate teleportation}\label{sec:multi-gate}

Gate teleportation uses a shared e-bit to link a control qubit with a remote communication qubit. The process creates a partially accessible clone of the control qubit at multiple locations, allowing compatible two-qubit gates to be performed using the communication qubit. \textit{Multi-gate} teleportation refers to the process of using a single e-bit to teleport multiple gates. This is based on the fact that certain operations kill the link between the control and the communication qubit, while certain others do not. A full multi-gate teleportation procedure can be broken up into three stages. 

\begin{figure}
  \includegraphics[width=\columnwidth,clip,trim=0.3cm 0.7cm 0.3cm 0]{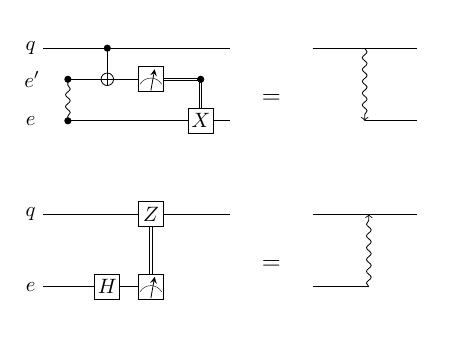}
    \caption{The starting and ending processes. The starting process $S_{q,e}$ is a linear map which maps the state of the root qubit $q$ onto a joint state between $q$ and an auxiliary qubit $e$ in another QPU. The ending process $E_{q,e}$ applies the inverse map, disentangling the auxiliary qubit from the root qubit, returning the original state to $q$. Diagrammatic notation from Ref. \cite{wuEntanglementefficientBipartitedistributedQuantum2023a}}
    \label{fig:starting_ending}
\end{figure}

First, a qubit is entangled with a communication qubit at a distance via a shared e-bit. Following this, a series of remote operations is performed using the communication qubit as control. When all compatible gates are complete, the qubit is then disentangled from the communication qubit. The first and final steps have been referred to as the \textit{cat-entangler} and \textit{cat-disentangler} primitives \cite{yimsiriwattanaDistributedQuantumComputing2004}, since the state of the communication qubit mirrors the state of the original qubit, a situation reminiscent of Schr\"{o}dinger's cat. For consistency with previous work, we adopt the terminology and notation introduced by Wu et al. \cite{wuEntanglementefficientBipartitedistributedQuantum2023a}, calling the multi-gate teleportation primitives the \textit{entanglement-assisted starting and ending processes}. These processes are illustrated in \cref{fig:starting_ending}.

Mathematically, the starting and ending processes are described as linear maps on general quantum states that compose to form the identity. The starting process $S_{q,e}$ maps an input state $\ket{\psi}_{q}$ for some root qubit $q$ on Hilbert space $\mathcal{H}_{q}$ onto a joint state between $q$ and an auxiliary qubit $e$ in another QPU:
\begin{equation}\label{eq:starting}
  \begin{aligned}
    S_{q,e}(\ket{\psi}) &= \bra{0}_{q}\ket{\psi}_{q}\ket{0}_{q}\ket{0}_{e} + \bra{1}_{q}\ket{\psi}_{q}\ket{1}_{q}\ket{1}_{e} \\
    &= CX_{q,e}\ket{\psi}_{q}\ket{0}_{e},
  \end{aligned}
\end{equation}
which has the effective action of a $CX$ gate between $q$ and $e$, causing a `fan-out' of the state of qubit $q$ onto $e$, facilitated by the shared e-bit. The symmetry of the resulting state in \cref{eq:starting} means that both the root qubit $q$ and the communication qubit $e$ behave the same when used as a control in a controlled-unitary operation. This symmetry is preserved until a non-diagonal single-qubit gate is applied to either qubit, that mixes the $\ket{0}$ and $\ket{1}$ basis components of the state. To ensure that we regain the correct state on $q$, the gate teleportation ending process must be applied before any gate that is not diagonal or anti-diagonal in the computational basis of the root qubit is applied \cite{wuEntanglementefficientBipartitedistributedQuantum2023a}. The ending process, $E_{q,e}$, applied to the state $ \ket{\psi '}_{q,e} = S_{q,e}(\ket{\psi}_{q})$, applies the inverse map of \cref{eq:starting}:
\begin{equation}\label{eq:ending}
    \begin{aligned}
        & E_{q,e}(\ket{\psi'}_{q,e}) \\ &= Tr_{e}(CX_{q,e} \ket{\psi'}_{q,e} \bra{\psi'}_{q,e} CX_{q,e}),
    \end{aligned}
\end{equation}
where $ Tr(\ket{i}\bra{j}) = \delta_{ij} $ and $Tr_{k}$ is the partial trace on qubit $k$, which removes it from the state description. As a result, $E_{q,e} \circ S_{q,e}(\ket{\psi}_{q}) = \ket{\psi}_{q}$ \cite{wuEntanglementefficientBipartitedistributedQuantum2023a}. This has the effect of a `fan-in' operation, disentangling the auxiliary qubit $e$ from the root qubit $q$, returning the original state to $q$. 

If we want to perform a controlled unitary $CU_{q,q'}$, where 
\begin{equation}
\begin{aligned}
  CU_{q,q'} &= \ket{0}_{q}\bra{0}_{q} \otimes I_{q'} + \ket{1}_{q}\bra{1}_{q} \otimes U_{q'} \\ &= \begin{pmatrix}
      I_{q'} & 0 \\
      0 & U_{q'}
  \end{pmatrix},
\end{aligned}
\end{equation}
this is equivalent to performing
\begin{equation}
E_{q,e} \circ  CU_{e,q'} \circ S_{q,e}(\ket{\psi}_{q}) = CU_{q,q'}\ket{\psi}_{q}.
\end{equation}

If two contiguous gate teleportation processes share the same root and auxiliary qubits, $q$ and $e$, then they can be merged into a single, multi-gate teleportation procedure that uses the same starting and ending processes. As described in Ref. \cite{andres-martinezDistributingCircuitsHeterogeneous2024}, this can be straightforwardly generalised to link multiple QPUs. 

A $k$-fold starting process $S_{q,\mathbb{E}}$ rooted on $q$ acting on a set of auxiliary qubits $\mathbb{E}$ spanning $k$ QPUs, can be described as:
\begin{equation}\label{eq:link_multi}
  \begin{aligned}
    S_{q,\mathbb{E}}(\ket{\psi}) &= \bra{0}_{q}\ket{\psi}_{q}\ket{0}_{q}\bigotimes_{e \in \mathbb{E}}\ket{0}_{e} \\ &+ \bra{1}_{q}\ket{\psi}_{q}\ket{1}_{q}\bigotimes_{e \in \mathbb{E}}\ket{1}_{e} \\
    &= \prod_{e \in \mathbb{E}} CX_{q,e}\ket{\psi}_{q}\bigotimes_{e \in \mathbb{E}}\ket{0}_{e},
  \end{aligned}
\end{equation}
which has the effect of fanning out the state of $q$ onto all auxiliary qubits in $\mathbb{E}$. 

Without any network constraints, $n$ e-bits allows $n$ QPUs to be linked to the root qubit $q$. Once the link state is active, it can be used to control any number of contiguous controlled-unitary operations in each linked QPU via merging the teleportation procedures: 
\begin{equation}\label{eq:multi_gate_tel}
  \begin{aligned}
    &\prod_{q' \in \mathbb{Q}} U_{q,q'}\ket{\Psi} = E_{q,\mathbb{E}}\circ ( \\ &\prod_{q', e \in \mathbb{Q} \times \mathbb{E}} (U_q \otimes I_e \otimes U_{q'})( I_q \otimes CU_{e,q'})) \circ S_{q,\mathbb{E}} (\ket{\Psi}),
  \end{aligned}
\end{equation}
where $\ket{\Psi}$ is the state across the full Hilbert space of the root qubit $q$ and all $q' \in \mathbb{Q}$, which are the \textit{receivers} in the gate teleportation. The starting and ending processes, however, still act only on the subspace of the root qubit and the auxiliary qubits. A \textit{compatible} two-qubit unitary is given by
\begin{equation}\label{eq:multi-CU}
\prod_{q' \in \mathbb{Q}} U_{q,q'} = \prod_{q', e \in \mathbb{Q} \times \mathbb{E}} (U_{q}\otimes U_{q'})CU_{q,q'},
\end{equation}
for $\theta = 2n\pi$ (diagonal) or $\theta = (2n+1)\pi$ (anti-diagonal) in the general single-qubit unitary $U(\theta,\phi,\lambda)$ \cref{eq:U_CP} on $q$. In the anti-diagonal case, $U_{e} = X_{e}$ is applied to account for the flipping of the basis components caused by $U_{q}$. Otherwise, $U_{e} = I_{e}$. \cref{eq:multi-CU} omits single-qubit and two-qubit gates acting on qubits in $Q$ which are independent of the root qubit $q$, since there are no restrictions on such gates for multi-gate teleportation. \cref{fig:starting} contrasts the coverage of a non-local operation using the starting and ending processes with a state teleportation procedure.

An important point to note is that the most effective method for covering non-local operations depends greatly on circuit structure, and the choice of method can have a significant impact on the resulting entanglement cost, quantified by the number of e-bits required. For example, methods that depend entirely on state teleportation, such as the fine-grained partitioning approach of Baker et al. \cite{bakerTimeslicedQuantumCircuit2020}, are very effective for quantum volume and arithmetic circuits, while they tend to be less effective for circuits with less regular connectivity and high potential for multi-gate teleportation \cite{burtGeneralisedCircuitPartitioning2024b,bakerTimeslicedQuantumCircuit2020}. Alternatively, multi-gate teleportation methods are effective for circuits densely packed with two-qubit gates, since many opportunities arise for merging teleportation procedures together \cite{andres-martinezDistributingCircuitsHeterogeneous2024}.

\subsubsection{Nested state teleportation}\label{sec:nested}

Since both state teleportation and gate teleportation are effective in different cases, it is desirable to exploit both methods. Beyond just treating them as separate options for covering non-local operations, we may also gain benefit from combining them into the same procedure.

\begin{figure}
  \centering
  \includegraphics[width=0.9\linewidth]{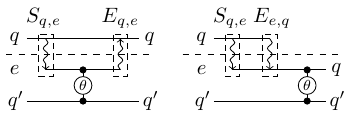}
  \caption{Methods for covering a non-local operation. On the left is the entanglement-assisted starting and ending processes $S_{q,e}$ and $E_{q,e}$, used for a gate teleportation. On the right, we flip the direction of the ending process to $E_{e,q}$, which allows us to teleport the state of $q$ onto $e$. The two circuits are equivalent up to relabelling of qubits after the ending process.}
  \label{fig:starting}
\end{figure}

\begin{figure*}
  \centering
  \includegraphics[width=0.8\textwidth]{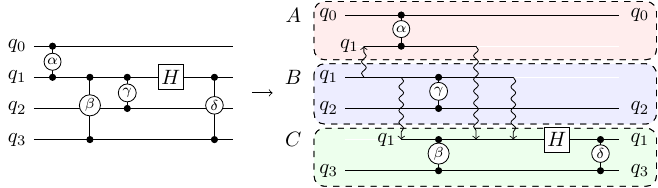}
\caption{The nested state teleportation procedure. A $2$-fold starting process is applied to entangle $q_1$, located in QPU $B$, with auxiliary qubits in $A$ and $C$. This fans the state out, allowing $CP(\theta)$ gates with $q_0$ and $q_3$ to be executed, as well as the local gate with $q_2$. The $H$ gate on $q_1$ requires the gate teleportation to be ended. Since there is another gate between $q_1$ and $q_3$, all ending processes should be re-routed towards $C$, collapsing the final state onto the auxiliary qubit. In fact, for each ending process, we may route the ending process to any QPU that still has an active link, provided the final ending process ends up at our chosen destination. The circuit on the right is equivalent to the one on the left up to a relabelling of qubits, as shown at the end of the wires.}
\label{fig:nested_state_tel}
\end{figure*}

First note an interesting feature of the above formulation of gate teleportation -- the starting process and the ending process are identical to those used in state teleportation, except the indices for the ending process are flipped. Accordingly, we may define a state teleportation procedure from $q$ to $e$ as:
\begin{equation}\label{eq:state_telep}
  E_{e,q} \circ S_{q,e}(\ket{\psi}_{q}) = \ket{\psi}_{e},
\end{equation}
where we have exchanged the indices $q$ and $e$, meaning that $q$ is measured, and the classically controlled operation is applied, such that the state of $q$ is teleported onto $e$. In \cref{fig:starting}, we use the same diagrammatic notation to indicate a state teleportation built out of the starting and ending processes. What is interesting about this is that it indicates we may always collapse a gate teleportation procedure into a state teleportation by redirecting the ending process away from the QPU of the root qubit. If we perform a $k$-fold starting process $S_{q,\mathbb{E}}$ on $q$ and a set of auxiliary qubits $\mathbb{E}$, we create the state in \cref{eq:link_multi}. To collapse the state onto $\tilde{e}$ we perform the following
\begin{equation}
  \begin{aligned}
    E_{\tilde{e}, \tilde{\mathbb{E}}} \circ S_{q,\mathbb{E}}(\ket{\psi}_{q}),
  \end{aligned}
\end{equation}
where $\tilde{\mathbb{E}} := (\mathbb{E} \setminus \tilde{e}) \cup \{q\}$, such that we replace $q$ with $\tilde{e}$ in the ending processes for all other auxiliary qubits, and interchange the indices $q$ and $\tilde{e}$ for the ending process corresponding to $S_{q,\tilde{e}}$. In \cref{fig:nested_state_tel}, we illustrate a situation where this is of use. We refer to this as \textit{nested state teleportation}, since we are effectively nesting a state teleportation inside a $k$-fold multi-gate teleportation process. If a qubit has been linked to many QPUs, ending processes can be successively composed to induce multiple nested state teleportations, until the final state is collapsed onto a single auxiliary qubit.

This tells us that choosing between state and gate teleportation reduces to choosing the direction of the ending processes of a multi-gate teleportation procedure. Covering non-local operations always requires a starting process, but the character of the teleportation is determined by the ending process.

\subsection{Generalised Circuit Partitioning}\label{sec:GCP}

Generalised circuit partitioning (GCP) was introduced in Burt et al. \cite{burtGeneralisedCircuitPartitioning2024b}, with the aim of creating an optimisation framework that considers both multi-gate and state teleportation methods when partitioning quantum circuits. Here we recap the basics and provide some extensions and clarifications on the notation. 

GCP uses a graphical picture of quantum circuits, with a spatial and a temporal component, which is extended to a hypergraph construction. 

Consider a circuit $\mathcal{C}$, consisting of $d$ layers of gates $\mathcal{L} = \bigcup_{t}^{d}\mathcal{L}^{(t)}$, where each $t$ corresponds to an independent time step of the circuit. Each $\mathcal{L}^{(t)}$ contains the single and two-qubit gates for time step $t$ of the circuit, where each gate is defined by the qubits on which it acts, which are elements of the set of logical qubits $Q_{L}$, and its parameters. The total number of logical qubits is $n_{q} = |Q_{L}|$.

The graph, $H(V,E)$, consists of $n_{q}d$ nodes, where each node is associated with a qubit and a time step. We can denote this using a time map $\tau : V \to \{1,...,d\}$, and a qubit map $\kappa : V \to Q_{L}$, where $d$ is the total depth of the circuit, i.e., the number of independent time steps, including initialisation and measurement. We use the shorthand $v_{q}^{(t)}$ to denote a node $v$ for which $\tau(v) = t$ and $\kappa (v) = q$. For convenience, we refer to the \textit{temporal} graph (or hypergraph), as $H(V,E;\tau,\kappa)$. The total set of nodes $V$ can be considered as the union of $d$ subsets $V^{(t)}$ of nodes
\begin{equation}\label{eq:V}
  V = \bigcup_{t = 1}^{d} V^{(t)},
\end{equation}
where
\begin{equation}
  V^{(t)} = \{v \mid \tau (v) = t, \forall i \in Q_{L} \}.
\end{equation}

The set of edges can be considered to be the union of two separate sets, the state edges $E_{s}$ and the gate edges $E_{g}$.
\begin{equation}\label{eq:E}    
  E = E_{s} \cup E_{g},
\end{equation}
\begin{multline}
  E_{s} = \{(v,u) \mid \kappa(v) = \kappa(u), \tau(u) = \tau(v) + 1 \\ \forall i \in Q_{L}, \forall t \in \{1,...,d-1\} \} 
\end{multline}
\begin{multline}
  E_{g} = \{(v,u) \mid (i,j) \in \mathcal{L}^{(t)}, \\ \forall t \in \{1,..., d\} \}
\end{multline}

For simplicity, the graph is built using a general gate set consisting of the following two gates:
\begin{gather}\label{eq:U_CP}
  U(\theta,\phi,\lambda) = 
  \begin{pmatrix}
    \cos(\theta/2) & -e^{i \lambda}\sin(\theta/2) \\
    e^{i (\phi + \lambda)}\sin(\theta/2) & e^{i\phi}\cos(\theta/2) 
  \end{pmatrix},
  \\
  CP(\theta) = \begin{pmatrix}
    1 & 0 & 0 & 0 \\
    0 & 1 & 0 & 0 \\
    0 & 0 & 1 & 0 \\
    0 & 0 & 0 & e^{i\theta} 
  \end{pmatrix}.
\end{gather}
\begin{figure}
  \centering
  \includegraphics[width=0.8\linewidth]{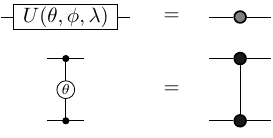}
  \caption{Correspondence between single and two-qubit gates and temporal hypergraphs. We represent any single-qubit gate as a grey node, storing the parameters $\theta, \phi, \lambda$ as node attributes. Two-qubit $CP(\theta)$ gates are represented as an edge between two black nodes, storing the phase parameter $\theta$ as an edge attribute.}
  \label{fig:U_CP}
\end{figure}

The gate set has a number of advantages. It allows us to compress many single-qubit gates together into one single $U$ -- giving us smaller problem instances -- while also allowing us to identify gates only by their respective parameters (three for each single-qubit gate and one for each two-qubit gate). Single-qubit gate parameters can be stored as node attributes while the two-qubit gate parameters can be stored as edge attributes, making the process of circuit extraction straightforward. Furthermore, the $CP(\theta)$ gate is symmetric, which means both qubits can play the role of the root in a gate teleportation procedure \cite{andres-martinezAutomatedDistributionQuantum2019}. In contrast, for a $CX$ gate, only the control qubit can be the root. This allows for many more gate grouping possibilities. 

Furthermore, common two-qubit gates such as $CZ$ and $CX$ can be constructed from a single $CP(\theta)$ gate and single-qubit unitaries, whereas an arbitrary $CP(\theta)$ gate requires two $CX$ gates and single-qubit unitaries. 

While it is possible to construct the graph using other gate sets, this often results in more two-qubit gates and smaller gate groups -- both likely contributors to higher e-bit cost. At present, the construction is not directly compatible with multi-qubit gates, such as the Toffoli, though we believe the extension would be straightforward. 

A circuit can be converted to and from our desired gate set using standard transpilation techniques \cite{javadi-abhariQuantumComputingQiskit2024,sivarajahT|ketRetargetableCompiler2020}. The correspondence between nodes and edges and gates is illustrated in \cref{fig:U_CP}.

\begin{figure}[ht]
  \centering
  \begin{subfigure}{0.465\textwidth} 
    \centering
    \includegraphics[width=\columnwidth]{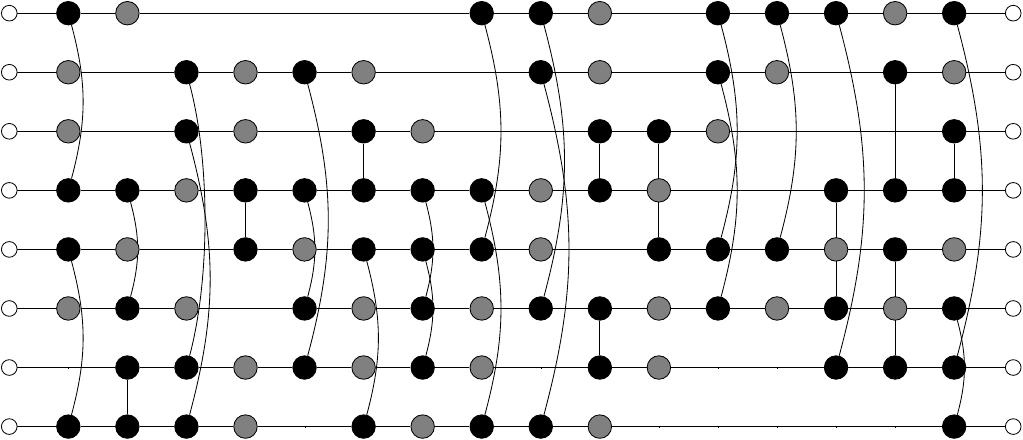}
    \caption{}
    \label{fig:CP5_8}
   \end{subfigure}
   ~
   \begin{subfigure}{0.48\textwidth}
    \centering  
    \includegraphics[width=\columnwidth]{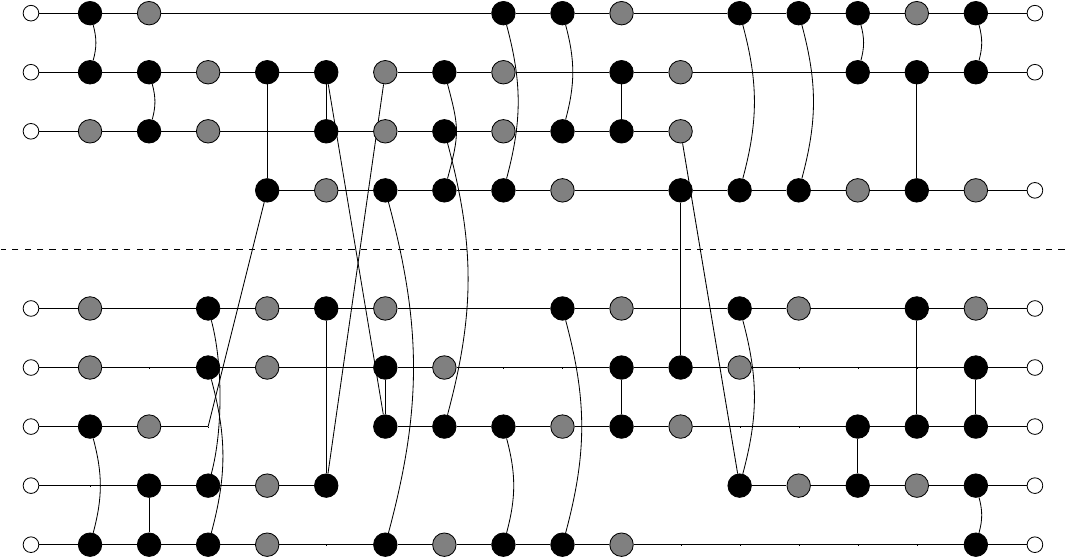}
    \caption{}
    \label{fig:CP5_8_FM}
  \end{subfigure}
  \caption{Base graph for a random, 8-qubit circuit. Grey nodes correspond to single-qubit gates, while black nodes correspond to qubits involved in $CP(\theta)$ gates, as illustrated in \cref{fig:U_CP}. State edges connect nodes representing the same qubit at different time steps. In \cref{fig:CP5_8_FM}, an optimised partitioning of the nodes is illustrated by dragging nodes vertically into different QPU regions. Cut state edges correspond to state teleportation, while cut gate edges correspond to gate teleportation.}
  \label{fig:CP5_graphs}
\end{figure}

An example of the base temporal graph is shown in \cref{fig:CP5_8}. At this stage, partitioning the nodes corresponds to choosing a set of state and gate teleportation operations to cover the non-local gates in the circuit, where each \textit{cut} edge indicates the use of an e-bit. A cut state-edge corresponds to state teleportation, and a cut gate-edge to gate teleportation (see \cref{fig:CP5_8_FM}). However, we would like to consider opportunities for multi-gate teleportation. Typically, circuit partitioning methods choose to allocate qubits before merging teleportation procedures, separating the problem into distinct components \cite{sundaramDistributionQuantumCircuits2022,wuEntanglementefficientBipartitedistributedQuantum2023a,wuAutoCommFrameworkEnabling2022}. An exception to this is the original hypergraph partitioning strategy of Andres-Martinez and Heunen \cite{andres-martinezAutomatedDistributionQuantum2019}, that is also used in Ref. \cite{andres-martinezDistributingCircuitsHeterogeneous2024}, which also chooses gate groups in the process of partitioning. Since we would like to partition qubits over time, but also consider multi-gate teleportation, we group gates in advance of partitioning, by choosing a subset of all possible gate groups that are merged into hyper-edges before partitioning. Grouped gates may end up as local or non-local after the partitioning. Thus, we partition the graph with prior knowledge of how gates will be grouped together if they are non-local, allowing us to efficiently determine the resulting e-bit cost. The limitation of this is that we must choose a strategy for grouping gates beforehand, which may turn out to be sub-optimal for the resulting partitions. We conjecture that this is still more effective than the reverse procedure, in which we choose the partition first, which may turn out to be sub-optimal for the best resulting set of gate teleportations. We note that it is only because we are using symmetric two-qubit gates that multiple grouping options exist -- the choice arises from the freedom in choosing which qubit is the root of the gate teleportation. We elaborate on this in the following \cref{sec:GG}.

To ensure that the partitioning objective directly corresponds to the e-bit cost, we need to design a unique cost function for the problem. This is also described in \cref{sec:GG}, ultimately resulting in \cref{eq:cost_GCP}.

\subsubsection{Gate grouping}\label{sec:GG}

Gate grouping requires conditions for teleportation compatibility, as described in \cref{sec:NLGC}, as well as a rule for choosing which gates should be added to which groups where there are multiple possibilities. Multiple gates can be considered a teleportation-compatible group if:

\begin{enumerate}
    \item Two-qubit gates share a common control qubit.
    \item Gates are contiguous on the common control qubit.
    \item Single-qubit gates on the common control are \textit{diagonal} or \textit{anti-diagonal}.
\end{enumerate}

\begin{figure}
  \centering
  \includegraphics[width=\linewidth]{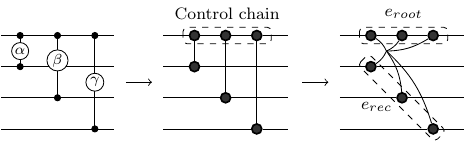}
  \caption{Identification of a compatible gate group. The $CP(\theta)$ (left) gates share a common control qubit, are contiguous on the control qubit, and there are no incompatible single-qubit gates on the control qubit between the two-qubit gates. The circuit is first converted to a temporal graph (middle), then the compatible-edges are grouped into a hyper-edge (right).}
  \label{fig:grouping_example}
\end{figure}

The common control qubit is assigned as the ``root'' qubit of the group, while the other qubits are assigned as ``receiver'' qubits. Recall that, since $CP(\theta)$ gates are symmetric, both qubits can be considered to be a control, making gate teleportation possible in both directions. This means that either qubit can serve as the root. However, once we have assigned an edge to a group as a receiver, we do not consider other possible groups rooted on that edge. This means that adding edges to groups is order-dependent, and we must define a strategy for choosing which group to add an edge to when multiple options exist. Note that this choice only arises due to the symmetric two-qubit gates. If instead we used an asymmetric two-qubit gate such as $CX$, the starting process can only be rooted on the control qubit. This means that maximising the size of the gate groups is trivial and can be achieved by merging all contiguous control chains together until there are no more possibilities. So while using symmetric gates increases the number of possible groupings, it also adds a new degree of freedom into the problem that requires a strategy to resolve.

To define the groups, we use the routine described in \cref{alg:group}. In essence, the algorithm works by scanning the graph from time step $t=0$ to $d$, identifying potential groupings, and adding identified gates to the largest existing group with which it is compatible. If instead we used the asymmetric $CX$ gate, there would only be one compatible group (rooted on the control qubit), so this choice would not arise. When encountering an incompatible single-qubit gate (non-diagonal) on the root of a group, the corresponding group must be closed. For each identified group, we replace all gate edges with a hyper-edge object identified by the \textit{first root} node (i.e., the node $v_{i}^{(t)}$, where $t$ is the time step of the first gate in the group).

The hyper-edge, in this case, is defined by two sets of nodes referred to as the \textit{root set} and the \textit{receiver set}. 

Formally, for each edge
\begin{equation}\label{eq:hyper-edge}
  e = e_{root} \cup e_{rec},
\end{equation}
where $e_{root}, e_{rec} \subset V$. This distinction makes the hyper-edges akin to \textit{directed} hyper-edges, where the root set plays the role of the ``tail'', and the receiver set plays the role of the ``head'' \cite{ausielloDirectedHypergraphsIntroduction2017}.

The root set contains each node associated with the root qubit from the time step of the first gate in the group to the time step of the last gate in the group, while the receiver set consists of all gate-like edges stemming from the root control node. Defined this way $|e_{root}| \geq |e_{rec}|$ for all groups, since the partition assignment of the root at each time step may affect the cost of the group, while for the receivers we are only concerned with the assignment at the time step of the gate. We show an example of the hyper-edge construction in \cref{fig:grouping_example}.

\begin{algorithm}[!htbp]
\caption{\textup{\textsc{GreedyGrouping}}}
\label{alg:group}
\KwIn{$H = (V,E; \tau, \kappa)$: Temporal hypergraph, with time map $\tau : V \to \{1,\ldots,d\}$ and qubit map $\kappa : V \to Q_L$.}
\KwOut{$H' = (V,E')$: A new hypergraph with merged hyper-edges.}

\BlankLine

$H' \gets H$, \ $E' \gets E$\;

\BlankLine
\tcp{Initialize prospective groups.}
\ForEach{$\ell \in Q_L$}{
  $R_\ell \gets \emptyset$,\quad $S_\ell \gets \emptyset$\;
  $\mathcal{G}[\ell] \gets (R_\ell, S_\ell)$\;
}

\BlankLine

\For{$t' \gets 1$ \KwTo $d$}{

  $V_{t'} \gets \{ v \in V \mid t(v) = t' \}$\;
  $E_{g, t'} \gets \{ e \in E_g \mid  t(u)=t(v)=t' \,\}$\;

  \BlankLine
  \tcp{2-qubit gates at time $t'$}
  \ForEach{$e=\{u,v\} \in E_{g,t'}$}{
    $\ell_u \gets q(u)$,\quad $\ell_v \gets q(v)$\;
    $E' \gets E' \setminus e$\;

    \BlankLine

    $s_u \gets |R_{\ell_u} \cup S_{\ell_u}|$\;
    $s_v \gets |R_{\ell_v} \cup S_{\ell_v}|$\;

    \uIf{$s_u \ge s_v$}{
      $S_{\ell_u} \gets S_{\ell_u} \cup \{v\}$\;
    }
    \Else{
      $S_{\ell_v} \gets S_{\ell_v} \cup \{u\}$\;
    }

    \BlankLine
    \tcp{Clear incompatible groups}
    \ForEach{$\ell \in Q_L$}{
      \If{$\neg \textup{\textsc{Compatible}}(\ell, \ell_u, \ell_v)$}{
        $R_\ell \gets \emptyset$\;
        $S_\ell \gets \emptyset$\;
      }
    }
  }

  \BlankLine
  \tcp{Single-qubit or idle nodes}
  $V^{(1)}_{t'} \gets \{\, v \in V_{t'} \mid v \text{ not in any } e \in E^{(2)}_{t'} \,\}$\;

  \ForEach{$v \in V^{(1)}_{t'}$}{
    $(\theta_v, \phi_v, \lambda_v) \gets$ parameters of $v$\;
    $\ell \gets q(v)$\;

    \uIf(\tcp*[f]{Non-diagonal: terminate group}){$\theta_v \notin \{0,\pi,2\pi\}$}{

      \If{$|S_\ell| > 0$}{
        $e' \gets (R_\ell, S_\ell)$\;
        $E' \gets E' \cup e'$ to $E'$\;
      }

      $R_\ell \gets \emptyset$\;
      $S_\ell \gets \emptyset$\;
    }
  }
}

\Return $(V,E')$\;

\end{algorithm}

\begin{figure*}[ht]
  \centering
  \begin{subfigure}{0.45\textwidth}
    \centering  
    \includegraphics[width=\columnwidth]{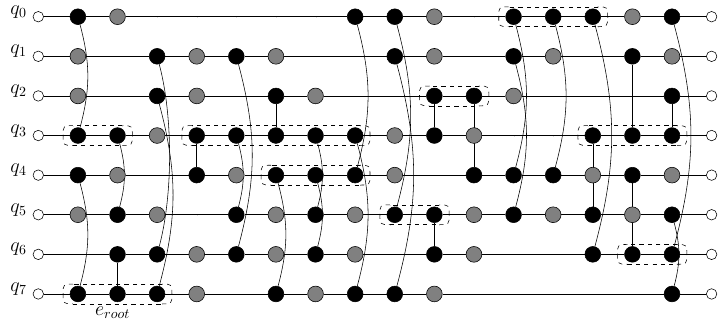}
    \caption{}
    \label{fig:CP5_8_merge}
  \end{subfigure}
  ~
  \begin{subfigure}{0.45\textwidth} 
    \centering
    \includegraphics[width=\columnwidth]{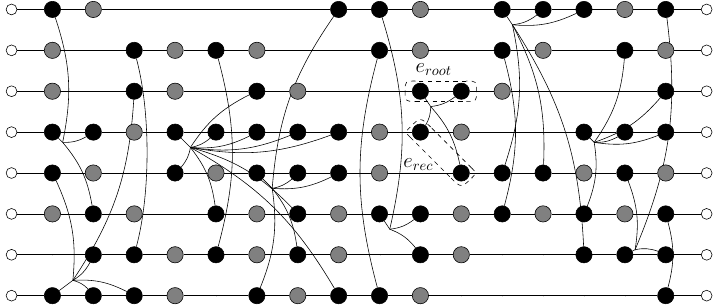}
    \caption{}
    \label{fig:CP5_8_grouped}
  \end{subfigure}
  \caption{Illustration of grouping gate edges into hyper-edges. \cref{fig:CP5_8_merge} shows the root sets identified using \cref{alg:group}. The root set is the set of nodes corresponding to the root qubit of the group, i.e. the qubit which will be linked to receiving QPUs via the starting process. \cref{fig:CP5_8_grouped} shows the resulting hyper-edges after grouping. Each hyper-edge contains a root set and a receiver set.}
  \label{fig:CP5_grouping}
\end{figure*}

The definition of the group in terms of the root and receiver set provides a simple mechanism for calculating the cost of a partitioned hyper-edge, despite the nodes representing different points in time. To see this we must first define an assignment function $\Phi : V \to Q$ mapping nodes to the set of QPUs, thus partitioning the nodes into disjoint subsets which, for the $i^{\text{th}}$ QPU, we may call $P_{i}$. The cost of each hyper-edge must correspond to the entanglement cost associated with covering all gates in the group given the assignment of the qubits. If the root qubit remains in its starting QPU for the duration of the group, then this corresponds to the standard hyper-cut, given by one less than the number of partitions spanned by the hyper-edge. However, if a state edge between two nodes in the root set is cut, this indicates a nested state teleportation. For each partition $k$ which appears in the set $\{\Phi(v) \mid v \in e_{rec}\}$, a starting process must be performed. However, for each partition in $\{\Phi(u)\mid u \in e_{root}\}$, we have a corresponding ending process routed to $\Phi(u)$ which collapses the starting process to $\Phi(u)$. If we have only one partition in the root set, then we simply collapse the state back to the root qubit at the end of the group, completing a standard multi-gate teleportation. For each additional partition in the root set, we collapse the state by re-routing the ending process to this partition. The e-bit cost is thus accounted for by the cut state-edge, rather than the hyper-edge.  

When the group is complete, the state of the root qubit will be located at $\Phi(u')$, where $u'$ corresponds to the final root node in the group. This allows us to define the hyper-edge cost as the length of the set of partitions in the receiver set that are not present in the root set, i.e. the number of partitions that are not accounted for by nested state teleportation. We write this as
\begin{equation}\label{eq:cost}
  c_{e}(\Phi) = | \{\Phi(v) \mid v \in e_{rec} \} \setminus \{\Phi(u) \mid u \in e_{root}\}|.
\end{equation}
Since the standard state and gate edges are special cases of hyper-edges where the distinction between root and receiver sets is arbitrary, we define all edges this way, ensuring that state and gate edges contain one node in each of the root and receiver sets. For state edges, we use the convention that the root set contains the node at time $t$ and the receiver set contains the node at time $t+1$. For symmetric gate edges, we assign the qubit with the lower index as the root. This allows us to use \cref{eq:cost} to calculate the cost of all edges in the graph.

This allows us to define the partitioning objective
\begin{equation}\label{eq:cost_GCP}
    \min_{\Phi} \sum_{e \in E} c_{e}(\Phi),
\end{equation}
that we refer to as the \textit{entanglement cost}, since it is designed to correspond to the number of e-bits required to partition the circuit. 

\cref{eq:cost_GCP} can be written in a number of different forms -- the form presented is appropriate under the assumption of all-to-all connectivity. In the case of limited connectivity, the cost function must be decomposed such that each contribution is scaled by a path dependent function which describes the auxiliary entanglement requirements for entanglement swapping and purification. We restrict to all-to-all connectivity as a base for demonstrating the capability of the framework, though we handle the more complex case in follow-up work, where we instead optimise over trees connecting root to receiver QPUs in the network \cite{burtEntanglementEfficientDistributionQuantum2025}. 

The constraints on $\Phi$ are determined by the qubit capacity of the QPUs. Since we have a set of nodes for each time step, we call the subset of nodes in partition $i$ at time $t$ $P_{i}^{(t)}$. Each $P_{i}^{(t)}$ should be a subset of $V^{(t)}$, giving us a sequence of partitions covering each $t$. This allows us to define the balance constraints according to the data qubit capacity of each QPU, which cannot be exceeded at any time step of the circuit.
\begin{equation}\label{eq:constraints}
  |P_{i}^{(t)}| \leq |Q_{i}|\text{,   } \forall i \in \{1,...,K\}\text{,   } 
  \\ \forall t \in \{1,...,d\},
\end{equation}
where $K = |Q|$ is the number of QPUs, corresponding to the number of partitions. In other words, there is a set of balance constraints to satisfy. For each time step $t$, there are $K$ constraints, one for each QPU.
\begin{figure}[ht]
  \centering
  \includegraphics[width=\columnwidth]{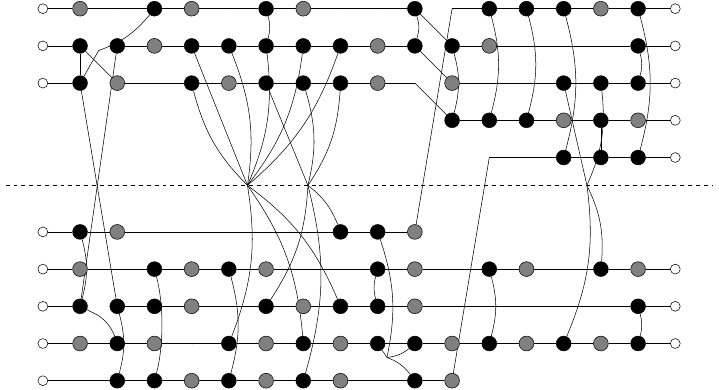}
  \caption{Graph from \cref{fig:CP5_8_grouped} after partitioning. For visual clarity, we return all local contributions to hyper-edges to their original gate edges, such that only non-trivial hyper-edges are shown. Note that some hyper-edge contributions which still appear to be local indicate that there has been a nested state teleportation. For such an edge, a gate teleportation procedure is started and then converted to a state teleportation after the final gate in the group.}
  \label{fig:CP5_grouped_FM}
\end{figure}

\subsection{Fiduccia-Mattheyses}

The Fiduccia-Mattheyses algorithm is an efficient, widely studied graph partitioning heuristic \cite{fiducciaLinearTimeHeuristicImproving1982}, which has a time complexity that is linear in the number of nodes and can effectively handle more complex structures such as hypergraphs. The algorithm leverages efficient data structures for storing and updating the net cost of moving nodes between partitions, avoiding redundant calculations. Balance constraints can be straightforwardly enforced by avoiding moves that violate the conditions or using a tolerance. Given an initial bipartitioning of the nodes, a single pass of the algorithm is executed as follows:

\begin{enumerate}
    \item Calculate the cost of moving each node to the other partition.
    \item Sort all moves into \textit{gain buckets}, where each bucket contains all moves resulting in a particular gain or loss in total cost.
    \item Choose a node from the highest (lowest) gain bucket and move it to the other partition.
    \item Remove the chosen node from the bucket and update the gains of all neighbouring nodes, moving them to their new buckets.
    \item Repeat steps (3-4) until no nodes are left in the bucket.
    \item Roll back to the partitioning at the iteration of maximum (minimum) net gain.
\end{enumerate}

The choice of maximising or minimising gain is a matter of convention, depending on whether gain is defined as a reduction or increase in cost. For the rest of this work, we use the negative gain definition, defining gain as the cost after the move minus the cost before the move.

\newcommand{\Cost}{\texttt{Cost}}
\newcommand{\ComputeGains}{\texttt{ComputeGains}}
\newcommand{\BuildGainBuckets}{\texttt{BuildGainBuckets}}
\newcommand{\BestMove}{\texttt{BestMove}}
\newcommand{\MoveNode}{\texttt{MoveNode}}
\newcommand{\UpdateGains}{\texttt{UpdateGains}}

\begin{algorithm*}[!htbp]
\caption{Fiduccia--Mattheyses for temporal hypergraphs}
\label{alg:FM-GCP}
\DontPrintSemicolon

\SetKwProg{FnFMPass}{FM}{:}{end}

\KwIn{\\
  $H = (V,E;\tau,\kappa)$: Temporal hypergraph extracted from a quantum circuit.\\
  $K$: Number of partitions (QPUs) \\
  $\Phi : V \to \{1,\ldots,K\}$: Initial partition assignment \\
  $\mathrm{Cap}[k]$: max qubit capacity in each partition $k$ \\
  \textnormal{Costs}: precomputed cost table for hyper-edge configurations
}
\KwOut{\\ 
Updated partition assignment $\widetilde{\Phi}$ with reduced entanglement cost $\tilde{c}$.}

\BlankLine


\FnFMPass{$(H,K,\Phi,\textup{Cap},\textup{Costs})$}{
  $\tilde{c} \gets \Cost(H,\Phi,\textnormal{Costs})$\;
  $\mathcal{X} \gets \emptyset$\tcp*{Lock set of already-moved nodes}
  $\Gamma \gets \ComputeGains(H,\Phi,K,\textnormal{Costs})$\tcp*{$\Gamma(v,p)$: gain of moving $v$ to partition $p$}
  $\mathcal{B} \gets \BuildGainBuckets(\Gamma)$\tcp*{Gain bucket structure built from $\Gamma$}

  \BlankLine

  $g_{cum} \gets 0$\tcp*{Cumulative gain}
  $\mathbf{g} \gets (g_{cum})$\tcp*{Sequence of cumulative gains}
  $\boldsymbol{\Phi} \gets (\Phi)$\tcp*{Sequence of assignments}

  \BlankLine

  \While{$|\mathcal{X}| < |V|$}{
    $(v^{\star},p^{\star}) \gets \BestMove(\mathcal{B},\mathcal{X},\mathrm{Capacity},\Phi)$\tcp*{Best admissible move}
    \If{$(v^{\star},p^{\star}) = \bot$}{
      \textbf{break}\tcp*{No valid moves remain in this pass}
    }

    $g \gets \Gamma(v^{\star},p^{\star})$\;
    $g_{cum} \gets g_{cum} + g$\;
    $\Phi \gets \MoveNode(\Phi, v^{\star}, p^{\star})$\tcp*{Update assignment by moving $v^\star$ to partition $p^\star$}
    $\mathbf{g} \gets \mathbf{g} \,\Vert\, (g_{cum})$\tcp*{Append current cumulative gain and assignment to the sequences}
    $\boldsymbol{\Phi} \gets \boldsymbol{\Phi} \,\Vert\, (\Phi)$\;

    $\mathcal{X} \gets \mathcal{X} \cup \{v^{\star}\}$\;

    \UpdateGains($H,\Gamma,\mathcal{B},v^{\star},p^{\star},\Phi,\textnormal{Costs}$)\tcp*{Update gain structure}
  }

  \BlankLine
  \tcp{Roll back to the best iteration in this pass (min cumulative gain)}
  $\textup{bestIter} \gets \arg\min_i \mathbf{g}_i$\;
  $\Phi \gets \boldsymbol{\Phi}_{bestIter}$\;
  $g_{min} \gets \mathbf{g}_{\textup{bestIter}}$\;
  
  \If{$g_{min} < 0$}{
    $\widetilde{\Phi} \gets \Phi$\;
    $\tilde{c} \gets \tilde{c} + g_{min}$\;
  }
\Return $\widetilde{\Phi}$\;
}
\end{algorithm*}

The pass is repeatedly executed up until no more reduction is achieved or a pass limit is reached. Storing the gains of all node exchanges and only updating the neighbours results in a time complexity $\mathcal{O}(n)$ which is linear in the number of nodes $n$, provided the gain updates are performed in constant time. This relies on the number of neighbours of each node being bounded by a constant. When dealing with problems requiring $k$ partitions, the algorithm can be modified by calculating gains for moving each node to each external partition, such that the number of node exchanges queried increases from $n$ to $n(k-1)$, resulting in a time complexity $\mathcal{O}(kn)$.

\section{Quantum Circuit Partitioning with Fiduccia-Mattheyses}\label{sec:GCP_FM}

The fundamental differences between the quantum circuit case and standard balanced min-cut graph partitioning lie in the balance constraints and the objective. Since FM extends naturally to hypergraphs and handles balance constraints easily, it is a strong candidate to be adapted for the temporal hypergraph construction. In particular, temporal graphs give balance constraints that are defined by the data qubit capacity of each QPU, which must be satisfied for each time step of the circuit, resulting in $d$ separate balance constraints per graph. 

The objective also differs: the cost function for quantum circuits is designed to minimise e-bit use, so the cost of moving nodes between partitions must be defined in terms of its effect on \cref{eq:cost_GCP}. As a result, although the overall structure of the FM algorithm remains largely unchanged, new conditions are required for enforcing balance, and the computation of move costs must be handled differently. Below we describe how move costs can be computed efficiently for the quantum circuit partitioning problem.

First we define $A(v) \subset E$ as the adjacency set of node $v$, containing all edges and hyper-edges in which $v$ is present. Contrast this with the set of neighbouring nodes, $N(v) \subset V$, which is the set of all nodes present in the edges in $A(v)$. Each edge in $A(v)$ has a contribution to the cost under assignment $\Phi(v)$. When a node is moved, $\Phi(v)$ updates to $\Phi(v)'$. The change in cost for each $e$  in $A(v)$ can be calculated simply as 
\begin{equation}
  \delta_{e}(\Phi,\Phi') = c_{e}(\Phi') - c_{e}(\Phi).
\end{equation}
We refer to this as the \textit{gain contribution} from $e$, adhering to the convention that a negative gain decreases the cost. The total \textit{gain} of moving node $v$ is thus the sum of the gain contributions from all adjacent edges:
\begin{equation}\label{eq:gain}
    g_{v}(\Phi,\Phi') = \sum_{e \in A(v)} \delta_{e}(\Phi,\Phi').
\end{equation}
Equally important is the change in gain after a particular node has been moved, since this must be computed for all neighbours $N(v)$ of the moved node $v$ after the action is taken. This is referred to as the \textit{delta gain} $\Delta g_{u,v}$, for node $u$ from moving node $v$. To retain the time complexity of FM, the delta gain must be computed in constant time and should avoid recomputing any auxiliary values where possible. To do this, we pre-compute the cost of all possible cost configurations of a hyper-edge. Following the approach of Huang and Khang \cite{huangPartitioningbasedStandardcellGlobal1997}, we denote the \textit{configuration} of a set of nodes $A$ assigned to partitions by $\Phi$ as a binary string of length $K = |Q|$:
\begin{equation}
    cfg^{(A)}_{i}(\Phi)= 
\begin{cases}
        1 & \text{if } \exists v \in A : \Phi(v) = i \\
        0 & \text{otherwise}
\end{cases}
\end{equation} 
where a $1$ at element $i$ indicates the presence of at least one node in $P_{i}$. Many different assignments may map to the same configuration, though the cost is completely determined by the configuration. Since we are interested in the e-bit count for each hyper-edge, we use the configurations of the root and receiver set to define the overall configuration of the edge. Entry $i$ of the configuration will be $1$ if there is at least one receiver node in $i$ and no root node in $i$. We can define this using the bitwise expression:
\begin{equation}\label{eq:conf_calc}
    cfg^{(e)}_{i}(\Phi) = cfg^{(e_{rec})}_{i}(\Phi) \wedge \neg cfg^{(e_{root})}_{i}(\Phi).
\end{equation}
The associated cost is the bit count, i.e., the number of ones, in the final configuration. There are $2^{K}-1$ possible configurations, and each configuration string identifies with an integer between $0$ and $2^{K}-1$. The cost of each configuration $c_{e}(\Phi)$ can be pre-computed and stored at the associated index of a vector $C$, such that it can be called in $\mathcal{O}(1)$ time. This is feasible for moderately sized $K$ ($\leq 25$) architectures, though for future large scale-systems this may not be possible. In such cases, the cost can be computed on-the-fly from the configuration string in $\mathcal{O}(K)$ time and cached for future use. In practice, many configurations will not appear in the partitioning process and each cost will only need to be computed once, making caching a more practical alternative in general. However, this is required for ensuring efficent delta gain calculations, which is only guaranteed if all configurations can be queried in constant time.

To efficiently calculate the delta gain, we need to store the configuration of the edge and determine the change in configuration after an action is made. Since any node move has a unique source and destination, finding the new configuration requires changing a maximum of two elements of the configuration, so can be done in constant time. The nature of the gate grouping routine results in each node being part of at most \textit{four} edges, since each node is part of at most two state edges and two gate edges. This is because a root node in a particular group may be part of the receiver set of another group. In this case, each node can be involved in a maximum of four edges and a minimum of one edge (note that, when the multilevel framework is introduced in \cref{sec:MLCP}, this will not be the case).

With this restriction, calculating $\Delta g_{u,v}(\Phi, \Phi', \tilde{\Phi}, \tilde{\Phi}')$ requires calculating the contribution from each edge adjacent to $u$, each of which requires \textit{at most} four configurations \cite{huangPartitioningbasedStandardcellGlobal1997} for each edge, giving a maximum of 16 configurations. To see this, consider node $v$ being moved from $P_{i}$ to $P_{j}$, updating $\Phi$ to $\Phi'$. Before the move, the action of moving node $u$ from $P_{k}$ to $P_{l}$ is $g_{u}(\Phi, \tilde{\Phi})$. Calling the new gain after both moves have been made $g_{u}(\Phi', \tilde{\Phi}')$, we note that the delta gain must obey the relation
\begin{equation}\label{eq:gain_delta_gain}
    g_{u}(\Phi', \tilde{\Phi}') = g_{u}(\Phi, \tilde{\Phi}) + \Delta g_{u,v}(\Phi, \Phi', \tilde{\Phi}, \tilde{\Phi}'),
\end{equation}
where $\tilde{\Phi}'$ corresponds to the assignment after both moves have been made. \cref{eq:gain_delta_gain} implies
\begin{align}\label{eq:delta_gain}
  &\Delta g_{u,v}(\Phi, \Phi', \tilde{\Phi}, \tilde{\Phi}') = g_{u}(\Phi', \tilde{\Phi}') - g_{u}(\Phi, \tilde{\Phi}) \nonumber \\
  &= \sum_{e \in A(u) \cap A(v)}[ \delta_{e}(\Phi',\tilde{\Phi}') -  \delta_{e}(\Phi,\tilde{\Phi})] \nonumber \\
  &= \sum_{e \in A(u) \cap A(v)}[ c_{e}(\Phi') - c_{e}(\tilde{\Phi}') -  c_{e}(\Phi) + c_{e}(\tilde{\Phi})].
\end{align}
Each assignment edge pair corresponds to a particular configuration, thus if a node saturates the maximum of $4$ edges, $16$ configurations must be queried. Additionally, since there exist edges in $A(u)$ that may not have been affected by moving node $v$, we need only consider contributions from edges that are adjacent to both $u$ and $v$. Thus, for each edge $e$ in $A(v)$, we update each node comprising $e$ using only the contributions from $e$. If, for any node $u$, there is more than one common edge between $u$ and $v$ i.e., $|A(u)\cap A(v)| > 1$,  then each contribution will be added separately to the delta gain. This must be repeated for each neighbouring node of the moved node $v$, for each destination.

To perform each gain update in $\mathcal{O}(1)$ time, we must store the configuration of each edge and the cost of each configuration. In addition, we need to store two auxiliary objects for each edge, which we call the \textit{root counts} and the \textit{receiver counts}. Each of these is simply a vector of length $|P|$, where each element corresponds to the number of nodes from the root/receiver set in a particular partition. Each move will increment either the root or the receiver counts by removing one from the source element and adding one to the destination element. The full configuration can be updated directly from the root and receiver counts, checking whether each of the counts has changed from zero to non-zero, or vice-versa. The cost can then be read from the pre-computed cost table.

Storage and retrieval of the counts and configurations of each edge adds an overhead, and for small graphs it may be quicker to directly compute the edge costs to find the delta gain. However, it is necessary for maintaining the time-complexity of the full algorithm, thus the benefits are clear for larger circuit sizes. The full algorithm logic is given in \cref{alg:FM-GCP}. Since much of the heavy lifting is done in the auxiliary routines for calculating gains and delta gains, the main logic remains similar to standard FM. We include pseudocode for initialisation routines, auxiliary routines and delta gain calculation in \cref{sec:fm_aux}, namely \cref{alg:InitFMGCP,alg:AuxFMGCP,alg:DeltaGainConfig}.

\section{Multilevel quantum circuit partitioning}\label{sec:MLCP}

In this section, we introduce multilevel partitioning, a powerful technique for improving partitioning quality on large-scale problems. We show how this can be applied to quantum circuit partitioning using temporal coarsening strategies that exploit the fixed temporal structure of the graphs.  
\subsection{Multilevel partitioning}

The multilevel paradigm is a crucial part of the success of state-of-the-art methods for large-scale graph partitioning, as demonstrated by partitioners such as METIS \cite{karypisMETISSoftwarePackage1997, karypisFastHighQuality1998} and KaHyPar \cite{schlagHighQualityHypergraphPartitioning2023}. While heuristic refinement methods, such as FM, have proved effective for small to medium problem sizes, the solution quality typically drops as problem sizes increase. This is expected as the solution space expands exponentially, making local search heuristics increasingly less effective at exploring the solution space. Multilevel methods employ \textit{coarsening} routines for transforming large-scale, complex problems into smaller problems, for which good results can be obtained. This is typically followed by iterative \textit{uncoarsening} and \textit{refinement} of the results into a solution for the larger problem. 

\begin{figure}[ht]
  \centering
  \includegraphics{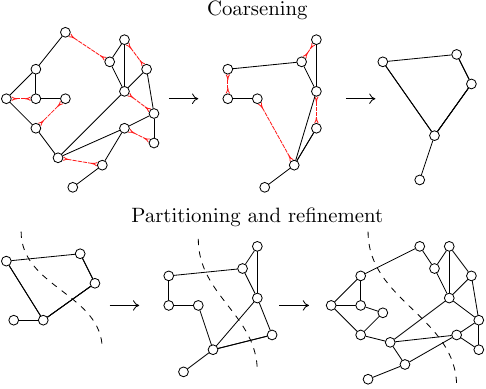}
  \caption{Illustration of multilevel partitioning for graphs. The graph is first coarsened by iteratively merging nodes together and contracting edges (red) between them. The coarsest graph is partitioned, and the solution is then uncoarsened level-by-level, applying heuristic refinement at each level to improve the solution.}
  \label{fig:MLCP_illustration}
\end{figure}

A graph can be coarsened by merging nodes together and contracting the edges between them. A good multilevel partitioner relies on effective coarsening, as this performs part of the heavy lifting of exploring the solution space. During uncoarsening, the solution from the previous level is used as a starting point for refinement at the new level. Applying a local search heuristic between each level allows the solution to be gradually improved as more degrees of freedom are reintroduced into the problem. An illustration of a basic multilevel procedure is shown in \cref{fig:MLCP_illustration}.

We describe our coarsening strategy in \cref{sec:TC}, followed by the partitioning and refinement procedure in \cref{sec:PR}. An illustration of the full multilevel procedure applied to temporal hypergraphs is found in \cref{fig:MLCP_illustration}. 

\subsection{Temporal coarsening}\label{sec:TC}

Many multilevel partitioners will employ coarsening routines which are agnostic to the graph structure, such as random or heavy-edge matching \cite{karypisFastHighQuality1998}. Alternatively, structural properties can be extracted and exploited using community-aware coarsening \cite{schlagHighQualityHypergraphPartitioning2023}. Since our hypergraphs have a fixed, temporal, structure, we propose coarsening strategies that exploit this.

Since each qubit is expanded into $d$ nodes, each associated with a particular time step, we propose \textit{temporal coarsening}. For each node $v$ such that $\tau(v) = t$ and $\kappa(v) = q$, the graph will always contain a corresponding node $u$ such that $\tau(u) = t+1$ and $\kappa(u) = q$, unless $t = d$, i.e., the final time step. This means that we can always merge nodes associated with the same qubit at adjacent time steps. Moreover, since each time step has exactly one node per qubit, we can merge full time steps, or intervals of time steps, while still ensuring that each merged node corresponds to a unique qubit. This makes the process of enforcing balance constraints straightforward, since there is no need to check which qubits are represented by merged nodes. 

The coarsening phase gives a sequence of graphs with decreasing temporal resolution, where the coarsest graph may have only a single time step. In this limit, we recover a \textit{static} partitioning problem, and the hyper-edge costs reduce to standard hyper-cuts, leaving no possibilities for state teleportation. The uncoarsening phase then gradually reintroduces temporal resolution, allowing state teleportation to be be considered at increasingly many time steps.

\begin{algorithm}[ht]
\caption{Single-layer time contraction of a temporal hypergraph}
\label{alg:contract}

\KwIn{\\
  $H = (V,E; \tau, \kappa)$: Temporal hypergraph, with time map $\tau : V \to \{1,\ldots,d\}$ and qubit map $\kappa : V \to Q_L$. \\
  $s$: The source time-layer to be merged. \\
  $t$: The target time-layer into which the nodes at $s$ will be merged.
}
\KwOut{
  A coarsened hypergraph $H'$ in which all nodes at time-layer $s$ have been merged into $t$. 
}

\SetKwProg{FuncContractTime}{ContractTime}{:}{end}
\FuncContractTime{$(H,s,t)$}{
  $H' \gets H$\;
  $V^{({s})} \gets \{\,v \in V \mid \tau(v) = s\}$\;
  $V^{({t})} \gets \{\,v \in V \mid \tau(v) = t\}$\;

  \ForEach{$v \in V^{(\mathrm{s})}$}{
    $u \gets w \in V : \kappa(w) = \kappa(v)$\;
    \tcp{Contract self-loop}
    $E \gets E \setminus (v,u)$\;
    \ForEach{$e \in E$ \textbf{such that} $v \in e$}{
      \If{$u \in e$}{
        $e \gets e \setminus v$\;
      }
      \ElseIf{}{
        $e \gets e \setminus v$\;
        $e \gets e \cup u$\;
      }
    }
  }
  \tcp{Remove original s-layer nodes}
  $V \gets V \setminus V^{(\mathrm{s})}$
  \Return $H'$\;
}

\end{algorithm}

After applying some initial partitioning, the graph is then uncoarsened by a single-level, using the solution from the previous level as a starting solution. When uncoarsening, we can transform the partitioning solution from the previous level into a starting solution for the new level by assigning each newly revealed node to the same partition as the merged node from the previous level. We can do this by storing a contraction history, which tells us which nodes have been merged between each level. Between phases of uncoarsening, the solution may be refined using the FM algorithm, or some other search-based heuristic, since each level introduces new degrees of freedom into the problem. This is repeated until the full temporal resolution is restored. 

Even restricting ourselves to temporal coarsening, there are still many different possible coarsening strategies, which differ in how time steps are merged together and how many merges are performed between each level. The only constraint that we place on the coarsening is full time steps of nodes must be merged in a single operation, which ensures that we always have the same number of nodes per time step as qubits.

We explore three different temporal basic coarsening strategies, referred to as \textit{window}, \textit{block} and \textit{recursive} coarsening. While each strategy has its advantages and disadvantages, we find the recursive coarsening to perform best overall, so we include the pseudocode \cref{alg:coarsen-recursive} in the main text. The other two strategies are included in \cref{app:tc_strats} for completeness (\cref{alg:coarsen-window,alg:coarsen-blocks}).

\subsubsection{Window coarsening}\label{sec:WC}

\begin{figure}[ht]
  \includegraphics[width=\linewidth]{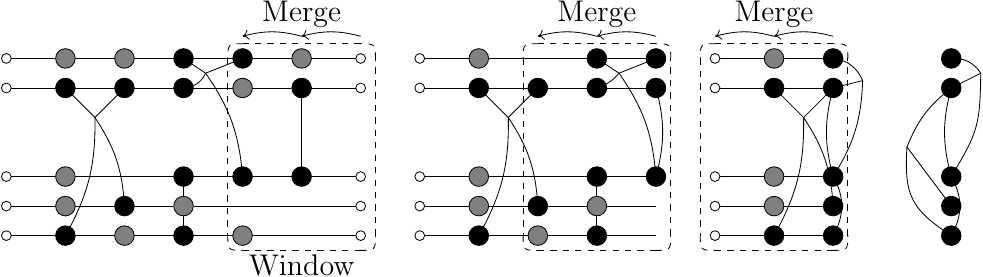}
  \caption{Illustration of the window-based coarsening procedure. A window size of $3$ is used, such that three time steps are contracted at each level.}
  \label{fig:window_coarse}
\end{figure}

The most basic uncoarsening procedure can be implemented by contracting time-intervals one-by-one from $d$ down to $0$. When uncoarsening, this is analogous to revealing the full temporal resolution one step at a time. At any given level, everything below a certain time step $t$ is fully resolved, while everything above $t$ is fully contracted. We still have information about the full graph, but this remains coarse for future time steps. We can generalise this procedure by revealing an interval of time steps at each level, as if sliding a window along the time-axis. Calling the size of each interval $w$, we can coarsen the graph by successively merging nodes in each interval into a single node, such that the number of time steps is reduced by $w$ at each level. This is repeated until only one time step remains, where the final interval may be smaller than $w$. After each interval has been contracted, a copy of the graph is stored for the refinement. This is referred to as \textit{window} coarsening. If we set the interval size to $w = 1$, then we reveal the graph one time step at a time. If instead we input the number of levels, we can calculate the required interval sizes as $w = d / n_{levels}$, and adjust for any remainder. The full window coarsening algorithm is given in \cref{alg:coarsen-window} in \cref{app:tc_strats} and illustrated in \cref{fig:window_coarse}.

\subsubsection{Block coarsening}\label{sec:BC}

\begin{figure}[ht]
  \includegraphics[width=\linewidth]{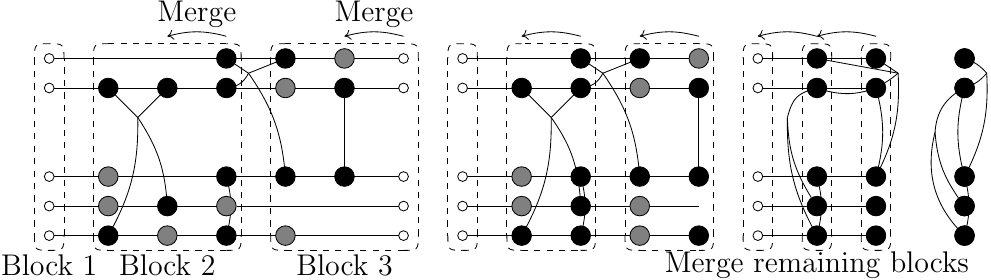}
  \caption{Illustration of the block-based coarsening procedure. A maximimum block size of $3$ is used. At each level, one time step is contracted in each block. Optionally, the final level may contract all remaining time steps into a single time step.}
  \label{fig:block_coarse}
\end{figure}

We may also coarsen the graph into blocks of nodes by performing window coarsening simultaneously on different regions of the graph. We call this block coarsening, since each region is coarsened down into a single time block. In this case, the size of each block $b$ corresponds to the number of levels, since we uncoarsen one time step per block at each level. We may also choose to coarsen all blocks down to a single time step. We can either set the number of blocks and calculate the block size from $b = d / n_{blocks}$ or choose the block size directly. The pseudocode is given in \cref{alg:coarsen-blocks} in \cref{app:tc_strats}. An illustration is shown in \cref{fig:block_coarse}, where the blocks are coarsened simultaneously, and the remaining time steps are merged into a single time step at the final level.

\subsubsection{Recursive coarsening}\label{sec:RC}

The final method considered is referred to as \textit{recursive} coarsening. We divide the graph into adjacent pairs of time steps then merge the nodes between them at each level. The number of nodes thus decreases exponentially with each level, making the number of levels automatically $\log_{2}(d)$, unless we choose to cap it earlier. Pseudocode for recursive coarsening is given in \cref{alg:coarsen-recursive}.

\begin{figure}[ht]
  \includegraphics[width=\linewidth]{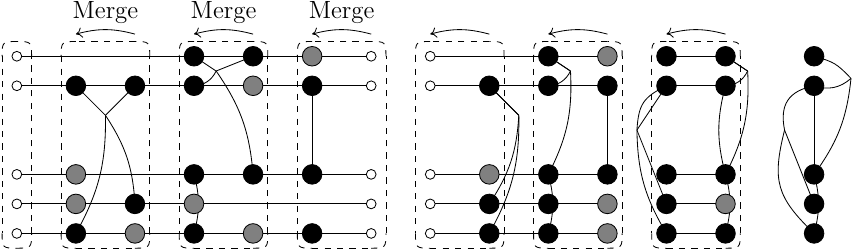}
  \caption{Illustration of the recursive coarsening procedure.}
  \label{fig:recursive_coarse}
\end{figure}

  
  
  
  
  
  
  

\begin{algorithm}[ht]
  \caption{Recursive coarsening}
  \label{alg:coarsen-recursive}
  
  \SetKwProg{FnCoarsenRecursive}{CoarsenRecursive}{:}{end}
  \SetKwFunction{FnContractTime}{ContractTime}
  
  \KwIn{\\
    $H_0 = (V_0,E_0,\tau,\kappa)$: temporal hypergraph with time-layers $1,\dots,d$.
  }
  \KwOut{\\
    Coarsening hierarchy $\mathcal{H} = (H_0,H_1,\dots,H_K)$.
  }
  
  \BlankLine
  
  \FnCoarsenRecursive{$(H_0)$}{
    $H \gets H_0$\;
    $\mathcal{H} \gets (H_0)$\;
    
    \While{$|\{\tau(v) : v \in V(H)\}|$ $> 1$}{
      \tcp{Sorted list of distinct time-layers present in $H$}
      $L \gets (\ell_1,\ell_2,\dots,\ell_m)$ with $\ell_1 < \ell_2 < \dots < \ell_m$\;
      
      
      \For{$i \gets 1$ \KwTo $m-1$ \textbf{\textup{step}} $2$}{
      
        $H \gets$ \FnContractTime{$H, \ell_{i+1}, \ell_i$}\;
      }
      
      $\mathcal{H} \gets \mathcal{H} \,\Vert\, H$\;
    }
    
    \Return $\mathcal{H}$\;
  }
\end{algorithm}

\begin{figure*}[htpb]
  \includegraphics[width=\linewidth]{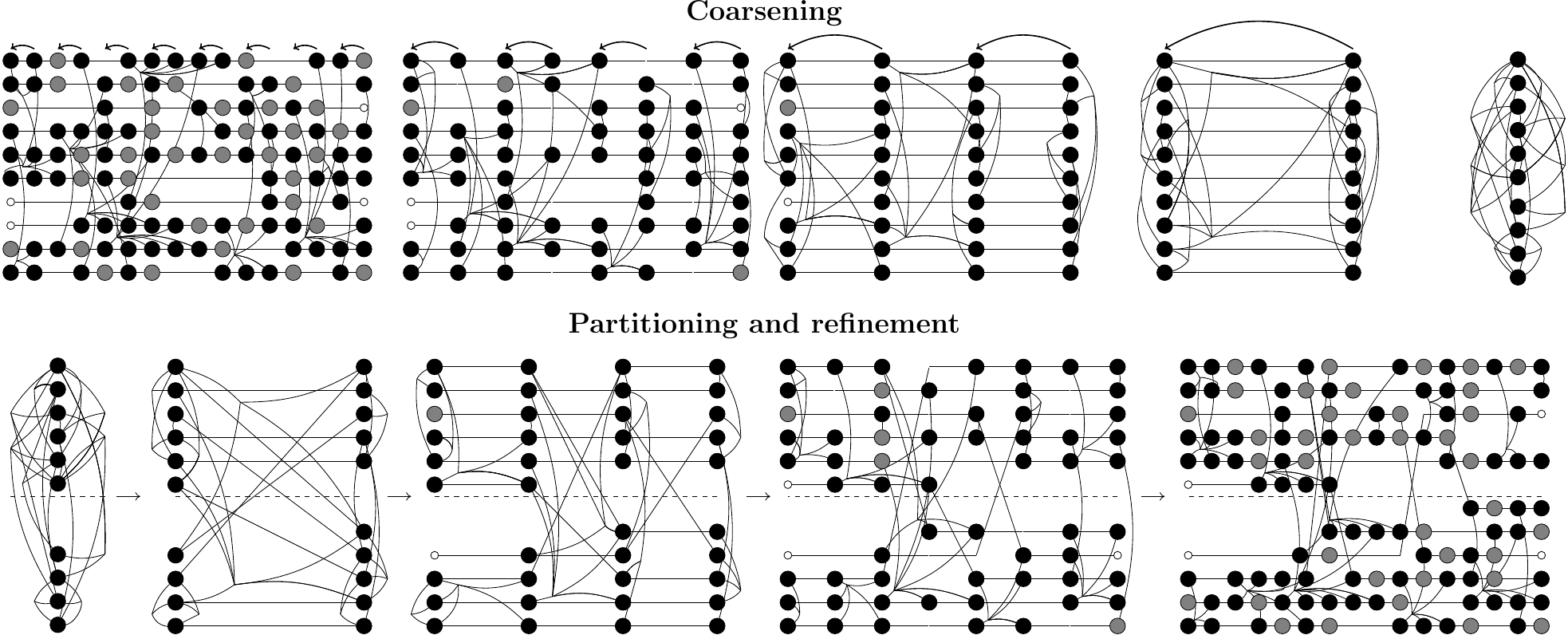}
  \caption{Full multilevel partitioning procedure using recursive coarsening. In the top sequence, the sequence of coarsened graphs is shown, starting from the full temporal resolution graph on the left, down to the coarsest graph on the right. Arrows indicate time steps being merged between levels. On the bottom, the partitioning and refinement procedure is illustrated. An initial partitioning is performed on the coarsest graph, which gives a single partition assignment for each qubit for the time of the full circuit. In the second level, the graph now splits each qubit into two time steps, corresponding to the first and second half of the circuit. Each time step is split into two at each new level, introducing new degrees of freedom for the partitioning. Between each level, FM refinement is able to improve the cost of the partitioning.}
  \label{fig:ml_part_full}
\end{figure*}

\subsection{Partitioning and refinement}\label{sec:PR}

The full multilevel partitioning procedure is illustrated in \cref{fig:ml_part_full}, using recursive coarsening. After coarsening the graph into a series of levels, we perform an initial partitioning on the coarsest graph. Though it is common in multilevel schemes to use a different partitioning algorithm we found that it was sufficient to start with a greedy initialisation followed by FM refinement. The greedy initialisation simply fills each partition up to capacity, iterating through the numerical indices of the qubits. This is followed by a number of FM passes. Between each uncoarsening phase, the solution must be transformed into a starting solution for the next level. This is straightforward, since merged nodes do not mix qubits, allowing us to simply expand the number of time steps by assigning each newly revealed node to the same partition as the node with which it was previously merged.

\section{Circuit extraction}\label{sec:circ_extr}

The final step in the workflow is to extract the circuit from the partitioned graph and the assignment function. Since the partitioned circuit will require entanglement distribution, we need to have a number of ancillary communication qubits in each QPU. If we have not used a gate grouping pass, communication qubit requirements are modest, since they are released immediately after each state and gate teleportation. However, if we have used gate grouping, there is likely to be a requirement for multiple communication qubits to be active at the same time, demanding larger communication qubit capacities. We choose to start with a predefined number of communication qubits, though we allow the possibility of dynamically adding communication qubits as needed. While this may not always be supported by physical architectures, spare data qubits may also be considered as pseudo communication qubits for links that are kept active over many time steps. Pseudo communication qubits may not be able to generate e-bits, but they can store states linked from starting processes to free up additional communication qubits. If this is not possible, then we may have to collapse and end the link with an early ending process. The previous teleportation procedure may then be restarted, at the cost of an additional e-bit. Naturally, this method will increase the depth of the resulting circuit and may introduce additional noise. We explore this trade-off further in \cref{sec:gg_res}. Assuming we have sufficient communication qubit capacity, we proceed as follows.

\begin{figure*}[htpb]
  \centering
  \includegraphics[width=0.78\linewidth]{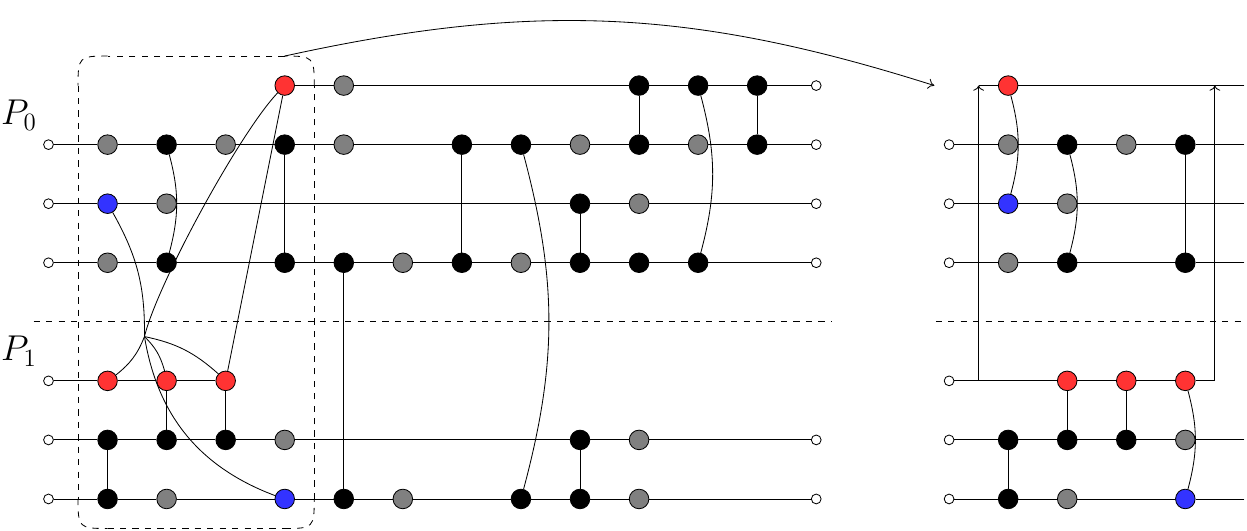}
  \caption{Partitioned graph with nested state teleportation. In the left graph, the highlighted hyper-edge has four root nodes (red) and two receiver nodes (blue). Both the receiver nodes and the root nodes span across $P_0$ and $P_1$, indicating the presence of a nested state teleportation. According to \cref{eq:cost}, the hyper-edge cost is $0$. The subgraph on the right shows how the edge configuration is converted to teleportation primitives. A starting process is performed from $P_1$ to $P_0$, but since the  final root node is assigned to $P_1$, the ending process is used to teleport the state to $P_1$. The `cost' is accounted for by the cut state-edge. We show the full extracted circuit in \cref{fig:nst_circ}.}
  \label{fig:nst_graph}
\end{figure*}

\begin{figure*}[htpb]
  \begin{subfigure}{\linewidth}
    \centering
      \includegraphics[width=0.7\linewidth]{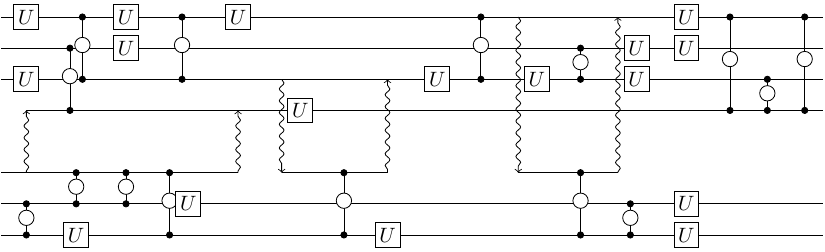}
      \caption{Partially extracted circuit}
      \label{fig:circ_partial}
  \end{subfigure}
  \begin{subfigure}{\linewidth}
    \centering
    \includegraphics[width=1.0\linewidth]{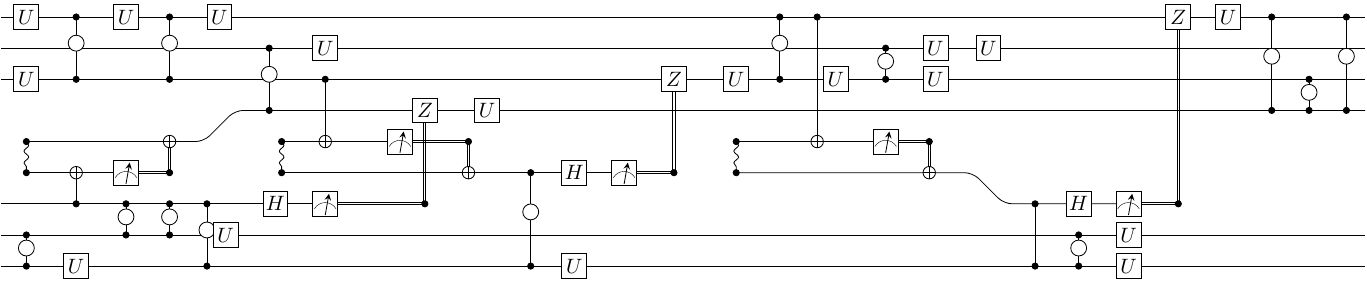}
    \caption{Fully extracted circuit}
    \label{fig:circ_full}
  \end{subfigure}
  \caption{Circuit extracted from the partitioned graph in \cref{fig:nst_graph}. In (a), the first stage of extraction is shown, where the hyper-edges are converted to the corresponding starting and ending processes. In (b) the starting and ending processes are decomposed down to standard gates, completing the circuit extraction. The result can then be passed into further hardware-specific compilation, provided the entanglement generation operations are left untouched. For simplicity, we omit the specific phase angle labels from the $CP$ gates. }
  \label{fig:nst_circ}

\end{figure*}

We start by creating a $d$-dimensional, empty list $L$, representing the time steps of the original circuit. We then pass through the gate edges and hyper-edges in the graph in order of time, appending operations to the corresponding entry in the list. For each cut state-edge, we schedule a state teleportation. For each gate edge, including hyper-edges, we check which partitions are spanned by both the root and receiver sets. If it is fully local, we schedule all gates as normal. If we have non-local components, we schedule a starting process for all partitions apparent in both the root and the receiver sets, and schedule all gates to their local copy of the root qubit. Since we scheduled starting processes across the root set as well as the receiver set, this means that the starting process for any nested state teleportations are already included. This means that we need to track when any linked communication qubits are no longer needed, and then schedule the corresponding ending process. If we encounter a cut-state edge while the root qubit is still active, we delay the ending process until all non-local gates in the partition have been executed, as shown in \cref{fig:nst_graph}. If we encounter an anti-diagonal single-qubit gate that occurs on the root qubit of an active group, we apply an $X$ gate to each linked communication qubit, as well as the root, as described in \cref{sec:multi-gate}. This laves us with a partially extracted circuit, where all hyper-edges have been converted to starting and ending processes, but not yet decomposed into standard gates. An example of a partially extracted circuit is shown in \cref{fig:circ_partial}.

The starting processes and ending processes are then decomposed into the sub-circuits in \cref{fig:starting_ending}, adding in the necessary e-bit generation and classically controlled operations. This gives the fully extracted circuit, as shown in \cref{fig:circ_full}.

The above procedure is sufficient for extracting a circuit with at the desired e-bit cost, though may still require further compilation within each sub-circuit to match the desired gate set or connectivity requirements. In future work, we aim to identify further avenues for optimisation, such as gate commutation within groups and detached gates, in which both controls of a symmetric two-qubit gate are linked to an external partition. 

An implementation of the circuit extraction is available in the \texttt{disqco} repository \cite{FelixburtDISQCO}. 

To ensure that the circuit extraction has succeeded, we can simulate the distributed circuit and the original circuit and compare the measurement probabilities. Within \texttt{disqco}, the extracted circuit is a Qiskit circuit object, containing custom blocks for entanglement generation and a classical register for communication. This can be simulated within Qiskit, or converted to a QASM string for use in other simulators. We use the Qiskit sampler primitive \cite{javadi-abhariQuantumComputingQiskit2024} to perform the simulation. In \cref{fig:dist_circ} we compare the output distribution of a random 8-qubit circuit with the output distribution of the circuit extracted from the partitioned graph. This also allows us to compare the depth of the original circuit with the extracted circuit, as well as the demand for communication qubits. Typically, if gate grouping is used, the resulting depth is reduced, since the entanglement requirements are lower, while the demand for communication qubits is higher. The extracted circuit can then be subject to further compilation passes for intra-QPU routing, gate decomposition or \textit{fault-tolerant compilation}, which we discuss briefly in \cref{sec:future_work}.

\begin{figure}[htpb]
  \centering
  \usepgfplotslibrary{statistics}
\begin{tikzpicture}
\begin{axis}[
    ybar, 
    xlabel = {Outcome}, 
    ylabel={Probability},
    xmin = 0,
    xmax = 63,
    ymin = 0,
    ymax = 0.12,
    xtick = {0,31,63},
    width = 0.9\linewidth,
    scaled y ticks = base 10:2,
]

\addplot+ [x = decimal,
y = probability,
  fill=red!30,
  draw=red,
  bar width=0.5,
  bar shift= 0.01
] table{
decimal probability
0 0.0087890625
1 0.00244140625
2 0.0029296875
3 0.004150390625
4 0.036865234375
5 0.004150390625
6 0.006591796875
7 0.00537109375
8 0.013427734375
9 0.05517578125
10 0.01220703125
11 0.05517578125
12 0.0888671875
13 0.040771484375
14 0.04248046875
15 0.058349609375
22 0.000244140625
25 0.000244140625
27 0.001708984375
30 0.0009765625
31 0.00244140625
32 0.000244140625
33 0.011474609375
34 0.00048828125
35 0.002685546875
36 0.00439453125
37 0.009521484375
39 0.004150390625
40 0.05517578125
41 0.0380859375
42 0.02392578125
43 0.051025390625
44 0.142333984375
45 0.064697265625
46 0.076904296875
47 0.064208984375
52 0.000244140625
53 0.000244140625
55 0.000244140625
56 0.00048828125
58 0.000732421875
59 0.001220703125
62 0.00146484375
63 0.002685546875
};

\addplot+ [x = decimal,
y = probability,
  fill=blue!30,
  draw=blue,
  bar width=0.5,
  bar shift= -0.01] table{
decimal probability
0 0.008056640625
1 0.001708984375
2 0.001953125
3 0.005615234375
4 0.037353515625
5 0.0029296875
6 0.006591796875
7 0.005126953125
8 0.01904296875
9 0.05029296875
10 0.006591796875
11 0.049560546875
12 0.094482421875
13 0.04931640625
14 0.042236328125
15 0.05859375
18 0.000244140625
27 0.00048828125
28 0.000244140625
29 0.000244140625
30 0.000732421875
31 0.000244140625
32 0.0009765625
33 0.009521484375
34 0.00048828125
35 0.001953125
36 0.006591796875
37 0.006591796875
39 0.00341796875
40 0.065185546875
41 0.0322265625
42 0.0234375
43 0.058349609375
44 0.13525390625
45 0.066162109375
46 0.07275390625
47 0.06884765625
50 0.000244140625
55 0.000244140625
58 0.000244140625
59 0.00146484375
61 0.000244140625
62 0.001953125
63 0.002197265625
};

\end{axis}
\end{tikzpicture}
  \caption{Comparison of the output distribution of the original circuit with that of the partitioned circuit, generated using the Qiskit Sampler. The partitioned graph is shown in \cref{fig:nst_graph} and the output circuit in \cref{fig:nst_circ}. The output distribution of the partitioned circuit matches the original when looking at the reduced subspace of only data qubits.}
  \label{fig:dist_circ}
\end{figure}

\section{Complexity analysis and optimisations}\label{sec:imp}

\subsection{Complexity of FM for quantum circuits}\label{sec:complexity}

The time complexity of the original FM algorithm, for bipartitioning, is linear in $n$ where $n = |V|$ is the number of nodes in the graph. This relies on the assumption that the gain updates can be performed in constant time and the degree (number of neighbours) of each node is bounded. Each pass requires the gain of each node to be queried once when initialising the buckets, for time $\mathcal{O}(n)$, after which each node is moved once, updating the gains of all neighbours. This means the gains of the neighbours must be updated $n$ times per pass. If there is no bound on the number of neighbours, this results in a time complexity $\mathcal{O}(n^{2})$, since each node has a maximum of $n-1$ neighbours. However, in many realistic scenarios, the maximum number of neighbours is bounded by a constant that is much smaller than the total number of nodes in the graph. If we have such a bound on the number of neighbours, and delta gain calculations are $\mathcal{O}(1)$ for each of these neighbours, leaving a time complexity $\mathcal{O}(n)$ for both the initialisation and the main loop, giving $\mathcal{O}(n)$ overall. When extending to multi-partitioning, the complexity increases linearly with the number of partitions, since the cost of moving each node must be calculated for each external partition. This makes the pass complexity $\mathcal{O}(kn)$, where $k$ is the number of partitions.

While the algorithm structure is similar for quantum circuit graphs, the number of nodes is in fact equal to the number of qubits $n_q$ times the depth of the circuit $d$, such that bipartitioning has complexity $\mathcal{O}(n_qd)$, making the pass complexity for $k$-partitioning $\mathcal{O}(kn_qd)$. \cref{sec:GCP} shows how to perform the gain updates in constant time for the temporal hypergraph, provided the costs of all hyper-edge configurations are pre-computed. We note that the complexity of the pre-computation is exponential in the number of partitions, since there are exponentially many possible edge configurations that must be pre-computed. However, since $k$ is typically much smaller than $n_q$, this has a negligible effect up to $20$ partitions. Scaling beyond this requires costs to be calculated on the fly and cached for future lookup. This is a more practical strategy even below $20$ partitions, though for large $k$ this cost calculation starts to become significant. We briefly highlight strategies for handling this in \cref{sec:precomp} and point the reader to the follow-up work for more details \cite{burtEntanglementEfficientDistributionQuantum2025}. 

Without gate grouping, the number of neighbours is bounded by $3$, so the time complexity is maintained. However, after gate grouping, the upper bound of the number of neighbours scales with $d$, since this is the maximum number of gates that could exist in a group. For most circuit types, the groups will be far from this threshold. However, circuit structures that lead to very large groups do exist. An example is the QFT, which contains control chains which span the depth of the circuit. Such circuits are more computationally intensive to optimise, though they typically result in a low proportion of e-bits to gates, as can be seen in the QFT result in \cref{fig:QFT_24}. As a result, they tend to converge to a solution earlier, requiring fewer passes to achieve a good result. The pass complexity, in this worst case, is $\mathcal{O}(kn_qd^{2})$. If a constant number of passes is used, the total complexity is $\mathcal{O}(kn_qd^{2})$.

\subsection{Improving efficiency of FM}\label{sec:efficiency}
When dealing with large circuits, it may be necessary to look for adaptations to increase the efficiency. Firstly, we need not move every node in each pass for large graphs, since typically the iteration of best gain is achieved relatively early (after $10-20\%$ of all nodes have been moved) in the pass. We can, therefore, limit the number of nodes that are moved in a pass. The limitation is that we become less likely to escape a local minimum. There are various possibilities for remedying this -- we choose to alternate between exploratory and exploitative passes. Exploitative passes are the normal passes defined in \cref{alg:FM-GCP}. Exploratory passes, on the other hand, still seek the best gain with each mvoe but \textit{do not} roll-back to the iteration of best cumulative gain at the end of the pass. As a result, the cost often increases after an exploratory pass, but tends to decrease on average. We find this to be effective provided we are limiting the number of moves to a small ($\simeq 0.1 n_qd$) subset of nodes, preventing exploratory passes from increasing the cost too much. We compare the effect of using exploration vs only exploiting in \cref{fig:explore}. Clearly, while the cost changes much more rapidly in exploratory runs, the minimum value is often significantly lower, despite the fact that fewer nodes are actually moved.

\begin{figure}
  \includegraphics[width=\linewidth]{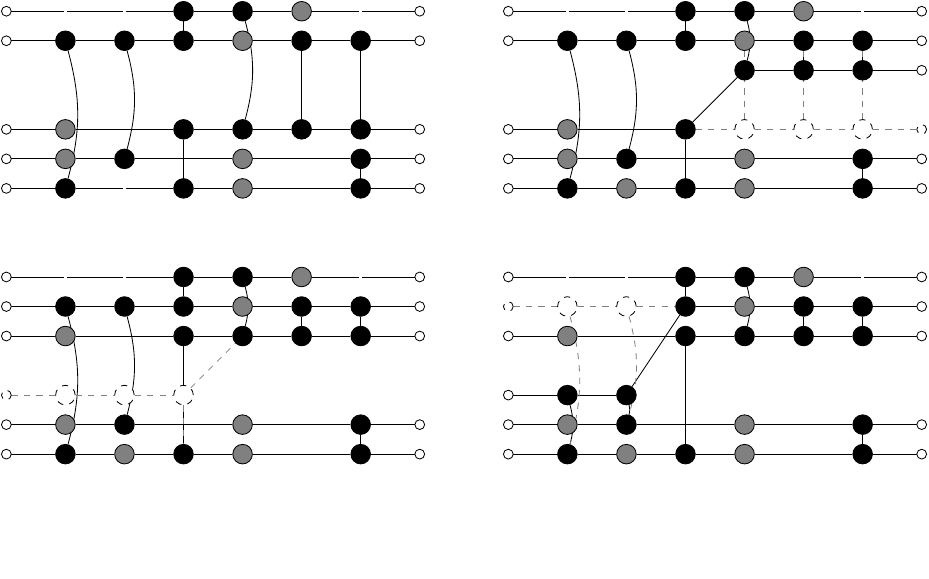}
  \caption{Illustration of a local minimum in the circuit optimization process. In the top left graph, there is no local action changing the assignment which improves the cut. A sequence of $4$ neighbouring nodes moved results in the top right graph. Another $4$ nodes are moved to achieve the bottom left. Finally, now that space has been made, $3$ more nodes can be moved to achieve the final partitioning at the bottom right, with just two cut edges.}
  \label{fig:loc_min}
\end{figure}

\subsection{Avoiding local minima}\label{sec:local_min}

Part of the reason the exploratory passes are effective is that they help to avoid local minima. In all cases, we start with a \textit{static} partitioning, meaning that $\Phi(v_{i}^{(t)}) = \Phi(v_{i}^{(1)})$ for all $t \in \{2,3,..., d\}$. In other words, all qubits remain in the same partition for the full duration of the circuit. The static assignment is a good starting point, since it ensures  all $n_{q}(d-1)$ state-edges are local. The downside is that this will always be a local minimum, since each node has two local state edges (except for nodes $v_{i}^{(1)}$ and $v_{i}^{(d)}$, which have one state edge but no gate edges) and a maximum of one gate edge, since no nodes have been merged. As a result, the best local action can only increase the total cost by $1$. If we consider the case after gate grouping, the same is true, since each node can be involved in a maximum of $4$ edges, two state edges and two gate edges, so in the best two state edges become non-local and two gate edges become local, resulting in no net change.

As is illustrated in \cref{fig:loc_min}, escaping such local minima often requires a long series of local moves to open up spaces for improvement. In standard FM, each series of `bad' moves can only move each node once due to the roll back at the end of the pass. This means that no move sequences that move multiple nodes more than once can be explored. With exploratory FM, however, this is possible. At the end of an exploratory pass, we continue the move sequence from the current configuration, unlocking all the nodes again for movement. In this way, we can explore long sequences of moves that may ultimately lead to a better configuration. By alternating between exploratory and exploitative passes, we tend to converge to lower costs, as shown in \cref{fig:explore}.

\subsection{Complexity of multilevel FM}\label{sec:MLFM_comp}

While we still perform partitioning at each level of the coarsening hierarchy, we gain a speed up from the fact that, at coarser levels, the number of nodes is reduced. In general, each pass has a complexity $\mathcal{O}(k|V| \max_{v \in V} |N(v)|)$, where $|V|$ is the number of nodes in the graph, and $\max_{v \in V} |N_H(v)|$ is the maximum degree of any node in the graph. When we coarsen down to a single time step, we decrease both the number of nodes in the graph from $n_q d$ to $n_q$, and the maximum degree from $\mathcal{O}(d)$ to $\mathcal{O}(n_q)$. However, the total complexity is not reduced if we still optimise at the finest level, as this level will dominate the scaling. This is demonstrated in \cref{fig:ML_comp_time}, where the slope of each of the multilevel methods steepens with each level until it reaches the finest.

There are approaches we can take to reduce the complexity of the finer levels. If the finest graph used is the original graph, and the number of levels is predefined, then there is no change to the complexity. When dealing with deep circuits, we may restrict the finest phase used to reduce the required computation. For example, if we temporally coarsen the graph down to $n_q$ nodes, we reduce the number of nodes as well as the maximum number of neighbours, which goes from $\mathcal{O}(d)$ to $\mathcal{O}(n_q)$. However, the number of edges each node is involved in, $A(v)$, has now gone from being bounded by $3$ (or $4$ after gate grouping) to being bounded by $d$. In this worst case scenario, where a node $v$ has an independent gate edge from each time step, then each of these must connect to a single other neighbour, i.e., they are all $2$-edges rather than hyper-edges. This means that for each of these, we only have one neighbour to update, for a maximum of $n_q$. Alternatively, if the node is involved in hyper-edges that extend over $d$ time steps, the same restriction applies, since each time step over which the hyper-edge spans constitutes a time step in which the node cannot be involved in another hyper-edge or gate-edge. Thus, even though we have coarsened the maximum number of neighbours to $n_q$, the maximum number of contributions to the delta gain is still $d$, so the pass complexity at the coarsest level is $\mathcal{O}(kn_qd)$. 

As we uncoarsen, we regain the factor of $d$ due to the increased number of time steps. However, due to the effectiveness of the coarser levels, we can put a firmer cap on the number of nodes moved at each pass. If we limit the number of nodes moved to $n_q$, we retain the pass complexity $\mathcal{O}(kn_qd)$, with an additional factor for the number of levels. In practice, the length of passes will still change between levels due to the initial gain computation, which scales with the number of nodes at the given level. We enforce the same number of levels for each coarsening strategy. In the recursive coarsening strategy, the resulting number of levels is logarithmic in the depth of the circuit. Thus, in all tests, we use $\log_{2}(d)$ coarsening levels. The resulting complexity of each of the multilevel methods is $\mathcal{O}(kn_qd\log_{2}(d))$. In \cref{sec:mlp_res}, we experimentally compare the speed and performance of each method. 

\section{Results} \label{sec:results}

We run several tests to benchmark the performance of the different workflows. Recalling the sequence of stages in the framework, we have \textit{transpilation}, \textit{gate grouping}, \textit{graph conversion}, \textit{coarsening}, \textit{partitioning}, \textit{uncoarsening/refinement} and \textit{circuit extraction}. Not all of these phases are varied in the benchmarking. Notably, transpilation, graph conversion and circuit extraction remain the same in all cases. We use the basic Qiskit transpiler into the gate set described in \cref{eq:U_CP}. For gate grouping, we investigate the \textit{greedy gate grouping} method, described by \cref{alg:group}, and the trivial case of no gate grouping. For coarsening, we investigate the basic FM with no coarsening, the window-based multilevel method, MLFM-W, the block-based method, MLFM-B, and the recursive method, MLFM-R. For the partitioning algorithm, we briefly compare the FM algorithm with its exploratory variant. For each of the $k$ QPUs used, we allow a qubit capacity $\lfloor n_q/k \rfloor + 1$, allowing an additional space to facilitate state teleportation. We make no assumptions on the number of communication qubits.

All tests are performed on a MacBook Pro M3 Pro with 18GB RAM and an 11-core CPU containing 6 performance cores and 5 efficiency cores, with performances cores clocked at 4.05 GHz and efficiency cores at 2.8 GHz. All code used to generate the results is available in the \texttt{disqco} repository \cite{FelixburtDISQCO}, with scripts to reproduce the results in the \texttt{benchmarking/} directory.

\subsection{Benchmark circuits}

The set of circuits to be used in benchmarking is chosen to cover a wide spectrum of circuit structures. This allows us to identify cases where existing methods perform well, and cases where they do not, while demonstrating the consistency of the proposed framework in all cases. This also allows us to draw some conclusions about which circuits are best suited to which methods of non-local gate coverage.

\subsubsection{CP-fraction}\label{sec:CPfraction}

$CP$-fraction circuits are a generalisation of the $CZ$-fraction circuits introduced by Sundaram et al. \cite{gsundaramEfficientDistributionQuantum2021}. They provide a useful tool for benchmarking the performance of circuit partitioning and distribution algorithms, since the proportion of two-qubit gates can be varied for a fixed depth and number of qubits. The $CP$-fraction variant was introduced in Ref. \cite{burtGeneralisedCircuitPartitioning2024b}, as a means of generalising to a universal gate set, since transpiled $CZ$-fraction circuits often lose depth due to commutation and cancellation between Hadamard gates and CZ gates. A $CP$-fraction circuit can be constructed as follows. For a given probability $p$, qubit count $n_q$ and depth $d$, single-qubit gates are applied to each qubit with probability $1-p$, while the remaining qubits are paired and acted on with two-qubit gates. This is repeated up to depth $d$. We use $U(\theta, \phi, \lambda)$ and $CP(\theta)$ gates with randomised parameters.

\subsubsection{Quantum Fourier transform}\label{sec:QFT}

The quantum Fourier transform (QFT) is a well-known subroutine that is used in various quantum algorithms. The prevalence of the QFT makes it an essential benchmark. The QFT can be built using only controlled-phase ($CP(\theta)$) gates and Hadamard $H$ gates. Long chains of contiguous $CP(\theta)$ gates lead to large hyper-edges, which favours methods permitting gate grouping. Furthermore, it is observed that in our framework, the lowest entanglement cost is achieved using the trivial, initial placement of qubits, indicating that the initial placement is a global optimal for this structure. We note that similar results have been found elsewhere \cite{chaModuleconditionedDistributionQuantum2025}. 

\begin{figure}[ht]
    \centering
    \begin{subfigure}{0.65\linewidth}
        \centering
        \includegraphics[width=0.95\linewidth]{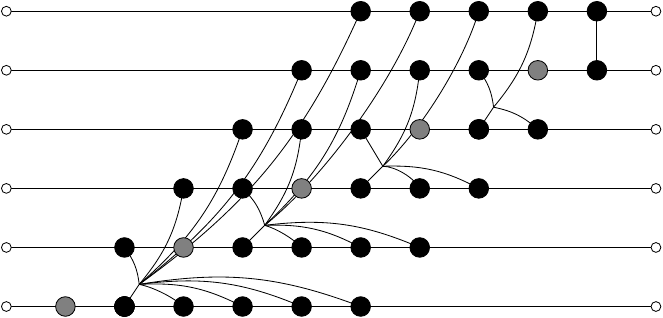}
        \caption{Unpartitioned QFT.}
        \label{fig:QFT_8_grouped}
    \end{subfigure}
    ~
    \begin{subfigure}{0.65\linewidth}
        \centering
        \includegraphics[width=0.95\linewidth]{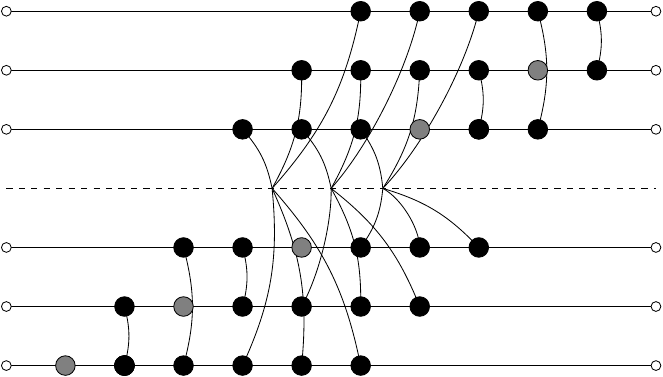}
        \caption{Partitioned QFT.}
        \label{fig:QFT_8_grouped_FM}
    \end{subfigure}
    \caption{6-qubit QFT graph after gate grouping. After partitioning, we remove all local contributions to hyper-edges. Note that, with gate grouping, the optimised result is the same as the initial placement.}
    \label{fig:enter-label}
\end{figure}

\subsubsection{Quantum volume}\label{sec:QV}

Quantum volume (QV) circuits, introduced by Cross et al. \cite{crossValidatingQuantumComputers2019}, are designed to benchmark the performance of quantum computers. QV circuits consist of layers of Haar random unitaries selected from the special unitary group $SU(4)$ applied to randomly chosen pairs of qubits. QV circuits provide a nice balance between structure and randomness, since the decomposed layers require the same qubits to interact for a number of time steps before the next layer is applied. Notably, since the unitaries are chosen randomly, decomposed two qubit-gates are unlikely to contain potential for gate grouping, since the resulting blocks do not contain contiguous $CP(\theta)$ gates, and any single-qubit gates are unlikely to be diagonal (or anti-diagonal). As a result, methods that are reliant on gate teleportation typically perform poorly, while methods employing state teleportation perform well, as observed in Ref. \cite{burtGeneralisedCircuitPartitioning2024b}. For all tests, we set the number of layers to be equal to the number of qubits.

\begin{figure}[ht]
    \centering
    \begin{subfigure}{0.8\linewidth}
        \centering
        \includegraphics[width=\linewidth]{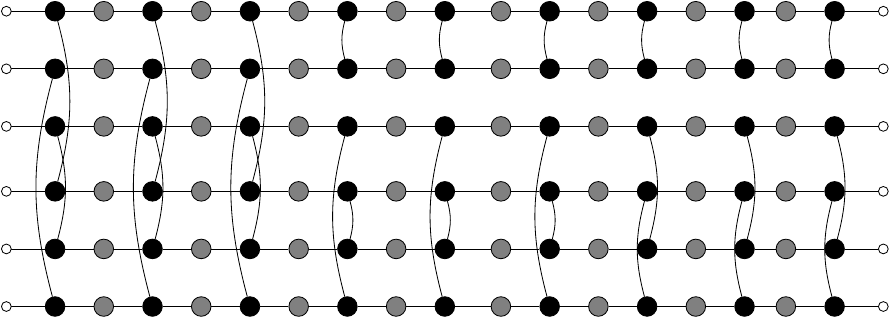}
        \caption{Unpartitioned QV.}
        \label{fig:QV_6}
    \end{subfigure}
    ~
    \begin{subfigure}{0.8\linewidth}
        \centering
        \includegraphics[width=\linewidth]{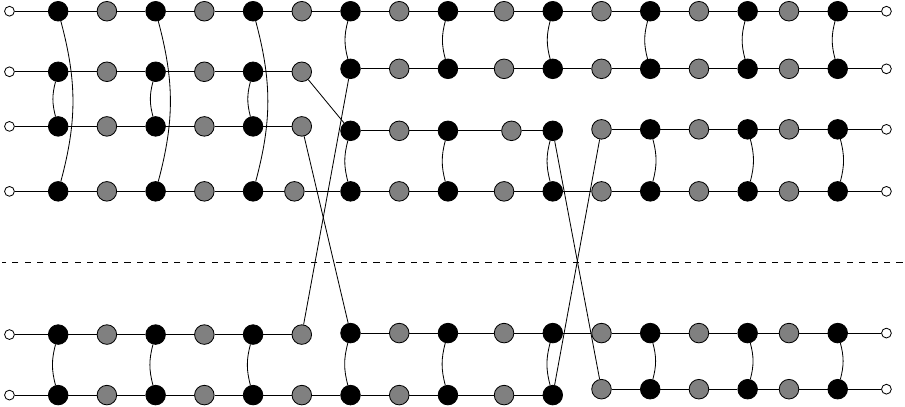}
        \caption{Partitioned QV.}
        \label{fig:QV_6_FM}
    \end{subfigure}
  \caption{Example graph from 6-qubit from QV circuit. Note that the grouping algorithm has no effect on QV circuits, since there are no contiguous $CP(\theta)$ gates and random unitaries are unlikely to decompose into diagonal or anti-diagonal single-qubit gates.}
  \label{fig:QV_graph}
\end{figure}

\subsubsection{Quantum approximate optimisation algorithm}\label{sec:QAOA}

The quantum approximate optimisation algorithm (QAOA) was introduced as a method for efficiently solving hard combinatorial optimisation problems, such as MaxCut and MaxSat \cite{farhiQuantumApproximateOptimization2014}. QAOA circuits comprise parameterised single- and two-qubit gates, for which the parameters are classically optimised to find an approximately optimal solution. We consider randomly initialised QAOA circuits for MaxCut problems over random graphs with an edge probability of 50\%. QAOA circuits, like QFT, permit large groups of gates for multi-gate teleportation.

\begin{figure}[ht]
  \centering
  \begin{subfigure}{0.8\linewidth}
      \centering
      \includegraphics[width=\linewidth]{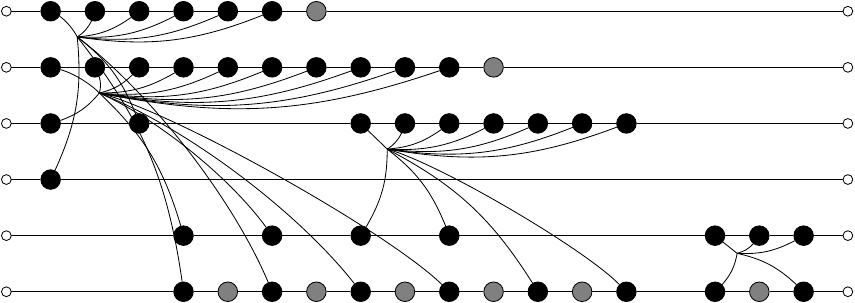}
      \label{fig:QAOA_6}
  \end{subfigure}
  ~
  \begin{subfigure}{0.8\linewidth}
      \centering
      \includegraphics[width=\linewidth]{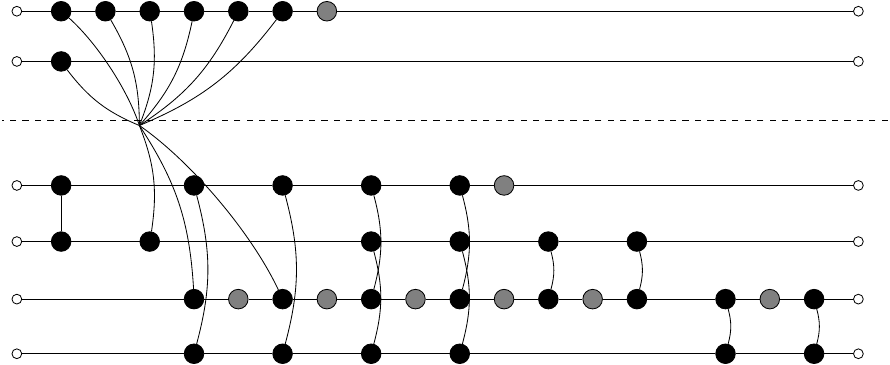}
      \caption{Partitioned QAOA.}
      \label{fig:QAOA_6_FM}
  \end{subfigure}
\caption{Example graph from 6-qubit QAOA circuit. Circuit is a randomly initialised instance of QAOA for MaxCut over a random graph with edge probability 50\%.}
\label{fig:QAOA_graph}
\end{figure}

\begin{figure*}[ht]
    \begin{subfigure}{0.3\linewidth}
        \usepgfplotslibrary{statistics}
\begin{tikzpicture}
\begin{axis}[
    ybar, 
    xlabel = {Minimum entanglement cost}, 
    ylabel={Frequency},
    xmin = 50,
    xmax = 100,
    xtick = {50,75,100},
    width = \linewidth,
]

  \addplot+ [
    hist={
      bins=20   
    },
    fill=blue!30,
    draw=blue
  ] table [y index=0] {
88
68
83
77
70
70
80
78
78
67
76
81
71
74
72
75
87
73
81
79
90
82
82
78
81
76
79
86
81
90
83
86
90
82
83
80
66
72
82
80
72
80
74
73
83
80
79
67
71
83
65
75
73
85
75
84
72
74
73
90
88
75
96
82
78
80
77
65
73
83
80
83
81
79
81
76
75
83
75
80
73
76
87
78
67
85
72
63
84
75
75
85
67
72
77
77
87
79
66
79
};

  \addplot+ [
    hist={
      bins=20   
    },
    fill=red!30,
    draw=red
  ] table [y index=0] {
78
65
69
70
66
68
73
66
66
64
73
73
63
66
65
66
70
58
70
62
71
71
75
63
70
59
77
71
80
73
72
81
77
81
74
72
60
72
67
71
68
69
67
58
69
65
73
63
74
69
66
64
73
78
62
71
69
72
64
86
73
68
74
69
64
65
71
64
62
70
71
66
63
72
69
75
62
73
67
62
70
67
71
70
62
75
64
66
74
67
65
74
64
69
76
68
84
66
61
78
};

\end{axis}
\end{tikzpicture}
        \caption{30\% CP gates}
        \label{fig:group_vs3}
    \end{subfigure}
    ~
    \begin{subfigure}{0.3\linewidth}
        \usepgfplotslibrary{statistics}
\begin{tikzpicture}
\begin{axis}[
    ybar, 
    xlabel={Minimum entanglement cost},
    ylabel={Frequency},
    xmin = 90,
    xmax = 150,
    xtick = {90,120,150},
    legend style={
      at={(0.5,1.1)},
      anchor=south,
      legend columns=-1
    },
    width = \linewidth,
]

  \addplot+ [
    hist={
      bins=20   
    },
    fill=blue!30,
    draw=blue
  ] table [y index=0] {
134
121
135
134
129
135
137
115
129
126
126
133
136
136
128
134
129
142
124
134
138
134
125
129
134
128
146
139
124
126
139
137
132
134
134
125
140
139
137
129
139
130
148
129
130
137
136
134
137
140
127
142
135
142
128
136
147
138
132
127
143
137
137
139
139
148
135
153
138
133
131
144
120
144
138
128
143
136
142
139
130
125
135
137
135
143
131
144
138
150
131
129
146
136
132
134
135
130
124
132
};

    \addlegendentry{No Grouping}

  \addplot+ [
    hist={
      bins=20   
    },
    fill=red!30,
    draw=red
  ] table [y index=0] {
121
107
111
113
113
110
114
106
107
109
106
104
112
110
101
110
108
124
95
108
100
117
109
98
108
108
115
110
107
101
109
121
112
107
113
111
107
103
111
107
116
120
120
107
108
119
109
113
120
120
106
106
105
120
110
112
110
119
114
108
127
116
115
111
108
121
117
110
114
102
112
110
107
105
112
113
119
109
109
110
105
114
107
106
118
112
112
99
113
111
110
112
108
108
110
106
110
109
105
103
};

    \addlegendentry{Grouping}

\end{axis}
\end{tikzpicture}
        \caption{50\% CP gates}
        \label{fig:group_vs5}
    \end{subfigure}
    ~
    \begin{subfigure}{0.3\linewidth}
        \usepgfplotslibrary{statistics}
\begin{tikzpicture}
\begin{axis}[
    ybar, 
    xlabel = {Minimum entanglement cost},
    ylabel = {Frequency},
    xmin = 120,
    xmax = 200,
    xtick = {120,160,200},
    width = \linewidth,
]

  \addplot+ [
    hist={
      bins=20   
    },
    fill=blue!30,
    draw=blue
  ] table [y index=0] {
188
192
202
186
177
186
192
188
194
189
189
187
190
197
181
199
193
194
180
177
180
183
197
185
202
187
185
200
186
187
180
192
190
180
185
176
178
195
188
181
196
187
189
198
193
191
191
192
190
181
185
175
199
178
199
198
189
193
182
182
191
183
186
192
198
201
197
185
174
186
192
196
192
185
195
192
188
189
184
190
202
182
177
203
193
188
192
189
189
206
194
196
190
185
187
187
186
189
189
181
};

  \addplot+ [
    hist={
      bins=20   
    },
    fill=red!30,
    draw=red
  ] table [y index=0] {
142
148
140
139
140
141
145
140
149
144
135
144
139
137
131
134
147
149
145
136
135
134
142
137
155
141
139
132
131
144
147
138
134
128
137
130
141
138
130
145
140
132
144
142
140
138
147
151
131
133
134
134
143
126
149
136
137
139
142
129
141
134
149
139
141
138
147
139
125
140
144
141
143
149
144
140
141
142
139
142
156
140
155
140
134
141
137
138
135
148
136
142
136
141
141
139
142
143
139
137
};

\end{axis}
\end{tikzpicture}
        \caption{70\% CP gates} 
        \label{fig:group_vs7}
    \end{subfigure}
    \caption{E-bit reduction from gate grouping. As the fraction of two-qubit gates increases, gates are grouped into larger hyper-edges, allowing for reduced entanglement costs via multi-gate teleportation. Plots produced from 100 runs of FM, on a 32-qubit $CP$-fraction circuit with depth 32, partitioned across 4 QPUs.}
    \label{fig:group_vs}
\end{figure*}

\begin{figure*}[ht]
    \centering
    \begin{subfigure}{0.45\textwidth}
        \begin{tikzpicture}[baseline]

\begin{axis}[
    width=\columnwidth,
    height=0.75\columnwidth,
    ymin=0,
    ymax=900,
    symbolic x coords={16, 24, 32, 40, 48}, 
    xtick=data,
    enlarge x limits=0.05,
    x tick style={draw=none},
    ylabel={Circuit depth},
    xlabel={Number of qubits},
    axis y line*=left,
    x tick label style={
        yshift=-5pt,
        xshift=9pt,
        anchor=east
    },
    grid=minor,
    minor grid style={line width=0.5pt, draw=gray!60, densely dotted},
    legend style={
        at={(0.5, 1.05)},
        anchor=south,
        legend columns=2,
        /tikz/every even column/.append style={column sep=1.5em},
        /tikz/every odd column/.append style={column sep=0.2em}
    }]

\pgfplotstableread[
  col sep=space,
  header=has colnames
]{
num_qubits	depth_ungrouped	depth_grouped	comm_qubits_ungrouped	comm_qubits_grouped
16	251.0	79.2	3.1	8.1
24	389.1	122.7	3.6	12.5
32	515.8	182.9	3.9	16.3
40	689.5	224.4	4.2	20.5
48	833.3	272.5	5.3	24.4
}\datatable

\addplot[color=blue,
fill opacity=.5,
mark= *,
mark size = 0.5pt
]
table [
  x=num_qubits,
  y=depth_ungrouped,
] {\datatable};

\addlegendentry{No grouping}

\pgfplotstableread[
  col sep=space,
  header=has colnames
]{
num_qubits	depth_ungrouped	depth_grouped	comm_qubits_ungrouped	comm_qubits_grouped
16	251.0	79.2	3.1	8.1
24	389.1	122.7	3.6	12.5
32	515.8	182.9	3.9	16.3
40	689.5	224.4	4.2	20.5
48	833.3	272.5	5.3	24.4
}\datatable
\addplot+ [
color=red,
fill opacity=.5,
mark= *,
mark size = 0.5pt
]
table [
  x=num_qubits,
  y=depth_grouped,
] {\datatable};
\addlegendentry{Grouping}

\end{axis}

\end{tikzpicture}
        \caption{Distributed circuit depth}
        \label{fig:group_vs_depth}
    \end{subfigure}
    ~
    \begin{subfigure}{0.45\textwidth}
        \begin{tikzpicture}[baseline]

\begin{axis}[
    width=\columnwidth,
    height=0.75\columnwidth,
    ymin=0,
    ymax=25,
    symbolic x coords={16, 24, 32, 40, 48}, 
    xtick=data,
    enlarge x limits=0.05,
    x tick style={draw=none},
    ylabel={Communication qubit count},
    xlabel={Number of qubits},
    axis y line*=left,
    x tick label style={
        yshift=-5pt,
        xshift=9pt,
        anchor=east
    },
    grid=minor,
    minor grid style={line width=0.5pt, draw=gray!60, densely dotted},
    legend style={
        at={(0.5, 1.05)},
        anchor=south,
        legend columns=2,
        /tikz/every even column/.append style={column sep=1.5em},
        /tikz/every odd column/.append style={column sep=0.2em}
    }]

\pgfplotstableread[
  col sep=space,
  header=has colnames
]{
num_qubits	depth_ungrouped	depth_grouped	comm_qubits_ungrouped	comm_qubits_grouped
16	251.0	79.2	3.1	8.1
24	389.1	122.7	3.6	12.5
32	515.8	182.9	3.9	16.3
40	689.5	224.4	4.2	20.5
48	833.3	272.5	5.3	24.4
}\datatable

\addplot[color=blue,
fill opacity=.5,
mark= *,
mark size = 0.5pt
]
table [
  x=num_qubits,
  y=comm_qubits_ungrouped,
] {\datatable};

\addlegendentry{No grouping}

\pgfplotstableread[
  col sep=space,
  header=has colnames
]{
num_qubits	depth_ungrouped	depth_grouped	comm_qubits_ungrouped	comm_qubits_grouped
16	251.0	79.2	3.1	8.1
24	389.1	122.7	3.6	12.5
32	515.8	182.9	3.9	16.3
40	689.5	224.4	4.2	20.5
48	833.3	272.5	5.3	24.4
}\datatable
\addplot+ [
color=red,
fill opacity=.5,
mark= *,
mark size = 0.5pt
]
table [
  x=num_qubits,
  y=comm_qubits_grouped,
] {\datatable};
\addlegendentry{Grouping}

\end{axis}

\end{tikzpicture}
        \caption{Communication qubit requirements}
        \label{fig:group_vs_comm}
    \end{subfigure}
    \caption{Comparison of circuit depth and communication qubit requirements with and without gate grouping. The mean results are shown from 10 runs of FM on QFT circuits, each partitioned over 4 QPUs. The circuit depth is significantly reduced when using gate grouping, though this is compensated by an increase in the number of communication qubits required.}
    \label{fig:group_vs_results}
\end{figure*}

\subsubsection{QASM benchmark suite}\label{sec:QASM}

In addition to the circuits described above, we test on a number of circuits provided by the QASM benchmark suite \cite{liQASMBenchLowlevelQASM2022}. We test on all circuits in the large category up to $100$ qubits and $1000$ time steps after transpilation, excluding QFT and QV since they are tested separately.

\begin{figure*}[ht]
    \centering
    \begin{subfigure}{0.4\textwidth}
      \centering
      \definecolor{one}{HTML}{c23b23}
\definecolor{two}{HTML}{f39a27}
\definecolor{three}{HTML}{eada52}
\definecolor{four}{HTML}{03c03c}

\begin{tikzpicture}[baseline]

\pgfplotstableread[
col sep=space,
header=has colnames
]{
x cost cost_expl
0 137 136
1 136 134
2 135 135
3 135 131
4 135 144
5 135 134
6 135 132
7 135 125
8 130 137
9 130 126
10 130 127
11 130 122
12 130 131
13 130 122
14 130 125
15 130 118
16 130 125
17 130 118
18 129 124
19 127 118
20 124 116
21 123 114
22 123 126
23 122 118
24 122 125
25 121 114
26 120 120
27 120 112
28 120 118
29 119 111
30 119 116
31 118 110
32 118 119
33 118 110
34 118 116
35 118 110
36 118 111
37 118 106
38 118 115
39 118 108
40 118 111
41 118 107
42 118 115
43 118 110
44 117 113
45 116 106
46 115 108
47 115 103
48 115 112
49 115 103
50 115 106
51 115 104
52 115 114
53 115 106
54 115 108
55 115 104
56 115 109
57 115 105
58 115 110
59 115 104
60 115 112
61 115 106
62 115 111
63 115 105
64 115 110
65 115 107
66 115 111
67 115 106
68 115 107
69 115 104
70 115 108
71 115 104
72 115 114
73 115 106
74 114 116
75 113 105
76 112 112
77 112 104
78 112 111
79 112 106
80 112 106
81 112 104
82 112 109
83 112 104
84 112 109
85 112 105
86 112 111
87 112 106
88 112 108
89 112 104
90 112 105
91 112 104
92 112 116
93 112 104
94 112 114
95 112 107
96 112 115
97 112 107
98 112 113
99 112 106
}\datatable

\begin{axis}[
    width=\columnwidth,
    height=0.75\columnwidth,
    ymin=100,
    ymax=140,
    xmin = 0,
    xmax = 100,
    xtick={0,20,40,60,80,100},
    enlarge x limits=0.05,
    x tick style={draw=none},
    ylabel={Entanglement cost},
    xlabel={Pass},
    axis y line*=left,
    x tick label style={
        yshift=-5pt,
        xshift=9pt,
        anchor=east
    },
    grid=minor,
    minor grid style={line width=0.5pt, draw=gray!60, densely dotted},
    legend style={
        at={(0.5, 1.07)},
        anchor=south,
        legend columns=2,
        /tikz/every even column/.append style={column sep=1.5em},
        /tikz/every odd column/.append style={column sep=0.2em}
    }
]

\addplot[color=red,
fill opacity=.5,
mark = *,
mark size = 0.5pt
]
table [
  x=x,
  y=cost,
] {\datatable};

\addlegendentry{Standard}

\addplot[
  color=blue,
  fill opacity=.5,
  mark = *,
  mark size = 0.5pt
]
table [
  x=x,
  y=cost_expl,
] {\datatable};

\addlegendentry{Exploratory}

\end{axis}

\end{tikzpicture}
      \caption{Cost vs. pass number}
      \label{fig:explore}
    \end{subfigure}
    ~
    \begin{subfigure}{0.4\textwidth}
      \centering
      \begin{tikzpicture}[baseline]

\pgfplotstableread[
col sep=space,
header=has colnames
]{
x y1 y2
1.0 66.10378813743591 197.0
0.95 66.5785698890686 184.0
0.9 51.374672174453735 168.0
0.85 41.00987768173218 167.0
0.8 33.98721718788147 171.0
0.75 28.602674961090088 148.0
0.7 24.609210729599 155.0
0.65 20.44732904434204 145.0
0.6 16.88066267967224 147.0
0.55 13.584961891174316 140.0
0.5 11.352663040161133 137.0
0.45 8.64962387084961 136.0
0.4 6.9262309074401855 132.0
0.35 5.601810932159424 119.0
0.3 4.563945055007935 120.0
0.25 3.4726991653442383 113.0
0.2 2.6330032348632812 115.0
0.15 2.047060966491699 115.0
0.1 1.6372349262237549 116.0
0.05 1.3310458660125732 137.0
}\datatable

\begin{axis}[
    width=\columnwidth,
    height=0.75\columnwidth,
    ymin=0,
    ymax=70,
    xmin = 0.05,
    xmax = 1.0,
    xtick={0,0.5,1.0},
    enlarge x limits=0.05,
    x tick style={draw=none},
    ylabel={Time taken (s)},
    xlabel={},
    axis y line*=left,
    x tick label style={
        yshift=-5pt,
        xshift=9pt,
        anchor=east
    },
    grid=minor,
    minor grid style={line width=0.5pt, draw=gray!60, densely dotted},
    legend style={
        at={(0.5, 0.8)},
        anchor=south,
        legend columns=1,
        /tikz/every even column/.append style={column sep=1.5em},
        /tikz/every odd column/.append style={column sep=0.2em}
    }]

\addplot[color=red,
fill opacity=.5,
mark= *,
mark size = 0.5pt
]
table [
  x=x,
  y=y1,
] {\datatable};

\addlegendentry{Time taken}

\end{axis}

\begin{axis}[
  width=\columnwidth,
  height=0.75\columnwidth,
  ymin=0,
  ymax=210,
  xmin = 0.05,
  xmax = 1.0,
  xtick={0.0, 0.5, 1.0},
  enlarge x limits=0.05,
  x tick style={draw=none},
  ylabel={Entanglement cost},
  xlabel={Proportion of nodes moved},
  axis y line*=right,
  x tick label style={
      yshift=-5pt,
      xshift=9pt,
      anchor=east
  },
  grid=minor,
  minor grid style={line width=0.5pt, draw=gray!60, densely dotted},
  legend style={
      at={(0.5, 1.03)},
      anchor=south,
      legend columns=2,
      /tikz/every even column/.append style={column sep=1.5em},
      /tikz/every odd column/.append style={column sep=0.2em}
  }
]

\addplot+ [
color=blue,
fill opacity=.5,
mark= *,
mark size = 0.5pt
]
table [
  x=x,
  y=y2,
] {\datatable};

\addlegendentry{Entanglement cost}

\end{axis}

\end{tikzpicture}
      \caption{Final cost vs. proportion of nodes moved.}
      \label{fig:curtail}
    \end{subfigure}
    \caption{Illustration of the exploratory process to avoid local minima in FM. \cref{fig:explore} shows the entanglement cost of the circuit after each pass. The cost in the exploratory method fluctuates significantly between passes but ultimately reaches lower cost regions. \cref{fig:curtail} indicates that the exploratory method is more effective when the number of moves per pass is limited. The optimal region is between 0.1 and 0.2 of the total number of nodes. Consequently, the exploratory method achieves lower costs and is faster than the standard FM algorithm.}
    \label{fig:explore_result}
  \end{figure*}

\subsection{Benefits and limitations of gate grouping}\label{sec:gg_res}

\subsubsection{Reduced entanglement requirements}\label{sec:ent_cost}

Here we provide a brief demonstration of the benefits and limitations of gate grouping. Using the $CP$ fraction benchmark circuits, we vary the fraction of two-qubit gates, for a constant depth and qubit count of $32$, partitioning the resulting circuits across $4$ QPUs. In \cref{fig:group_vs}, we plot the distributions of the entanglement costs achieved for $100$ randomly generated $CP$ fraction circuits. While the effects are less noticeable for low fractions, the benefits of gate grouping become apparent as the fraction of two-qubit gates increases, where the distribution of the results with gate grouping is shifted towards lower cost. 
However, this can come at an extra cost in time as the fraction approaches $1$, since the gain updates must be performed over very large hyper-edges. Additionally, as discussed in \cref{sec:circ_extr}, the demand for communication qubits increases as the fraction of two-qubit gates increases. Without multi-gate teleportation, communication qubits are released immediately after use, while multi-gate teleportation requires communication qubits to be kept alive for the duration of the group. Conversely, a large communication qubit capacity reduces the additional communication depth, since entanglement can be generated in parallel.

\subsubsection{Increased communication qubit requirements}\label{sec:comm_qubit}

A notable limitation of the current work is that no constraints are placed on the number of communication qubits permitted. This is of significance when considering instances of hypergraph with large hyper-edges, since a large cut hyper-edge corresponds to a communication qubit being used over a large number of time steps. Facilitating multiple linked root qubits concurrently may require a large number of communication qubits. The benefit of not using cat-entanglements is that we only require a fewer communication qubits, since each gate teleportation and each state teleportation results in communication qubits being released immediately after use, unlike for multi-gate teleportation. The only time we may require additional communication qubits is if we need to handle multiple state teleportations at the same time step, though any additional communication qubits are still freed up directly after use. The trade-off is that we will likely end up with a deeper circuit, requiring more entanglement distribution. We briefly numerically analyse this trade-off with the following experiment on QFT circuits. For each circuit, we run the standard FM algorithm, with and without using the gate-grouping pass. When extracting the circuit, we plot both the resulting depth and the communication qubit requirements. It can be seen that the number of communication qubits required is constant when not using gate grouping, while the circuits are significantly deeper. On the other hand, the gate-grouping pass gives us a much shallower circuit but with an increasing communication qubit requirement. The choice of whether or not to use gate grouping should be made on the basis of additional qubit capacity. Additionally, we hope the lower e-bit costs from gate grouping serves to guide future DQC architectures to not under-allocate communication qubits, since they permit significantly reduced circuit depth.

\subsection{Exploratory vs exploitative FM}\label{sec:exp_vs}

Here we demonstrate the effectiveness of exploratroy FM described in \cref{sec:efficiency} for avoiding local optima. To recall, the method involves alternating between exploratory and exploitative passes, where the former does not roll-back to the iteration of best gain at the end of the pass. This allows the algorithm to explore larger sequences of moves which may allow the same node to be moved multiple times before a roll-back. We find that this is most effective when the number of nodes moved in each pass is limited, since the algorithm is less likely to escape a local minimum. This has an added benefit of providing a reduction of the pass time. However, we typically use a constant fraction of the number of nodes, so this reduction is only constant. The results are shown in \cref{fig:explore_result}.

\subsection{Multilevel partitioning}\label{sec:mlp_res}

\begin{figure*}[!ht]
  \centering
  \begin{subfigure}{0.4\linewidth}
    \input{plots/Coarse/coarseners_cost_full_exploit.tex}
    \caption{Cost after each pass with no move cap.}
    \label{fig:coarseners_cost_full}
  \end{subfigure}
  \begin{subfigure}{0.4\linewidth}
    \input{plots/Coarse/coarseners_time_full_exploit.tex}
    \caption{Time of each pass with no move cap.}
    \label{fig:coarseners_time_full}
  \end{subfigure}

  \begin{subfigure}{0.4\linewidth}
    \input{plots/Coarse/coarseners_cost_limit_explore.tex}
    \caption{Cost after each pass with exploration and move cap.}
    \label{fig:coarseners_cost_explore}
  \end{subfigure}
  \begin{subfigure}{0.4\linewidth}
    \input{plots/Coarse/coarseners_time_limit_explore.tex}
    \caption{Time of each pass with exploration and move cap.}
    \label{fig:coarseners_time_explore}
  \end{subfigure}
  \caption{Performance results of the coarsening methods. All results are produced from 10 runs on a 32-qubit $CP$-fraction circuit with depth 32 and 50\% two-qubit gates, partitioned across 4 QPUs. In \cref{fig:coarseners_cost_full,fig:coarseners_time_full}, we implement no limit on the number of nodes moved per pass, such that the time of the pass scales with the number of nodes in the graph. In \cref{fig:coarseners_cost_explore,fig:coarseners_time_explore}, we place a cap of $n_{q}$ on the number of nodes moved per pass for the ML methods, as discussed in \cref{sec:MLFM_comp}.} 
  \label{fig:coarsening_performance}
\end{figure*}

We compare the cost and time performance of the coarsening routines described in \cref{sec:MLCP}. Partitioning is performed using no coarsening (FM), window coarsening (MLFM-W) \cref{sec:WC}, block coarsening (MLFM-B) \cref{sec:BC} and recursive coarsening (MLFM-R) \cref{sec:RC}. In \cref{fig:coarsening_performance}, we show the entanglement cost and time taken as function of the number of passes for each method. For each of the multilevel methods, we run $10$ passes at each level, and for FM we run $10 \lceil \log_{2}(d) \rceil $ passes, such that the total number of passes is the same in all cases. \cref{fig:coarseners_cost_full} shows the entanglement cost after each pass, demonstrating the improved performance of the multilevel methods. After each 10 passes, the entanglement cost plateaus, indicating that the methods are reaching a local optimum for the given level, which is then quickly improved upon at the next level. \cref{fig:coarseners_time_full} shows the time taken for each pass, demonstrating the improvement in run time of coarser passes. Notably, the last $10$ passes are performed at the finest level of granularity, such that there is no improvement on the pass time. Overall, time taken is significantly reduced from the coarser levels. \cref{fig:coarseners_cost_explore,fig:coarseners_time_explore} employ the exploratory method with a cap on the number of nodes moved per pass. The cap is set to $0.125n_{q}d$ (the optimal proportion of nodes to move from \cref{fig:curtail}) for the fine-grained FM method and $n_{q}$ for the multilevel methods, since the performance is retained for fewer node moves. We see similar results for the entanglement costs, while the time taken is much lower for all coarsened methods. In particular, the recursive method is both the quickest and achieves the lowest entanglement costs.

\begin{figure*}
  \begin{subfigure}{0.48\linewidth}
    \begin{tikzpicture}
\begin{axis}[
    ybar=15pt,
    ymin=0,
    ymax=750,
    symbolic x coords={16,24,32,40,48,56,64},
    xtick=data,
    enlarge x limits=0.15,
    x tick style={draw=none},
    ylabel={Entanglement cost},
    xlabel={Number of qubits},
    x tick label style={
        yshift=-5pt,
        xshift=9pt,
        anchor=east
    },
    grid=minor,
    minor grid style={line width=0.5pt, draw=gray!60, densely dotted},
    legend style={
        at={(0.22,0.6)},
        anchor=south,
        legend columns=1,
        /tikz/every even column/.append style={column sep=1.5em},
        /tikz/every odd column/.append style={column sep=0.2em}
    },
    ymajorgrids = true,
    major grid style = { dotted, draw=gray!70 },
    bar width=3pt,
    error bars/y dir=both,
    error bars/y explicit,
    error bars/error mark options={
        rotate=90,
        mark size=1pt
    }
]
\pgfplotstableread[
  col sep=space,
  header=has colnames
]{
num_qubits f_mean f_min f_max w_mean w_min w_max b_mean b_min b_max r_mean r_min r_max
16 13.4 11.0 18.0 12.0 10.0 14.0 12.4 10.0 15.0 11.0 10.0 13.0
24 50.2 47.0 58.0 43.0 40.0 47.0 41.0 37.0 44.0 40.6 37.0 43.0
32 116.2 112.0 122.0 101.4 92.0 107.0 96.2 88.0 102.0 94.2 89.0 98.0
40 208.0 199.0 218.0 198.0 189.0 206.0 177.6 173.0 183.0 171.4 166.0 177.0
48 328.0 317.0 336.0 320.8 302.0 349.0 283.0 275.0 295.0 272.8 264.0 283.0
56 486.0 470.0 502.0 447.0 423.0 464.0 415.4 406.0 424.0 397.0 383.0 408.0
64 655.8 640.0 674.0 616.4 596.0 630.0 580.6 567.0 598.0 555.6 541.0 568.0
}\datatable
\addplot+ [
  color=red,
fill opacity=.3,
bar shift = -6pt
]
table [
  x=num_qubits,
  y=f_mean,
  y error plus expr=\thisrow{f_max}-\thisrow{f_mean},
  y error minus expr=\thisrow{f_mean}-\thisrow{f_min}
] {\datatable};

\addlegendentry{FM}

\pgfplotstableread[
  col sep=space,
  header=has colnames
]{
num_qubits f_mean f_min f_max w_mean w_min w_max b_mean b_min b_max r_mean r_min r_max
16 13.4 11.0 18.0 12.0 10.0 14.0 12.4 10.0 15.0 11.0 10.0 13.0
24 50.2 47.0 58.0 43.0 40.0 47.0 41.0 37.0 44.0 40.6 37.0 43.0
32 116.2 112.0 122.0 101.4 92.0 107.0 96.2 88.0 102.0 94.2 89.0 98.0
40 208.0 199.0 218.0 198.0 189.0 206.0 177.6 173.0 183.0 171.4 166.0 177.0
48 328.0 317.0 336.0 320.8 302.0 349.0 283.0 275.0 295.0 272.8 264.0 283.0
56 486.0 470.0 502.0 447.0 423.0 464.0 415.4 406.0 424.0 397.0 383.0 408.0
64 655.8 640.0 674.0 616.4 596.0 630.0 580.6 567.0 598.0 555.6 541.0 568.0
}\datatable
\addplot+ [
  color=blue,
fill opacity=.3,
bar shift = -2pt,
]
table [
  x=num_qubits,
  y=w_mean,
  y error plus expr=\thisrow{w_max}-\thisrow{w_mean},
  y error minus expr=\thisrow{w_mean}-\thisrow{w_min}
] {\datatable};

\addlegendentry{MLFM-W}

\pgfplotstableread[
  col sep=space,
  header=has colnames
]{
num_qubits f_mean f_min f_max w_mean w_min w_max b_mean b_min b_max r_mean r_min r_max
16 13.4 11.0 18.0 12.0 10.0 14.0 12.4 10.0 15.0 11.0 10.0 13.0
24 50.2 47.0 58.0 43.0 40.0 47.0 41.0 37.0 44.0 40.6 37.0 43.0
32 116.2 112.0 122.0 101.4 92.0 107.0 96.2 88.0 102.0 94.2 89.0 98.0
40 208.0 199.0 218.0 198.0 189.0 206.0 177.6 173.0 183.0 171.4 166.0 177.0
48 328.0 317.0 336.0 320.8 302.0 349.0 283.0 275.0 295.0 272.8 264.0 283.0
56 486.0 470.0 502.0 447.0 423.0 464.0 415.4 406.0 424.0 397.0 383.0 408.0
64 655.8 640.0 674.0 616.4 596.0 630.0 580.6 567.0 598.0 555.6 541.0 568.0
}\datatable
\addplot+ [
  color=orange,
fill opacity=.3,
bar shift = 2pt,
]
table [
  x=num_qubits,
  y=b_mean,
  y error plus expr=\thisrow{b_max}-\thisrow{b_mean},
  y error minus expr=\thisrow{b_mean}-\thisrow{b_min}
] {\datatable};

\addlegendentry{MLFM-B}

\pgfplotstableread[
  col sep=space,
  header=has colnames
]{
num_qubits f_mean f_min f_max w_mean w_min w_max b_mean b_min b_max r_mean r_min r_max
16 13.4 11.0 18.0 12.0 10.0 14.0 12.4 10.0 15.0 11.0 10.0 13.0
24 50.2 47.0 58.0 43.0 40.0 47.0 41.0 37.0 44.0 40.6 37.0 43.0
32 116.2 112.0 122.0 101.4 92.0 107.0 96.2 88.0 102.0 94.2 89.0 98.0
40 208.0 199.0 218.0 198.0 189.0 206.0 177.6 173.0 183.0 171.4 166.0 177.0
48 328.0 317.0 336.0 320.8 302.0 349.0 283.0 275.0 295.0 272.8 264.0 283.0
56 486.0 470.0 502.0 447.0 423.0 464.0 415.4 406.0 424.0 397.0 383.0 408.0
64 655.8 640.0 674.0 616.4 596.0 630.0 580.6 567.0 598.0 555.6 541.0 568.0
}\datatable
\addplot+ [
  color=teal,
fill opacity=.3,
bar shift = -10000pt,
]
table [
  x=num_qubits,
  y=,
] {\datatable};

\pgfplotstableread[
  col sep=space,
  header=has colnames
]{
num_qubits f_mean f_min f_max w_mean w_min w_max b_mean b_min b_max r_mean r_min r_max
16 13.4 11.0 18.0 12.0 10.0 14.0 12.4 10.0 15.0 11.0 10.0 13.0
24 50.2 47.0 58.0 43.0 40.0 47.0 41.0 37.0 44.0 40.6 37.0 43.0
32 116.2 112.0 122.0 101.4 92.0 107.0 96.2 88.0 102.0 94.2 89.0 98.0
40 208.0 199.0 218.0 198.0 189.0 206.0 177.6 173.0 183.0 171.4 166.0 177.0
48 328.0 317.0 336.0 320.8 302.0 349.0 283.0 275.0 295.0 272.8 264.0 283.0
56 486.0 470.0 502.0 447.0 423.0 464.0 415.4 406.0 424.0 397.0 383.0 408.0
64 655.8 640.0 674.0 616.4 596.0 630.0 580.6 567.0 598.0 555.6 541.0 568.0
}\datatable
\addplot+ [
  color=teal,
fill opacity=.3,
bar shift=6pt
]
table [
  x=num_qubits,
  y=r_mean,
  y error plus expr=\thisrow{r_max}-\thisrow{r_mean},
  y error minus expr=\thisrow{r_mean}-\thisrow{r_min}
] {\datatable};

\addlegendentry{MLFM-R}

\end{axis}
\end{tikzpicture}
    \caption{Entanglement cost}
    \label{fig:ML_comp_cost}
  \end{subfigure}
  ~
  \begin{subfigure}{0.48\linewidth}
    \input{plots/FM/ML/ml_comp_time.tex}
    \caption{Time taken}
    \label{fig:ML_comp_time}
  \end{subfigure}
  
      \caption{Comparison of the coarsening methods in terms of entanglement cost and time. Tests use a $CP$-fraction circuit with $50\%$ two-qubit gates, using the same total number of passes for each method.}
  \label{fig:ML_comp}
\end{figure*}

The results in \cref{fig:ML_comp} are generated using the same parameters as \cref{fig:coarseners_cost_explore,fig:coarseners_time_explore}. The recursive method maintains the lowest entanglement costs and is significantly faster than the other methods.

\subsection{Benchmark algorithms}\label{sec:bench_algs}

To compare the performance of the multilevel framework, we compare the results with the best performing methods from the literature. We use a number of workflows from Pytket-DQC \cite{andres-martinezCQCLPytketdqc2024}, which are the subject of References \cite{andres-martinezAutomatedDistributionQuantum2019,wuEntanglementefficientBipartitedistributedQuantum2023a,andres-martinezDistributingCircuitsHeterogeneous2024}, and the fine-grained partitioning method from Ref. \cite{bakerTimeslicedQuantumCircuit2020}. We use the best performing pipeline from our methods, which uses gate grouping, recursive coarsening and exploratory FM, referred to as \textit{MLFM-R}. The number of moves per pass is capped at $n_q$, as discussed in \cref{sec:MLFM_comp}, and $10$ passes are performed at each level. Below we describe the methods used for comparison.

\subsubsection{Pytket-DQC}\label{sec:pytket-dqc}

Pytket-DQC is a freely available python library within Tket, Quanntinuum's open-source quantum computing toolkit \cite{sivarajahT|ketRetargetableCompiler2020}, containing various methods for quantum circuit optimisation. Pytket-DQC is an extension that contains a number of tools for distributing quantum circuits over multiple QPUs. The optimisation techniques available are the subject of References \cite{andres-martinezAutomatedDistributionQuantum2019,wuEntanglementefficientBipartitedistributedQuantum2023a,andres-martinezDistributingCircuitsHeterogeneous2024}. The most basic workflow available, referred to as \textit{Partition} (P), is a static hypergraph partitioning framework \cite{andres-martinezAutomatedDistributionQuantum2019} that employs a third-party software, KaHyPar \cite{schlagHighQualityHypergraphPartitioning2023}, to obtain low cost assignments of qubits to QPUs and leverages \textit{detached gates} to reduce entanglement costs \cite{andres-martinezDistributingCircuitsHeterogeneous2024}. The methods of Ref. \cite{wuEntanglementefficientBipartitedistributedQuantum2023a} are integrated into Pytket-DQC, using minimum vertex cover and embedding techniques to reduce the entanglement costs of the circuit. These methods are combined in Ref. \cite{andres-martinezDistributingCircuitsHeterogeneous2024}, with additional workflows for refining the results. The best performing workflow, based on the results in Ref. \cite{andres-martinezDistributingCircuitsHeterogeneous2024}, is the \textit{CoverEmbedSteinerDetached} (ESD) method, which uses various methods to refine an initial partition, though the combination of techniques means it is also the slowest. The fastest method is Partition (P), since KaHyPar is a highly optimised library written in C$++$. Despite being the quickest, the performance of Partition is often comparable with the more advanced techniques. A method for refining the result of Partition to incorporate embedding is also available, referred to as PartitionEmbed (PE). This method is slower than Partition, but faster than EmbedSteinerDetached, and tends to achieve better results than Partition. None of the methods available use state teleportation, resulting in certain cases where all techniques underperform, notably the Quantum Volume benchmarks shown in \cref{fig:QV_24}. This was highlighted by the authors in Ref. \cite{andres-martinezDistributingCircuitsHeterogeneous2024}, though not attributed to the absence of state teleportation capabilities. We compare results with Partition, PartitionEmbed and EmbedSteinerDetached where applicable. We note that Pytket-DQC requires its own transpilation pass to be used before distribution. This gate-set also contains controlled-phase gates, so two-qubit gate counts should be similar though we cannot guarantee identical circuits. We pass the circuits as transpiled by Qiskit directly to Pytket-DQC, which performs its own transpilation internally.

\begin{figure*}[!htbp]
    \centering
    \begin{subfigure}{0.48\textwidth}
      \centering
      \begin{tikzpicture}
\begin{axis}[
    width=\columnwidth,
    height=0.75\columnwidth,
    ybar=15pt,
    ymin=0,
    ymax=1600,
    symbolic x coords={16,24,32,40,48,56,64,72,80,88,96},
    xtick=data,
    enlarge x limits=0.05,
    x tick style={draw=none},
    ylabel={Entanglement cost},
    xlabel={Number of qubits},
    x tick label style={
        yshift=-5pt,
        xshift=9pt,
        anchor=east
    },
scaled y ticks = base 10:-3,
grid=minor,
    minor grid style={line width=0.5pt, draw=gray!60, densely dotted},
    legend style={
        at={(0.25,0.5)},
        anchor=south,
        legend columns=1,
        /tikz/every even column/.append style={column sep=1.5em},
        /tikz/every odd column/.append style={column sep=0.2em}
    },
    ymajorgrids = true,
    major grid style = { dotted, draw=gray!70 },
    bar width=2.5pt,
    error bars/y dir=both,
    error bars/y explicit,
    error bars/error mark options={
        rotate=90,
        mark size=1pt
    }
]
\pgfplotstableread[
  col sep=space,
  header=has colnames
]{
num_qubits fgp_mean fgp_min fgp_max
16 14.4 12.0 16.0
24 49.4 45.0 55.0
32 104.8 99.0 109.0
40 187.4 183.0 194.0
48 279.2 263.0 294.0
56 407.8 397.0 413.0
64 586.0 571.0 598.0
72 739.6 712.0 771.0
80 951.4 919.0 989.0
88 1176.4 1154.0 1233.0
96 1464.4 1443.0 1492.0
}\datatable
\addplot+ [
  color=one,
fill opacity=.3,
bar shift = -6pt
]
table [
  x=num_qubits,
  y=fgp_mean,
  y error plus expr=\thisrow{fgp_max}-\thisrow{fgp_mean},
  y error minus expr=\thisrow{fgp_mean}-\thisrow{fgp_min}
] {\datatable};

\addlegendentry{FGP-rOEE}

\pgfplotstableread[
  col sep=space,
  header=has colnames
]{
num_qubits r_mean r_min r_max
16 5.8 3.0 9.0
24 25.0 22.0 27.0
32 53.5 49.0 57.0
40 105.0 98.0 114.0
48 167.0 160.0 173.0
56 242.6 236.0 250.0
64 334.4 315.0 351.0
72 461.4 438.0 478.0
80 566.7 544.0 593.0
88 717.4 706.0 742.0
96 864.9 836.0 897.0
}\datatable
\addplot+ [
  color=teal,
fill opacity=.3,
bar shift=-2pt
]
table [
  x=num_qubits,
  y=r_mean,
  y error plus expr=\thisrow{r_max}-\thisrow{r_mean},
  y error minus expr=\thisrow{r_mean}-\thisrow{r_min}
] {\datatable};

\addlegendentry{MLMF-R}

\pgfplotstableread[
  col sep=space,
  header=has colnames
]{
num_qubits part_mean part_min part_max embed_mean embed_min embed_max
16 6.7 4.0 8.0 6.7 4.0 8.0
24 26.6 23.0 30.0 26.6 23.0 30.0
32 60.2 50.0 66.0 60.2 50.0 66.0
40 111.4 104.0 120.0 111.4 104.0 120.0
48 180.4 171.0 189.0 180.4 171.0 189.0
56 269.1 256.0 290.0 269.1 256.0 290.0
64 380.9 374.0 392.0 380.9 374.0 392.0
72 493.0 474.0 521.0 493.0 474.0 521.0
80 644.0 623.0 658.0 644.0 623.0 658.0
88 804.2 792.0 821.0 804.2 792.0 821.0
96 983.6 965.0 999.0 983.6 965.0 999.0
}\datatable
\addplot+ [
  color=three,
fill opacity=.3,
bar shift = 2pt
]
table [
  x=num_qubits,
  y=embed_mean,
  y error plus expr=\thisrow{embed_max}-\thisrow{embed_mean},
  y error minus expr=\thisrow{embed_mean}-\thisrow{embed_min}
] {\datatable};

\addlegendentry{Pytket PE}

\pgfplotstableread[
  col sep=space,
  header=has colnames
]{
num_qubits part_mean part_min part_max embed_mean embed_min embed_max
16 0.024157619476318358 0.012878179550170898 0.06699991226196289 0.0037047863006591797 0.002553224563598633 0.005204916000366211
24 0.06192600727081299 0.05439305305480957 0.07361888885498047 0.01784491539001465 0.014481782913208008 0.024126768112182617
32 0.13731591701507567 0.10930895805358887 0.16730999946594238 0.05933806896209717 0.04245805740356445 0.06825017929077148
40 0.305842924118042 0.19381999969482422 0.47350502014160156 0.16313061714172364 0.1390089988708496 0.1957230567932129
48 0.40891027450561523 0.34348487854003906 0.504971981048584 0.38114893436431885 0.3269612789154053 0.43102216720581055
56 0.7670342683792114 0.5716879367828369 1.0328338146209717 0.7469875574111938 0.6814920902252197 0.9160311222076416
64 0.9975391626358032 0.8889648914337158 1.1579101085662842 1.4262382745742799 1.2987189292907715 1.5510120391845703
72 1.5333998441696166 1.266780138015747 1.7970762252807617 2.252950739860535 2.0781257152557373 2.4784860610961914
80 2.1506540536880494 1.7897520065307617 3.1388137340545654 3.743028450012207 3.3874850273132324 4.046696186065674
88 3.3019176959991454 2.5279300212860107 4.600592136383057 5.579227256774902 5.293071985244751 6.003679037094116
96 4.167246556282043 3.30957293510437 4.7729339599609375 8.22713918685913 7.720250844955444 8.80504584312439
}\datatable
\addplot+ [
  color=two,
fill opacity=.3,
bar shift = -10000pt
]
table [
  x=num_qubits,
  y=,
] {\datatable};

\pgfplotstableread[
  col sep=space,
  header=has colnames
]{
num_qubits init_mean init_min init_max vc_mean vc_min vc_max steiner_mean steiner_min steiner_max detached_mean detached_min detached_max
16 6.25 4.0 8.0 6.25 4.0 8.0 6.25 4.0 8.0 6.25 4.0 8.0
24 24.5 19.0 28.0 25.0 19.0 30.0 25.0 19.0 30.0 24.5 19.0 28.0
32 61.5 56.0 66.0 63.7 57.0 69.0 63.7 57.0 69.0 61.0 55.0 65.0
40 114.2 107.0 123.0 117.2 109.0 127.0 117.2 109.0 127.0 113.7 106.0 124.0
48 182.3 171.0 200.0 188.2 180.0 206.0 188.2 180.0 206.0 180.7 170.0 198.0
56 273.4 261.0 287.0 282.2 268.0 300.0 282.2 268.0 300.0 274.7 261.0 292.0
64 380.1 362.0 400.0 390.8 372.0 406.0 390.8 372.0 406.0 379.0 362.0 398.0
72 500.3 479.0 510.0 515.3 491.0 528.0 515.3 491.0 528.0 497.5 477.0 504.0
80 664.0 639.0 680.0 681.0 657.0 703.0 681.0 657.0 703.0 657.5 633.0 676.0
88 826.0 808.0 850.0 845.6 821.0 864.0 845.6 821.0 864.0 817.6 797.0 840.0
96 999.6 972.0 1032.0 1021.6 985.0 1061.0 1021.6 985.0 1061.0 993.4 958.0 1028.0
}\datatable
\addplot+ [
  color=two,
fill opacity=.3,
bar shift= 6pt
]
table [
  x=num_qubits,
  y=detached_mean,
  y error plus expr=\thisrow{detached_max}-\thisrow{detached_mean},
  y error minus expr=\thisrow{detached_mean}-\thisrow{detached_min}
] {\datatable};

\addlegendentry{Pytket ESD}

\end{axis}
\end{tikzpicture}
    \end{subfigure}
    \begin{subfigure}{0.48\textwidth}
      \centering
      \begin{tikzpicture}
\begin{axis}[
    width=\columnwidth,
    height=0.75\columnwidth,
    ybar,
    bar width=15pt,
    ymin=0,
    ymax=130,
    symbolic x coords={16,24,32,40,48,56,64,72,80,88,96},
    xtick=data,
    enlarge x limits=0.05,
    x tick style={draw=none},
    ylabel={{Time taken (s)}},
    xlabel={Number of qubits},
    x tick label style={
        yshift=-5pt,
        xshift=9pt,
        anchor=east
    },
    scaled y ticks = base 10:-2,
    scaled y ticks = base 10:-2,
grid=minor,
    minor grid style={line width=0.5pt, draw=gray!60, densely dotted},
    legend style={
        at={(0.25,0.5)},
        anchor=south,
        legend columns=1,
        /tikz/every even column/.append style={column sep=1.5em},
        /tikz/every odd column/.append style={column sep=0.2em}
    },
    ymajorgrids = true,
    major grid style = { dotted, draw=gray!70 },
    bar width=2.5pt,
    error bars/y dir=both,
    error bars/y explicit,
    error bars/error mark options={
        rotate=90,
        mark size=1pt
    }
]
\pgfplotstableread[
  col sep=space,
  header=has colnames
]{
num_qubits fgp_mean fgp_min fgp_max
16 0.0027686595916748048 0.0024042129516601562 0.0031087398529052734
24 0.013534975051879884 0.011714935302734375 0.01625800132751465
32 0.041688156127929685 0.0389409065246582 0.04461216926574707
40 0.10578303337097168 0.10220599174499512 0.11059713363647461
48 0.21045346260070802 0.19304800033569336 0.22524809837341309
56 0.4152510643005371 0.39061522483825684 0.4384582042694092
64 0.7538611888885498 0.7282040119171143 0.7717339992523193
72 1.1559597492218017 1.094834804534912 1.2700598239898682
80 1.8362872123718261 1.7163100242614746 1.9521780014038086
88 2.666691017150879 2.623861074447632 2.7422242164611816
96 4.194344758987427 3.882456064224243 4.686321020126343
}\datatable
\addplot+ [
  color=one,
fill opacity=.3,
bar shift = -6pt
]
table [
  x=num_qubits,
  y=fgp_mean,
  y error plus expr=\thisrow{fgp_max}-\thisrow{fgp_mean},
  y error minus expr=\thisrow{fgp_mean}-\thisrow{fgp_min}
] {\datatable};

\addlegendentry{FGP-rOEE}

\pgfplotstableread[
  col sep=space,
  header=has colnames
]{
num_qubits r_mean r_min r_max
16 0.0427 0.0352 0.0607
24 0.1507 0.1435 0.1573
32 0.3616 0.3421 0.3774
40 0.8064 0.7672 0.8582
48 1.4766 1.4452 1.5031
56 2.4390 2.3107 2.5032
64 3.8598 3.7041 3.9626
72 6.5888 6.4218 6.7132
80 9.3105 9.0760 9.5154
88 13.3251 13.0375 13.6966
96 18.0502 17.5014 18.4345
}\datatable
\addplot+ [
  color=teal,
fill opacity=.3,
bar shift=-2pt
]
table [
  x=num_qubits,
  y=r_mean,
  y error plus expr=\thisrow{r_max}-\thisrow{r_mean},
  y error minus expr=\thisrow{r_mean}-\thisrow{r_min}
] {\datatable};

\addlegendentry{MLFM-R}

\pgfplotstableread[
  col sep=space,
  header=has colnames
]{
num_qubits part_mean part_min part_max embed_mean embed_min embed_max
16 0.024157619476318358 0.012878179550170898 0.06699991226196289 0.0037047863006591797 0.002553224563598633 0.005204916000366211
24 0.06192600727081299 0.05439305305480957 0.07361888885498047 0.01784491539001465 0.014481782913208008 0.024126768112182617
32 0.13731591701507567 0.10930895805358887 0.16730999946594238 0.05933806896209717 0.04245805740356445 0.06825017929077148
40 0.305842924118042 0.19381999969482422 0.47350502014160156 0.16313061714172364 0.1390089988708496 0.1957230567932129
48 0.40891027450561523 0.34348487854003906 0.504971981048584 0.38114893436431885 0.3269612789154053 0.43102216720581055
56 0.7670342683792114 0.5716879367828369 1.0328338146209717 0.7469875574111938 0.6814920902252197 0.9160311222076416
64 0.9975391626358032 0.8889648914337158 1.1579101085662842 1.4262382745742799 1.2987189292907715 1.5510120391845703
72 1.5333998441696166 1.266780138015747 1.7970762252807617 2.252950739860535 2.0781257152557373 2.4784860610961914
80 2.1506540536880494 1.7897520065307617 3.1388137340545654 3.743028450012207 3.3874850273132324 4.046696186065674
88 3.3019176959991454 2.5279300212860107 4.600592136383057 5.579227256774902 5.293071985244751 6.003679037094116
96 4.167246556282043 3.30957293510437 4.7729339599609375 8.22713918685913 7.720250844955444 8.80504584312439
}\datatable
\addplot+ [
  color=three,
fill opacity=.3,
bar shift = 2pt
]
table [
  x=num_qubits,
  y=embed_mean,
  y error plus expr=\thisrow{embed_max}-\thisrow{embed_mean},
  y error minus expr=\thisrow{embed_mean}-\thisrow{embed_min}
] {\datatable};

\addlegendentry{Pytket PE}

\pgfplotstableread[
  col sep=space,
  header=has colnames
]{
num_qubits part_mean part_min part_max embed_mean embed_min embed_max
16 0.024157619476318358 0.012878179550170898 0.06699991226196289 0.0037047863006591797 0.002553224563598633 0.005204916000366211
24 0.06192600727081299 0.05439305305480957 0.07361888885498047 0.01784491539001465 0.014481782913208008 0.024126768112182617
32 0.13731591701507567 0.10930895805358887 0.16730999946594238 0.05933806896209717 0.04245805740356445 0.06825017929077148
40 0.305842924118042 0.19381999969482422 0.47350502014160156 0.16313061714172364 0.1390089988708496 0.1957230567932129
48 0.40891027450561523 0.34348487854003906 0.504971981048584 0.38114893436431885 0.3269612789154053 0.43102216720581055
56 0.7670342683792114 0.5716879367828369 1.0328338146209717 0.7469875574111938 0.6814920902252197 0.9160311222076416
64 0.9975391626358032 0.8889648914337158 1.1579101085662842 1.4262382745742799 1.2987189292907715 1.5510120391845703
72 1.5333998441696166 1.266780138015747 1.7970762252807617 2.252950739860535 2.0781257152557373 2.4784860610961914
80 2.1506540536880494 1.7897520065307617 3.1388137340545654 3.743028450012207 3.3874850273132324 4.046696186065674
88 3.3019176959991454 2.5279300212860107 4.600592136383057 5.579227256774902 5.293071985244751 6.003679037094116
96 4.167246556282043 3.30957293510437 4.7729339599609375 8.22713918685913 7.720250844955444 8.80504584312439
}\datatable
\addplot+ [
  color= two,
fill opacity=.3,
bar shift = -10000pt
]
table [
  x=num_qubits,
  y=,
] {\datatable};

\pgfplotstableread[
  col sep=space,
  header=has colnames
]{
num_qubits init_mean init_min init_max vc_mean vc_min vc_max steiner_mean steiner_min steiner_max detached_mean detached_min detached_max
16 0.013805210590362549 0.010658979415893555 0.019384145736694336 0.027864181995391847 0.021682024002075195 0.0333709716796875 0.038712596893310545 0.029275894165039062 0.04650402069091797 0.041128683090209964 0.03129315376281738 0.05013585090637207
24 0.03777413368225098 0.03282880783081055 0.04334211349487305 0.10465791225433349 0.07748579978942871 0.15558600425720215 0.15915956497192382 0.11261367797851562 0.20579791069030762 0.17001755237579347 0.11971569061279297 0.2155768871307373
32 0.08184146881103516 0.07485508918762207 0.08854103088378906 0.3298435926437378 0.29152822494506836 0.3563687801361084 0.5300977945327758 0.45431041717529297 0.5871968269348145 0.5597260236740113 0.4772024154663086 0.6202888488769531
40 0.14204142093658448 0.13611125946044922 0.1459648609161377 0.8989493131637574 0.8191120624542236 0.9943761825561523 1.47152841091156 1.3181302547454834 1.6695342063903809 1.5323419570922852 1.3763141632080078 1.7364392280578613
48 0.23918240070343016 0.22136998176574707 0.277144193649292 2.2398776531219484 1.9962711334228516 2.602031946182251 3.63671019077301 3.2663421630859375 4.143260717391968 3.744936156272888 3.3692092895507812 4.26572060585022
56 0.35585846900939944 0.3344860076904297 0.37777018547058105 4.809450149536133 4.43488335609436 5.24679970741272 7.712965202331543 7.112765312194824 8.418497562408447 7.889635848999023 7.277104139328003 8.607675313949585
64 0.5230571031570435 0.4821479320526123 0.5932719707489014 9.401302313804626 8.784233093261719 10.063905239105225 14.90531415939331 13.824879169464111 16.234150171279907 15.178289031982422 14.063387393951416 16.60283398628235
72 0.7118887662887573 0.6542227268218994 0.7804450988769531 16.880827593803406 15.404985666275024 17.93718910217285 26.78094048500061 24.218093633651733 28.994656085968018 27.181564331054688 24.57467746734619 29.485709190368652
80 1.0257100820541383 0.944176197052002 1.0989477634429932 31.045921754837035 29.686490058898926 32.14656138420105 49.28761746883392 45.66091799736023 51.843602657318115 49.892007780075076 46.178879261016846 52.49196267127991
88 1.3468218326568604 1.2417099475860596 1.5783250331878662 49.81621789932251 47.680004835128784 51.45695424079895 78.42892694473267 74.91080713272095 81.346351146698 79.21183824539185 75.65958023071289 82.1537070274353
96 1.7355512142181397 1.6435668468475342 1.8118529319763184 75.23851752281189 69.17584657669067 81.24712204933167 116.43294997215271 107.01752471923828 125.4558801651001 117.42242770195007 107.9608826637268 126.54821014404297
}\datatable
\addplot+ [
  color= two,
fill opacity=.3,
bar shift=6pt
]
table [
  x=num_qubits,
  y=detached_mean,
  y error plus expr=\thisrow{detached_max}-\thisrow{detached_mean},
  y error minus expr=\thisrow{detached_mean}-\thisrow{detached_min}
] {\datatable};

\addlegendentry{Pytket ESD}

\end{axis}
\end{tikzpicture}
    \end{subfigure}
    \caption{Circuit with 30\% two-qubit gates. Starting with two QPUs for a 16 qubit circuit, for each increase in number of qubits we add a new QPU to the system. Each QPU has 9 qubit slots, which leaves a potential free data slot for teleportation on each QPU.}
    \label{fig:CP_30}
  \end{figure*}

  \begin{figure*}[!htbp]
    \centering
    \begin{subfigure}{0.48\textwidth}
      \centering
      \begin{tikzpicture}
\begin{axis}[
    width=\columnwidth,
    height=0.75\columnwidth,
    ybar=15pt,
    ymin=0,
    ymax=2600,
    symbolic x coords={16,24,32,40,48,56,64,72,80,88, 96},
    xtick=data,
    enlarge x limits=0.05,
    x tick style={draw=none},
    ylabel={Entanglement cost},
    xlabel={Number of qubits},
    x tick label style={
        yshift=-5pt,
        xshift=9pt,
        anchor=east
    },
scaled y ticks = base 10:-3,
grid=minor,
    minor grid style={line width=0.5pt, draw=gray!60, densely dotted},
    legend style={
        at={(0.25,0.5)},
        anchor=south,
        legend columns=1,
        /tikz/every even column/.append style={column sep=1.5em},
        /tikz/every odd column/.append style={column sep=0.2em}
    },
    ymajorgrids = true,
    major grid style = { dotted, draw=gray!70 },
    bar width=2.5pt,
    error bars/y dir=both,
    error bars/y explicit,
    error bars/error mark options={
        rotate=90,
        mark size=1pt
    }
]
\pgfplotstableread[
  col sep=space,
  header=has colnames
]{
num_qubits fgp_mean fgp_min fgp_max
16 23.8 22.0 26.0
24 75.2 70.0 79.0
32 167.4 161.0 175.0
40 298.2 286.0 313.0
48 470.4 459.0 492.0
56 686.4 665.0 712.0
64 920.2 899.0 945.0
72 1206.2 1182.0 1246.0
80 1549.2 1526.0 1572.0
88 1938.4 1913.0 1971.0
96 2376.6 2338.0 2422.0
}\datatable
\addplot+ [
  color=one,
fill opacity=.3,
bar shift = -6pt
]
table [
  x=num_qubits,
  y=fgp_mean,
  y error plus expr=\thisrow{fgp_max}-\thisrow{fgp_mean},
  y error minus expr=\thisrow{fgp_mean}-\thisrow{fgp_min}
] {\datatable};

\addlegendentry{FGP-rOEE}

\pgfplotstableread[
  col sep=space,
  header=has colnames
]{
num_qubits r_mean r_min r_max
16 10.8 8.0 13.0
24 41.3 37.0 44.0
32 92.2 84.0 98.0
40 174.5 169.0 181.0
48 270.5 261.0 282.0
56 400.4 390.0 417.0
64 554.0 543.0 567.0
72 742.3 733.0 751.0
80 914.3 901.0 927.0
88 1137.7 1117.0 1177.0
96 1400.0 1369.0 1425.0
}\datatable
\addplot+ [
  color=teal,
fill opacity=.3,
bar shift=-2pt
]
table [
  x=num_qubits,
  y=r_mean,
  y error plus expr=\thisrow{r_max}-\thisrow{r_mean},
  y error minus expr=\thisrow{r_mean}-\thisrow{r_min}
] {\datatable};

\addlegendentry{MLFM-R}

\pgfplotstableread[
  col sep=space,
  header=has colnames
]{
num_qubits part_mean part_min part_max embed_mean embed_min embed_max
16 11.5 9.0 15.0 11.5 9.0 15.0
24 44.2 42.0 47.0 44.2 42.0 47.0
32 97.1 91.0 103.0 97.1 91.0 103.0
40 177.9 171.0 187.0 177.9 171.0 187.0
48 286.5 275.0 302.0 286.5 275.0 302.0
56 421.0 407.0 440.0 421.0 407.0 440.0
64 575.5 564.0 599.0 575.5 564.0 599.0
72 772.6 755.0 806.0 772.6 755.0 806.0
80 991.1 954.0 1014.0 991.1 954.0 1014.0
88 1231.6 1213.0 1250.0 1231.6 1213.0 1250.0
96 1502.4 1474.0 1547.0 1502.4 1474.0 1547.0
}\datatable
\addplot+ [
  color=three,
fill opacity=.3,
bar shift = 2pt
]
table [
  x=num_qubits,
  y=embed_mean,
  y error plus expr=\thisrow{embed_max}-\thisrow{embed_mean},
  y error minus expr=\thisrow{embed_mean}-\thisrow{embed_min}
] {\datatable};

\addlegendentry{Pytket PE}

\pgfplotstableread[
  col sep=space,
  header=has colnames
]{
num_qubits part_mean part_min part_max embed_mean embed_min embed_max
16 0.02773442268371582 0.02064204216003418 0.034748077392578125 0.009195232391357422 0.007353782653808594 0.010781288146972656
24 0.09533805847167968 0.08246684074401855 0.1142270565032959 0.055626511573791504 0.048445940017700195 0.06445598602294922
32 0.2141348123550415 0.18077325820922852 0.26163506507873535 0.18966774940490722 0.1647169589996338 0.21106815338134766
40 0.45887031555175783 0.373913049697876 0.5584170818328857 0.5571754693984985 0.5039520263671875 0.6013350486755371
48 0.6928657293319702 0.6370103359222412 0.8819680213928223 1.3001453399658203 1.1822409629821777 1.4405572414398193
56 1.1322642326354981 1.054063081741333 1.2124481201171875 2.6391883850097657 2.4270999431610107 2.875708818435669
64 1.7580196857452393 1.6767950057983398 1.887073040008545 4.700679063796997 4.570685148239136 4.989161968231201
72 2.574975872039795 2.4533729553222656 2.699486017227173 8.099718713760376 7.439749002456665 8.944865942001343
80 3.912350821495056 3.4602701663970947 4.777190923690796 12.911817073822021 11.658912897109985 13.744668006896973
88 5.2263283967971805 4.850888013839722 6.178251028060913 19.50987355709076 18.88469409942627 20.084751844406128
96 7.168324327468872 6.664013147354126 8.473192930221558 29.28718400001526 28.053810834884644 30.995959043502808
}\datatable
\addplot+ [
  color=two,
fill opacity=.3,
bar shift = -10000pt
]
table [
  x=num_qubits,
  y=,
] {\datatable};

\pgfplotstableread[
  col sep=space,
  header=has colnames
]{
num_qubits init_mean init_min init_max vc_mean vc_min vc_max steiner_mean steiner_min steiner_max detached_mean detached_min detached_max
16 11.65 8.0 13.0 11.65 8.0 13.0 11.65 8.0 13.0 11.65 8.0 13.0
24 42.3 37.0 47.0 45.6 40.0 51.0 45.6 40.0 51.0 42.5 37.0 47.0
32 97.9 94.0 104.0 104.1 100.0 114.0 104.1 100.0 114.0 97.4 93.0 103.0
40 179.6 169.0 192.0 193.3 184.0 201.0 193.3 184.0 201.0 178.4 170.0 190.0
48 296.1 282.0 312.0 316.7 302.0 331.0 316.7 302.0 331.0 291.9 281.0 308.0
56 432.1 419.0 444.0 465.3 449.0 481.0 465.3 449.0 481.0 427.6 410.0 442.0
64 599.4 592.0 618.0 643.9 628.0 674.0 643.9 628.0 674.0 591.2 580.0 605.0
72 793.9 782.0 811.0 855.2 835.0 876.0 855.2 835.0 876.0 782.7 767.0 798.0
80 1022.2 996.0 1041.0 1112.0 1084.0 1137.0 1112.0 1084.0 1137.0 1017.5 1001.0 1035.0
88 1254.8 1231.0 1269.0 1371.2 1352.0 1385.0 1371.2 1352.0 1385.0 1259.6 1243.0 1271.0
96 1539.0 1513.0 1554.0 1697.4 1663.0 1715.0 1697.4 1663.0 1715.0 1550.4 1524.0 1574.0
}\datatable
\addplot+ [
  color=two,
fill opacity=.3,
bar shift=6pt
]
table [
  x=num_qubits,
  y=detached_mean,
  y error plus expr=\thisrow{detached_max}-\thisrow{detached_mean},
  y error minus expr=\thisrow{detached_mean}-\thisrow{detached_min}
] {\datatable};

\addlegendentry{Pytket ESD}

\end{axis}
\end{tikzpicture}
    \end{subfigure}
    \begin{subfigure}{0.48\textwidth}
      \centering
      \begin{tikzpicture}
\begin{axis}[
    width=\columnwidth,
    height=0.75\columnwidth,
    ybar=15pt,
    ymin=0,
    ymax=400,
    symbolic x coords={16,24,32,40,48,56,64,72,80,88, 96},
    xtick=data,
    enlarge x limits=0.05,
    x tick style={draw=none},
    ylabel={{Time taken (s)}},
    xlabel={Number of qubits},
    x tick label style={
        yshift=-5pt,
        xshift=9pt,
        anchor=east
    },
    scaled y ticks = base 10:-2,
grid=minor,
    minor grid style={line width=0.5pt, draw=gray!60, densely dotted},
    legend style={
        at={(0.25,0.5)},
        anchor=south,
        legend columns=1,
        /tikz/every even column/.append style={column sep=1.5em},
        /tikz/every odd column/.append style={column sep=0.2em}
    },
    ymajorgrids = true,
    major grid style = { dotted, draw=gray!70 },
    bar width=2.5pt,
    error bars/y dir=both,
    error bars/y explicit,
    error bars/error mark options={
        rotate=90,
        mark size=1pt
    }
]
\pgfplotstableread[
  col sep=space,
  header=has colnames
]{
num_qubits fgp_mean fgp_min fgp_max
16 0.003966760635375976 0.0038008689880371094 0.004383087158203125
24 0.01966705322265625 0.017395973205566406 0.022716999053955078
32 0.07690587043762206 0.06716394424438477 0.11214828491210938
40 0.17305350303649902 0.14946675300598145 0.21014904975891113
48 0.35167632102966306 0.34048986434936523 0.3651139736175537
56 0.7180802345275878 0.6806490421295166 0.7869570255279541
64 1.1940353870391847 1.1528949737548828 1.2284841537475586
72 1.8929708003997803 1.8510158061981201 1.9608261585235596
80 2.959031343460083 2.881956100463867 3.101538896560669
88 4.610352611541748 4.340718746185303 4.946233034133911
96 6.561051654815674 6.318596839904785 6.798417091369629
}\datatable
\addplot+ [
  color=one,
fill opacity=.3,
bar shift = -6pt
]
table [
  x=num_qubits,
  y=fgp_mean,
  y error plus expr=\thisrow{fgp_max}-\thisrow{fgp_mean},
  y error minus expr=\thisrow{fgp_mean}-\thisrow{fgp_min}
] {\datatable};

\addlegendentry{FGP-rOEE}

\pgfplotstableread[
  col sep=space,
  header=has colnames
]{
num_qubits r_mean r_min r_max
16 0.0525 0.0502 0.0557
24 0.2061 0.1912 0.2322
32 0.4937 0.4638 0.5191
40 1.1093 1.0571 1.1439
48 1.9776 1.9147 2.0783
56 3.2693 3.1890 3.3610
64 5.1469 5.0502 5.2660
72 8.3910 8.2984 8.5631
80 12.1262 11.9594 12.2945
88 16.7062 16.5423 16.9746
96 22.5105 22.1376 22.8552
}\datatable
\addplot+ [
  color=teal,
fill opacity=.3,
bar shift=-2pt
]
table [
  x=num_qubits,
  y=r_mean,
  y error plus expr=\thisrow{r_max}-\thisrow{r_mean},
  y error minus expr=\thisrow{r_mean}-\thisrow{r_min}
] {\datatable};

\addlegendentry{MLFM-R}

\pgfplotstableread[
  col sep=space,
  header=has colnames
]{
num_qubits part_mean part_min part_max embed_mean embed_min embed_max
16 0.02773442268371582 0.02064204216003418 0.034748077392578125 0.009195232391357422 0.007353782653808594 0.010781288146972656
24 0.09533805847167968 0.08246684074401855 0.1142270565032959 0.055626511573791504 0.048445940017700195 0.06445598602294922
32 0.2141348123550415 0.18077325820922852 0.26163506507873535 0.18966774940490722 0.1647169589996338 0.21106815338134766
40 0.45887031555175783 0.373913049697876 0.5584170818328857 0.5571754693984985 0.5039520263671875 0.6013350486755371
48 0.6928657293319702 0.6370103359222412 0.8819680213928223 1.3001453399658203 1.1822409629821777 1.4405572414398193
56 1.1322642326354981 1.054063081741333 1.2124481201171875 2.6391883850097657 2.4270999431610107 2.875708818435669
64 1.7580196857452393 1.6767950057983398 1.887073040008545 4.700679063796997 4.570685148239136 4.989161968231201
72 2.574975872039795 2.4533729553222656 2.699486017227173 8.099718713760376 7.439749002456665 8.944865942001343
80 3.912350821495056 3.4602701663970947 4.777190923690796 12.911817073822021 11.658912897109985 13.744668006896973
88 5.2263283967971805 4.850888013839722 6.178251028060913 19.50987355709076 18.88469409942627 20.084751844406128
96 7.168324327468872 6.664013147354126 8.473192930221558 29.28718400001526 28.053810834884644 30.995959043502808
}\datatable
\addplot+ [
  color=three,
fill opacity=.3,
bar shift = 2pt
]
table [
  x=num_qubits,
  y=embed_mean,
  y error plus expr=\thisrow{embed_max}-\thisrow{embed_mean},
  y error minus expr=\thisrow{embed_mean}-\thisrow{embed_min}
] {\datatable};

\addlegendentry{Pytket PE}

\pgfplotstableread[
  col sep=space,
  header=has colnames
]{
num_qubits part_mean part_min part_max embed_mean embed_min embed_max
16 0.02773442268371582 0.02064204216003418 0.034748077392578125 0.009195232391357422 0.007353782653808594 0.010781288146972656
24 0.09533805847167968 0.08246684074401855 0.1142270565032959 0.055626511573791504 0.048445940017700195 0.06445598602294922
32 0.2141348123550415 0.18077325820922852 0.26163506507873535 0.18966774940490722 0.1647169589996338 0.21106815338134766
40 0.45887031555175783 0.373913049697876 0.5584170818328857 0.5571754693984985 0.5039520263671875 0.6013350486755371
48 0.6928657293319702 0.6370103359222412 0.8819680213928223 1.3001453399658203 1.1822409629821777 1.4405572414398193
56 1.1322642326354981 1.054063081741333 1.2124481201171875 2.6391883850097657 2.4270999431610107 2.875708818435669
64 1.7580196857452393 1.6767950057983398 1.887073040008545 4.700679063796997 4.570685148239136 4.989161968231201
72 2.574975872039795 2.4533729553222656 2.699486017227173 8.099718713760376 7.439749002456665 8.944865942001343
80 3.912350821495056 3.4602701663970947 4.777190923690796 12.911817073822021 11.658912897109985 13.744668006896973
88 5.2263283967971805 4.850888013839722 6.178251028060913 19.50987355709076 18.88469409942627 20.084751844406128
96 7.168324327468872 6.664013147354126 8.473192930221558 29.28718400001526 28.053810834884644 30.995959043502808
}\datatable
\addplot+ [
  color=two,
fill opacity=.3,
bar shift = -10000pt
]
table [
  x=num_qubits,
  y=,
] {\datatable};

\pgfplotstableread[
  col sep=space,
  header=has colnames
]{
num_qubits init_mean init_min init_max vc_mean vc_min vc_max steiner_mean steiner_min steiner_max detached_mean detached_min detached_max
16 0.02125202417373657 0.018611907958984375 0.027553081512451172 0.049959218502044676 0.040674686431884766 0.10684013366699219 0.08501414060592652 0.0661928653717041 0.13727903366088867 0.09219602346420289 0.07236099243164062 0.1448981761932373
24 0.0560227632522583 0.052960872650146484 0.06146597862243652 0.20382201671600342 0.1774899959564209 0.23929905891418457 0.3878135919570923 0.3385288715362549 0.4572148323059082 0.41967942714691164 0.3689577579498291 0.4921610355377197
32 0.13009073734283447 0.12040400505065918 0.14791107177734375 0.7700612783432007 0.7145299911499023 0.8787050247192383 1.476627230644226 1.3436329364776611 1.6720681190490723 1.5611873626708985 1.4253108501434326 1.765242099761963
40 0.23126046657562255 0.22086811065673828 0.2515387535095215 2.2761496543884276 2.056283712387085 2.4380056858062744 4.361180830001831 3.939086675643921 4.734943866729736 4.533787083625794 4.111143589019775 4.91733980178833
48 0.4201726198196411 0.3733401298522949 0.47313380241394043 5.952672839164734 5.253054141998291 6.376796245574951 11.256665110588074 9.85902714729309 12.284243822097778 11.575715565681458 10.147588014602661 12.628304719924927
56 0.6502074718475341 0.602808952331543 0.6957352161407471 13.094058012962341 12.202373027801514 13.601558208465576 24.468363666534422 22.33546280860901 26.61510729789734 24.990031266212462 22.801990032196045 27.160502433776855
64 0.9858702898025513 0.9397289752960205 1.059920072555542 25.45898833274841 24.482951164245605 27.75055432319641 47.84508681297302 45.62173819541931 52.869731187820435 48.67315726280212 46.420767307281494 53.78958010673523
72 1.4132241487503052 1.3626360893249512 1.4592149257659912 46.74429295063019 44.67621088027954 48.54746413230896 86.37684705257416 81.44859170913696 90.48327302932739 87.59926443099975 82.6563367843628 91.73335599899292
80 2.1223999738693236 1.9866011142730713 2.241624116897583 85.58975570201873 81.70053911209106 88.0958788394928 158.22104659080506 151.3307991027832 164.45687747001648 160.03656816482544 153.06363201141357 166.28980135917664
88 2.8402683258056642 2.6351962089538574 3.1394689083099365 136.22631855010985 131.49615693092346 140.6138472557068 248.35370421409607 239.46786189079285 255.27637124061584 250.83041644096375 241.7867078781128 257.9123034477234
96 3.752411699295044 3.6431920528411865 3.8656787872314453 211.48433394432067 203.2017903327942 215.1980917453766 387.14081902503966 370.6390326023102 398.562130689621 390.48248357772826 373.7858245372772 402.08690762519836
}\datatable
\addplot+ [
  color=two,
fill opacity=.3,
bar shift=6pt
]
table [
  x=num_qubits,
  y=detached_mean,
  y error plus expr=\thisrow{detached_max}-\thisrow{detached_mean},
  y error minus expr=\thisrow{detached_mean}-\thisrow{detached_min}
] {\datatable};

\addlegendentry{Pytket ESD}

\end{axis}
\end{tikzpicture}
    \end{subfigure}
    \caption{Circuit with 50\% two-qubit gates.}
    \label{fig:CP_50}
  \end{figure*}
  
  \begin{figure*}[!htbp]
    \centering
    \begin{subfigure}{0.48\textwidth}
      \centering
      \begin{tikzpicture}
\begin{axis}[
    width=\columnwidth,
    height=0.75\columnwidth,
    ybar=15pt,
    ymin=0,
    ymax=3500,
    symbolic x coords={16,24,32,40,48,56,64,72,80,88, 96},
    xtick=data,
    enlarge x limits=0.05,
    x tick style={draw=none},
    ylabel={Entanglement cost},
    xlabel={Number of qubits},
    x tick label style={
        yshift=-5pt,
        xshift=9pt,
        anchor=east
    },
    scaled y ticks = base 10:-3,
grid=minor,
    minor grid style={line width=0.5pt, draw=gray!60, densely dotted},
    legend style={
        at={(0.25,0.5)},
        anchor=south,
        legend columns=1,
        /tikz/every even column/.append style={column sep=1.5em},
        /tikz/every odd column/.append style={column sep=0.2em}
    },
    ymajorgrids = true,
    major grid style = { dotted, draw=gray!70 },
    bar width=2.5pt,
    error bars/y dir=both,
    error bars/y explicit,
    error bars/error mark options={
        rotate=90,
        mark size=1pt
    }
]
\pgfplotstableread[
  col sep=space,
  header=has colnames
]{
num_qubits fgp_mean fgp_min fgp_max
16 33.0 28.0 36.0
24 111.2 108.0 113.0
32 237.2 228.0 251.0
40 415.8 412.0 421.0
48 658.8 636.0 674.0
56 961.0 947.0 978.0
64 1302.4 1285.0 1340.0
72 1708.2 1680.0 1728.0
80 2182.2 2141.0 2256.0
88 2686.6 2645.0 2716.0
96 3247.2 3227.0 3278.0
}\datatable
\addplot+ [
  color=one,
fill opacity=.3,
bar shift = -6pt
]
table [
  x=num_qubits,
  y=fgp_mean,
  y error plus expr=\thisrow{fgp_max}-\thisrow{fgp_mean},
  y error minus expr=\thisrow{fgp_mean}-\thisrow{fgp_min}
] {\datatable};

\addlegendentry{FGP-rOEE}

\pgfplotstableread[
  col sep=space,
  header=has colnames
]{
num_qubits r_mean r_min r_max
16 14.1 13.0 16.0
24 50.1 44.0 55.0
32 114.9 108.0 126.0
40 208.2 198.0 221.0
48 339.3 327.0 349.0
56 508.9 492.0 524.0
64 701.9 679.0 727.0
72 956.6 943.0 974.0
80 1159.7 1129.0 1177.0
88 1462.3 1431.0 1485.0
96 1785.0 1772.0 1830.0
}\datatable
\addplot+ [
  color=teal,
fill opacity=.3,
bar shift=-2pt
]
table [
  x=num_qubits,
  y=r_mean,
  y error plus expr=\thisrow{r_max}-\thisrow{r_mean},
  y error minus expr=\thisrow{r_mean}-\thisrow{r_min}
] {\datatable};

\addlegendentry{MLFM-R}

\pgfplotstableread[
  col sep=space,
  header=has colnames
]{
num_qubits part_mean part_min part_max embed_mean embed_min embed_max
16 13.6 12.0 15.0 13.6 12.0 15.0
24 47.5 41.0 51.0 47.5 41.0 51.0
32 108.7 104.0 118.0 108.7 104.0 118.0
40 200.4 193.0 208.0 200.4 193.0 208.0
48 328.2 321.0 344.0 328.2 321.0 344.0
56 479.7 467.0 489.0 479.7 467.0 489.0
64 654.5 642.0 664.0 654.5 642.0 664.0
72 866.2 850.0 886.0 866.2 850.0 886.0
80 1116.5 1082.0 1147.0 1116.5 1082.0 1147.0
88 1397.4 1377.0 1421.0 1397.4 1377.0 1421.0
96 1691.3 1661.0 1724.0 1691.3 1661.0 1724.0
}\datatable
\addplot+ [
  color=three,
fill opacity=.3,
bar shift = 2pt
]
table [
  x=num_qubits,
  y=embed_mean,
  y error plus expr=\thisrow{embed_max}-\thisrow{embed_mean},
  y error minus expr=\thisrow{embed_mean}-\thisrow{embed_min}
] {\datatable};

\addlegendentry{Pytket PE}

\pgfplotstableread[
  col sep=space,
  header=has colnames
]{
num_qubits part_mean part_min part_max embed_mean embed_min embed_max
16 0.03419475555419922 0.023661136627197266 0.04221487045288086 0.014299321174621581 0.012793779373168945 0.01608896255493164
24 0.14852614402770997 0.10640096664428711 0.21245503425598145 0.08178679943084717 0.0760047435760498 0.08944582939147949
32 0.330162787437439 0.2701258659362793 0.4044609069824219 0.3309386491775513 0.3072190284729004 0.35906291007995605
40 0.6864403247833252 0.5557730197906494 0.8604509830474854 0.9229176998138428 0.8329761028289795 1.0153987407684326
48 1.2418020486831665 1.009756088256836 1.6725561618804932 2.2355015516281127 2.0204880237579346 2.393527030944824
56 2.0419219732284546 1.7812631130218506 2.2060978412628174 4.591543030738831 4.40931510925293 4.954710006713867
64 3.2909801959991456 2.6754097938537598 3.6469991207122803 8.222691750526428 7.6230058670043945 8.871248960494995
72 5.17089877128601 4.606496810913086 5.794766187667847 14.016770911216735 13.00545597076416 14.87307333946228
80 7.394749617576599 5.772594928741455 8.142525911331177 22.545075035095216 21.68347716331482 23.704101085662842
88 11.027815580368042 9.344386100769043 12.013834238052368 35.090847516059874 33.07154989242554 37.46260690689087
96 16.13992247581482 14.833529949188232 17.00726819038391 54.3115871667862 52.79211378097534 55.8539400100708
}\datatable
\addplot+ [
  color=two,
fill opacity=.3,
bar shift = -10000pt
]
table [
  x=num_qubits,
  y=,
] {\datatable};

\pgfplotstableread[
  col sep=space,
  header=has colnames
]{
num_qubits init_mean init_min init_max vc_mean vc_min vc_max steiner_mean steiner_min steiner_max detached_mean detached_min detached_max
16 13.3 12.0 15.0 13.3 12.0 15.0 13.3 12.0 15.0 13.3 12.0 15.0
24 47.4 43.0 52.0 51.8 48.0 58.0 51.8 48.0 58.0 47.8 44.0 53.0
32 113.5 111.0 118.0 123.1 120.0 128.0 123.1 120.0 128.0 112.1 108.0 115.0
40 206.8 199.0 211.0 234.6 224.0 243.0 234.6 224.0 243.0 204.4 197.0 211.0
48 336.8 326.0 349.0 382.3 373.0 395.0 382.3 373.0 395.0 330.0 320.0 343.0
56 495.6 486.0 507.0 566.9 554.0 576.0 566.9 554.0 576.0 490.7 482.0 507.0
64 665.0 651.0 679.0 795.9 785.0 812.0 795.9 785.0 812.0 671.9 659.0 688.0
72 882.4 851.0 898.0 1065.6 1026.0 1087.0 1065.6 1026.0 1087.0 900.4 877.0 918.0
80 1140.7 1115.0 1167.0 1380.9 1338.0 1403.0 1380.9 1338.0 1403.0 1167.2 1136.0 1192.0
88 1414.25 1379.0 1431.0 1732.25 1715.0 1752.0 1732.25 1715.0 1752.0 1446.75 1423.0 1470.0
96 1714.8 1697.0 1731.0 2137.4 2117.0 2153.0 2137.4 2117.0 2153.0 1778.6 1768.0 1786.0
}\datatable
\addplot+ [
  color=two,
fill opacity=.3,
bar shift=6pt
]
table [
  x=num_qubits,
  y=detached_mean,
  y error plus expr=\thisrow{detached_max}-\thisrow{detached_mean},
  y error minus expr=\thisrow{detached_mean}-\thisrow{detached_min}
] {\datatable};

\addlegendentry{Pytket ESD}

\end{axis}
\end{tikzpicture}
    \end{subfigure}
    \begin{subfigure}{0.48\textwidth}
      \centering
      \begin{tikzpicture}
\begin{axis}[
    width=\columnwidth,
    height=0.75\columnwidth,
    ybar=15pt,
    ymin=0,
    ymax=800,
    symbolic x coords={16,24,32,40,48,56,64,72,80,88, 96},
    xtick=data,
    enlarge x limits=0.05,
    x tick style={draw=none},
    ylabel={{Time taken (s)}},
    xlabel={Number of qubits},
    x tick label style={
        yshift=-5pt,
        xshift=9pt,
        anchor=east
    },
    scaled y ticks = base 10:-2,
grid=minor,
    minor grid style={line width=0.5pt, draw=gray!60, densely dotted},
    legend style={
        at={(0.25,0.5)},
        anchor=south,
        legend columns=1,
        /tikz/every even column/.append style={column sep=1.5em},
        /tikz/every odd column/.append style={column sep=0.2em}
    },
    ymajorgrids = true,
    major grid style = { dotted, draw=gray!70 },
    bar width=2.5pt,
    error bars/y dir=both,
    error bars/y explicit,
    error bars/error mark options={
        rotate=90,
        mark size=1pt
    }
]
\pgfplotstableread[
  col sep=space,
  header=has colnames
]{
num_qubits fgp_mean fgp_min fgp_max
16 0.005234813690185547 0.0047838687896728516 0.005610942840576172
24 0.03825278282165527 0.026273012161254883 0.07964205741882324
32 0.09218869209289551 0.08646774291992188 0.09483885765075684
40 0.24344067573547362 0.2288062572479248 0.2623608112335205
48 0.5168071746826172 0.4884660243988037 0.5361661911010742
56 1.0263038158416748 0.977888822555542 1.1057941913604736
64 1.7255557537078858 1.657433271408081 1.8431282043457031
72 2.7922242164611815 2.7170569896698 2.891957998275757
80 4.429377126693725 4.242176055908203 4.575012922286987
88 6.697661066055298 6.414860010147095 6.951433897018433
96 9.124274063110352 8.865937948226929 9.566908359527588
}\datatable
\addplot+ [
  color=one,
fill opacity=.3,
bar shift = -6pt
]
table [
  x=num_qubits,
  y=fgp_mean,
  y error plus expr=\thisrow{fgp_max}-\thisrow{fgp_mean},
  y error minus expr=\thisrow{fgp_mean}-\thisrow{fgp_min}
] {\datatable};

\addlegendentry{FGP-rOEE}

\pgfplotstableread[
  col sep=space,
  header=has colnames
]{
num_qubits r_mean r_min r_max
16 0.0642 0.0597 0.0685
24 0.2565 0.2425 0.2827
32 0.6461 0.6028 0.7094
40 1.4082 1.3441 1.4960
48 2.5476 2.4841 2.5947
56 4.2377 4.0896 4.4087
64 6.6949 6.4779 6.8342
72 10.8181 10.4468 11.1664
80 15.3976 15.1714 15.6535
88 21.4883 21.1026 21.7966
96 28.8951 28.6799 29.2527
}\datatable
\addplot+ [
  color=teal,
fill opacity=.3,
bar shift=-2pt
]
table [
  x=num_qubits,
  y=r_mean,
  y error plus expr=\thisrow{r_max}-\thisrow{r_mean},
  y error minus expr=\thisrow{r_mean}-\thisrow{r_min}
] {\datatable};

\addlegendentry{MLFM-R}

\pgfplotstableread[
  col sep=space,
  header=has colnames
]{
num_qubits part_mean part_min part_max embed_mean embed_min embed_max
16 0.03419475555419922 0.023661136627197266 0.04221487045288086 0.014299321174621581 0.012793779373168945 0.01608896255493164
24 0.14852614402770997 0.10640096664428711 0.21245503425598145 0.08178679943084717 0.0760047435760498 0.08944582939147949
32 0.330162787437439 0.2701258659362793 0.4044609069824219 0.3309386491775513 0.3072190284729004 0.35906291007995605
40 0.6864403247833252 0.5557730197906494 0.8604509830474854 0.9229176998138428 0.8329761028289795 1.0153987407684326
48 1.2418020486831665 1.009756088256836 1.6725561618804932 2.2355015516281127 2.0204880237579346 2.393527030944824
56 2.0419219732284546 1.7812631130218506 2.2060978412628174 4.591543030738831 4.40931510925293 4.954710006713867
64 3.2909801959991456 2.6754097938537598 3.6469991207122803 8.222691750526428 7.6230058670043945 8.871248960494995
72 5.17089877128601 4.606496810913086 5.794766187667847 14.016770911216735 13.00545597076416 14.87307333946228
80 7.394749617576599 5.772594928741455 8.142525911331177 22.545075035095216 21.68347716331482 23.704101085662842
88 11.027815580368042 9.344386100769043 12.013834238052368 35.090847516059874 33.07154989242554 37.46260690689087
96 16.13992247581482 14.833529949188232 17.00726819038391 54.3115871667862 52.79211378097534 55.8539400100708
}\datatable
\addplot+ [
  color=three,
fill opacity=.3,
bar shift = 2pt
]
table [
  x=num_qubits,
  y=embed_mean,
  y error plus expr=\thisrow{embed_max}-\thisrow{embed_mean},
  y error minus expr=\thisrow{embed_mean}-\thisrow{embed_min}
] {\datatable};

\addlegendentry{Pytket PE}

\pgfplotstableread[
  col sep=space,
  header=has colnames
]{
num_qubits part_mean part_min part_max embed_mean embed_min embed_max
16 0.03419475555419922 0.023661136627197266 0.04221487045288086 0.014299321174621581 0.012793779373168945 0.01608896255493164
24 0.14852614402770997 0.10640096664428711 0.21245503425598145 0.08178679943084717 0.0760047435760498 0.08944582939147949
32 0.330162787437439 0.2701258659362793 0.4044609069824219 0.3309386491775513 0.3072190284729004 0.35906291007995605
40 0.6864403247833252 0.5557730197906494 0.8604509830474854 0.9229176998138428 0.8329761028289795 1.0153987407684326
48 1.2418020486831665 1.009756088256836 1.6725561618804932 2.2355015516281127 2.0204880237579346 2.393527030944824
56 2.0419219732284546 1.7812631130218506 2.2060978412628174 4.591543030738831 4.40931510925293 4.954710006713867
64 3.2909801959991456 2.6754097938537598 3.6469991207122803 8.222691750526428 7.6230058670043945 8.871248960494995
72 5.17089877128601 4.606496810913086 5.794766187667847 14.016770911216735 13.00545597076416 14.87307333946228
80 7.394749617576599 5.772594928741455 8.142525911331177 22.545075035095216 21.68347716331482 23.704101085662842
88 11.027815580368042 9.344386100769043 12.013834238052368 35.090847516059874 33.07154989242554 37.46260690689087
96 16.13992247581482 14.833529949188232 17.00726819038391 54.3115871667862 52.79211378097534 55.8539400100708
}\datatable
\addplot+ [
  color=two,
fill opacity=.3,
bar shift = -10000pt
]
table [
  x=num_qubits,
  y=,
] {\datatable};

\pgfplotstableread[
  col sep=space,
  header=has colnames
]{
num_qubits init_mean init_min init_max vc_mean vc_min vc_max steiner_mean steiner_min steiner_max detached_mean detached_min detached_max
16 0.02297499179840088 0.022466182708740234 0.023786067962646484 0.055050992965698244 0.05130791664123535 0.05892205238342285 0.11540286540985108 0.10608196258544922 0.12679624557495117 0.12831199169158936 0.11588406562805176 0.14458417892456055
24 0.06984786987304688 0.06655716896057129 0.07550883293151855 0.2810733079910278 0.25336289405822754 0.31363487243652344 0.6602291345596314 0.6088700294494629 0.7051002979278564 0.7229100465774536 0.6577701568603516 0.7782890796661377
32 0.16675896644592286 0.16014480590820312 0.17229509353637695 1.1220108985900878 1.074319839477539 1.188506841659546 2.6034271001815794 2.4494497776031494 2.742461919784546 2.7814005851745605 2.6117498874664307 2.9184396266937256
40 0.3272660255432129 0.3218519687652588 0.3323521614074707 3.5581010818481444 3.346644163131714 3.7016210556030273 8.135493087768555 7.860360860824585 8.383244276046753 8.525331830978393 8.238452911376953 8.793975830078125
48 0.5989335775375366 0.5636558532714844 0.657783031463623 9.270846843719482 8.86327052116394 10.193022012710571 20.755881905555725 19.732389450073242 23.461267948150635 21.458326196670534 20.448893308639526 24.21415686607361
56 0.9678317785263062 0.9241571426391602 1.065490961074829 20.702145838737486 20.058820009231567 21.316049337387085 45.55994622707367 44.62418985366821 48.67478537559509 46.76969091892242 45.75250172615051 49.95874834060669
64 1.5069351196289062 1.4760479927062988 1.5526351928710938 41.859218096733095 40.833412170410156 42.683711767196655 91.8092610836029 89.0262622833252 93.62563967704773 93.78969964981079 90.732351064682 95.95378875732422
72 2.320264983177185 2.2181198596954346 2.6060378551483154 79.56894557476043 76.34138989448547 81.7564549446106 173.39698889255524 166.7985680103302 179.41522908210754 176.56229190826417 170.43360686302185 182.63322806358337
80 3.3024064540863036 3.193856954574585 3.4633591175079346 139.0046290874481 133.40762877464294 143.23986268043518 299.25269610881804 290.1907937526703 309.99761271476746 303.6747828245163 295.2237358093262 314.33749556541443
88 4.572441935539246 4.471622943878174 4.725454807281494 232.23154711723328 226.48475098609924 236.12219333648682 499.51616299152374 490.1397647857666 505.2299690246582 505.5903716683388 496.34189558029175 510.6568658351898
96 6.40035924911499 6.126535177230835 6.6452271938323975 361.15179772377013 353.8287272453308 368.5010280609131 766.9568595409394 745.9088091850281 780.4633054733276 775.5569538593293 753.8309440612793 789.6226325035095
}\datatable
\addplot+ [
  color=two,
fill opacity=.3,
bar shift=6pt
]
table [
  x=num_qubits,
  y=detached_mean,
  y error plus expr=\thisrow{detached_max}-\thisrow{detached_mean},
  y error minus expr=\thisrow{detached_mean}-\thisrow{detached_min}
] {\datatable};

\addlegendentry{Pytket ESD}

\end{axis}
\end{tikzpicture}
    \end{subfigure}
    \caption{Circuit with 70\% two-qubit gates.}
    \label{fig:CP_70}
  \end{figure*}
  
  \begin{figure*}[!htbp]
    \begin{subfigure}{0.48\textwidth}
      \centering
      \begin{tikzpicture}
\begin{axis}[
    width=\columnwidth,
    height=0.75\columnwidth,
    ybar=15pt,
    ymin=0,
    ymax=4500,
    symbolic x coords={16,24,32,40,48,56,64,72,80,88, 96},
    xtick=data,
    enlarge x limits=0.05,
    x tick style={draw=none},
    ylabel={Entanglement cost},
    xlabel={Number of qubits},
    x tick label style={
        yshift=-5pt,
        xshift=9pt,
        anchor=east
    },
    scaled y ticks = base 10:-3,
grid=minor,
    minor grid style={line width=0.5pt, draw=gray!60, densely dotted},
    legend style={
        at={(0.25,0.5)},
        anchor=south,
        legend columns=1,
        /tikz/every even column/.append style={column sep=1.5em},
        /tikz/every odd column/.append style={column sep=0.2em}
    },
    ymajorgrids = true,
    major grid style = { dotted, draw=gray!70 },
    bar width=2.5pt,
    error bars/y dir=both,
    error bars/y explicit,
    error bars/error mark options={
        rotate=90,
        mark size=1pt
    }
]
\pgfplotstableread[
  col sep=space,
  header=has colnames
]{
num_qubits fgp_mean fgp_min fgp_max
16 42.0 41.0 44.0
24 146.4 127.0 164.0
32 304.0 300.0 307.0
40 541.8 531.0 557.0
48 827.8 797.0 863.0
56 1223.0 1185.0 1255.0
64 1668.2 1635.0 1705.0
72 2154.6 2124.0 2199.0
80 2775.8 2720.0 2885.0
88 3455.0 3388.0 3494.0
96 4179.2 4057.0 4284.0
}\datatable
\addplot+ [
  color=one,
fill opacity=.3,
bar shift = -6pt
]
table [
  x=num_qubits,
  y=fgp_mean,
  y error plus expr=\thisrow{fgp_max}-\thisrow{fgp_mean},
  y error minus expr=\thisrow{fgp_mean}-\thisrow{fgp_min}
] {\datatable};

\addlegendentry{FGP-rOEE}

\pgfplotstableread[
  col sep=space,
  header=has colnames
]{
num_qubits r_mean r_min r_max
16 12.5 10.0 15.0
24 42.2 38.0 44.0
32 95.9 85.0 110.0
40 181.5 164.0 191.0
48 290.3 266.0 304.0
56 445.8 426.0 474.0
64 626.3 606.0 643.0
72 864.6 824.0 915.0
80 1088.5 1051.0 1121.0
88 1382.4 1355.0 1431.0
96 1726.5 1686.0 1754.0
}\datatable
\addplot+ [
  color=teal,
fill opacity=.3,
bar shift=-2pt
]
table [
  x=num_qubits,
  y=r_mean,
  y error plus expr=\thisrow{r_max}-\thisrow{r_mean},
  y error minus expr=\thisrow{r_mean}-\thisrow{r_min}
] {\datatable};

\addlegendentry{MLFM-R}

\pgfplotstableread[
  col sep=space,
  header=has colnames
]{
num_qubits part_mean part_min part_max embed_mean embed_min embed_max
16 10.9 9.0 12.0 10.9 9.0 12.0
24 38.0 36.0 42.0 38.0 36.0 42.0
32 83.7 71.0 90.0 83.7 71.0 90.0
40 147.3 130.0 160.0 147.3 130.0 160.0
48 223.2 202.0 239.0 223.2 202.0 239.0
56 312.4 286.0 331.0 312.4 286.0 331.0
64 439.6 421.0 462.0 439.6 421.0 462.0
72 564.5 534.0 583.0 564.5 534.0 583.0
80 707.3 653.0 753.0 707.3 653.0 753.0
88 844.7 822.0 888.0 844.7 822.0 888.0
96 1031.3 993.0 1098.0 1031.3 993.0 1098.0
}\datatable
\addplot+ [
  color=three,
fill opacity=.3,
bar shift = 2pt
]
table [
  x=num_qubits,
  y=embed_mean,
  y error plus expr=\thisrow{embed_max}-\thisrow{embed_mean},
  y error minus expr=\thisrow{embed_mean}-\thisrow{embed_min}
] {\datatable};

\addlegendentry{Pytket PE}

\pgfplotstableread[
  col sep=space,
  header=has colnames
]{
num_qubits part_mean part_min part_max embed_mean embed_min embed_max
16 0.03419475555419922 0.023661136627197266 0.04221487045288086 0.014299321174621581 0.012793779373168945 0.01608896255493164
24 0.14852614402770997 0.10640096664428711 0.21245503425598145 0.08178679943084717 0.0760047435760498 0.08944582939147949
32 0.330162787437439 0.2701258659362793 0.4044609069824219 0.3309386491775513 0.3072190284729004 0.35906291007995605
40 0.6864403247833252 0.5557730197906494 0.8604509830474854 0.9229176998138428 0.8329761028289795 1.0153987407684326
48 1.2418020486831665 1.009756088256836 1.6725561618804932 2.2355015516281127 2.0204880237579346 2.393527030944824
56 2.0419219732284546 1.7812631130218506 2.2060978412628174 4.591543030738831 4.40931510925293 4.954710006713867
64 3.2909801959991456 2.6754097938537598 3.6469991207122803 8.222691750526428 7.6230058670043945 8.871248960494995
72 5.17089877128601 4.606496810913086 5.794766187667847 14.016770911216735 13.00545597076416 14.87307333946228
80 7.394749617576599 5.772594928741455 8.142525911331177 22.545075035095216 21.68347716331482 23.704101085662842
88 11.027815580368042 9.344386100769043 12.013834238052368 35.090847516059874 33.07154989242554 37.46260690689087
96 16.13992247581482 14.833529949188232 17.00726819038391 54.3115871667862 52.79211378097534 55.8539400100708
}\datatable
\addplot+ [
  color=two,
fill opacity=.3,
bar shift = -10000pt
]
table [
  x=num_qubits,
  y=,
] {\datatable};

\pgfplotstableread[
  col sep=space,
  header=has colnames
]{
num_qubits init_mean init_min init_max vc_mean vc_min vc_max steiner_mean steiner_min steiner_max detached_mean detached_min detached_max
16 11.2 10.0 12.0 11.1 10.0 12.0 11.1 10.0 12.0 11.1 10.0 12.0
24 39.7 37.0 44.0 42.2 37.0 51.0 42.2 37.0 51.0 39.0 33.0 47.0
32 88.7 80.0 99.0 101.0 93.0 112.0 101.0 93.0 112.0 90.6 84.0 105.0
40 158.6 148.0 169.0 185.3 171.0 199.0 185.3 171.0 199.0 156.7 141.0 174.0
48 236.3 221.0 248.0 306.7 294.0 335.0 306.7 294.0 335.0 251.5 238.0 267.0
56 331.4 308.0 350.0 452.9 425.0 484.0 452.9 425.0 484.0 362.0 347.0 380.0
64 455.0 433.0 487.0 658.6 601.0 686.0 658.6 601.0 686.0 511.1 468.0 530.0
72 592.6 571.0 610.0 886.3 811.0 925.0 886.3 811.0 925.0 672.2 631.0 706.0
80 757.2 735.0 780.0 1173.2 1100.0 1213.0 1173.2 1100.0 1213.0 868.8 833.0 902.0
88 905.2 886.0 937.0 1474.8 1438.0 1514.0 1474.8 1438.0 1514.0 1071.8 1026.0 1123.0
96 1127.4 1114.0 1153.0 1838.2 1822.0 1874.0 1838.2 1822.0 1874.0 1331.4 1275.0 1363.0
}\datatable
\addplot+ [
  color=two,
fill opacity=.3,
bar shift = 6pt
]
table [
  x=num_qubits,
  y=detached_mean,
  y error plus expr=\thisrow{detached_max}-\thisrow{detached_mean},
  y error minus expr=\thisrow{detached_mean}-\thisrow{detached_min}
] {\datatable};

\addlegendentry{Pytket ESD}

\end{axis}
\end{tikzpicture}
    \end{subfigure}
    \begin{subfigure}{0.48\textwidth}
      \centering
      \begin{tikzpicture}
\begin{axis}[
    width=\columnwidth,
    height=0.75\columnwidth,
    ybar=15pt,
    ymin=0,
    ymax=1000,
    symbolic x coords={16,24,32,40,48,56,64,72,80,88, 96},
    xtick=data,
    enlarge x limits=0.05,
    x tick style={draw=none},
    ylabel={Time taken (s)},
    xlabel={Number of qubits},
    x tick label style={
        yshift=-5pt,
        xshift=9pt,
        anchor=east
    },
    scaled y ticks = base 10:-2,
grid=minor,
    minor grid style={line width=0.5pt, draw=gray!60, densely dotted},
    legend style={
        at={(0.25,0.5)},
        anchor=south,
        legend columns=1,
        /tikz/every even column/.append style={column sep=1.5em},
        /tikz/every odd column/.append style={column sep=0.2em}
    },
    ymajorgrids = true,
    major grid style = { dotted, draw=gray!70 },
    bar width=2.5pt,
    error bars/y dir=both,
    error bars/y explicit,
    error bars/error mark options={
        rotate=90,
        mark size=1pt
    }
]
\pgfplotstableread[
  col sep=space,
  header=has colnames
]{
num_qubits fgp_mean fgp_min fgp_max
16 0.006081771850585937 0.005761861801147461 0.0062520503997802734
24 0.041736698150634764 0.03324079513549805 0.05220508575439453
32 0.2031567096710205 0.12992000579833984 0.34691476821899414
40 0.3506504535675049 0.3303558826446533 0.3716471195220947
48 0.7998756885528564 0.7278158664703369 0.8552548885345459
56 1.65540771484375 1.5417349338531494 1.8472418785095215
64 2.8926740169525145 2.6083059310913086 3.1477460861206055
72 4.863784646987915 4.4808430671691895 5.630013942718506
80 7.7091165542602536 7.072470188140869 8.281652927398682
88 12.05374584197998 11.508241176605225 13.407248258590698
96 17.2431782245636 13.73129391670227 19.012799978256226
}\datatable
\addplot+ [
  color=one,
fill opacity=.3,
bar shift = -6pt
]
table [
  x=num_qubits,
  y=fgp_mean,
  y error plus expr=\thisrow{fgp_max}-\thisrow{fgp_mean},
  y error minus expr=\thisrow{fgp_mean}-\thisrow{fgp_min}
] {\datatable};

\addlegendentry{FGP-rOEE}

\pgfplotstableread[
  col sep=space,
  header=has colnames
]{
num_qubits r_mean r_min r_max
16 0.0814 0.0732 0.0903
24 0.3602 0.3365 0.3888
32 0.9985 0.8877 1.0779
40 2.2227 2.1233 2.4137
48 4.2226 3.9063 4.4958
56 7.0250 6.5248 7.6515
64 11.1219 10.4169 11.6532
72 17.3891 15.8775 18.2770
80 25.4668 24.0533 26.4226
88 35.8380 34.4886 36.8210
96 48.2697 46.6446 49.8950
}\datatable
\addplot+ [
  color=teal,
fill opacity=.3,
bar shift=-2pt
]
table [
  x=num_qubits,
  y=r_mean,
  y error plus expr=\thisrow{r_max}-\thisrow{r_mean},
  y error minus expr=\thisrow{r_mean}-\thisrow{r_min}
] {\datatable};

\addlegendentry{MLFM-R}

\pgfplotstableread[
  col sep=space,
  header=has colnames
]{
num_qubits part_mean part_min part_max embed_mean embed_min embed_max
16 0.04667329788208008 0.04251503944396973 0.052423954010009766 0.01642429828643799 0.013830900192260742 0.029082775115966797
24 0.18525230884552002 0.17531371116638184 0.20311903953552246 0.07707314491271973 0.07307696342468262 0.0821981430053711
32 0.47943720817565916 0.40457868576049805 0.5793416500091553 0.27271740436553954 0.2585480213165283 0.29368019104003906
40 1.0899725914001466 0.925239086151123 1.2354631423950195 0.7539139986038208 0.7112808227539062 0.7784173488616943
48 2.4459378242492678 2.002049207687378 3.1868348121643066 1.6757470846176148 1.5782709121704102 1.7441339492797852
56 3.9337402105331423 3.496495008468628 4.480030298233032 3.4874715328216555 3.3928611278533936 3.567575693130493
64 6.821216797828674 5.326786041259766 8.253294944763184 6.31878182888031 6.028108835220337 6.57309103012085
72 10.75671980381012 8.088166952133179 12.75140118598938 11.000451350212098 10.438180923461914 11.462172031402588
80 15.253899192810058 12.911457061767578 17.297257900238037 18.31858186721802 17.124468088150024 19.775872945785522
88 20.784938502311707 17.136726140975952 25.021101713180542 28.494950580596925 27.182917833328247 29.732630014419556
96 28.429611253738404 23.526342153549194 36.589191198349 43.81736562252045 41.6746621131897 45.17979598045349
}\datatable
\addplot+ [
  color=three,
fill opacity=.3,
bar shift = 2pt
]
table [
  x=num_qubits,
  y=embed_mean,
  y error plus expr=\thisrow{embed_max}-\thisrow{embed_mean},
  y error minus expr=\thisrow{embed_mean}-\thisrow{embed_min}
] {\datatable};

\addlegendentry{Pytket PE}

\pgfplotstableread[
  col sep=space,
  header=has colnames
]{
num_qubits part_mean part_min part_max embed_mean embed_min embed_max
16 0.03419475555419922 0.023661136627197266 0.04221487045288086 0.014299321174621581 0.012793779373168945 0.01608896255493164
24 0.14852614402770997 0.10640096664428711 0.21245503425598145 0.08178679943084717 0.0760047435760498 0.08944582939147949
32 0.330162787437439 0.2701258659362793 0.4044609069824219 0.3309386491775513 0.3072190284729004 0.35906291007995605
40 0.6864403247833252 0.5557730197906494 0.8604509830474854 0.9229176998138428 0.8329761028289795 1.0153987407684326
48 1.2418020486831665 1.009756088256836 1.6725561618804932 2.2355015516281127 2.0204880237579346 2.393527030944824
56 2.0419219732284546 1.7812631130218506 2.2060978412628174 4.591543030738831 4.40931510925293 4.954710006713867
64 3.2909801959991456 2.6754097938537598 3.6469991207122803 8.222691750526428 7.6230058670043945 8.871248960494995
72 5.17089877128601 4.606496810913086 5.794766187667847 14.016770911216735 13.00545597076416 14.87307333946228
80 7.394749617576599 5.772594928741455 8.142525911331177 22.545075035095216 21.68347716331482 23.704101085662842
88 11.027815580368042 9.344386100769043 12.013834238052368 35.090847516059874 33.07154989242554 37.46260690689087
96 16.13992247581482 14.833529949188232 17.00726819038391 54.3115871667862 52.79211378097534 55.8539400100708
}\datatable
\addplot+ [
  color=two,
fill opacity=.3,
bar shift = -10000pt
]
table [
  x=num_qubits,
  y=,
] {\datatable};

\pgfplotstableread[
  col sep=space,
  header=has colnames
]{
num_qubits init_mean init_min init_max vc_mean vc_min vc_max steiner_mean steiner_min steiner_max detached_mean detached_min detached_max
16 0.022627663612365723 0.021724939346313477 0.02363896369934082 0.055414128303527835 0.05337405204772949 0.0567479133605957 0.16479928493499757 0.1444079875946045 0.1890697479248047 0.19114937782287597 0.17142176628112793 0.21599745750427246
24 0.07406561374664307 0.07114386558532715 0.07671785354614258 0.2709683656692505 0.251964807510376 0.29201197624206543 0.8598877429962158 0.7757754325866699 0.9602367877960205 0.9827005863189697 0.9013304710388184 1.079939842224121
32 0.1892146348953247 0.1730940341949463 0.24834465980529785 1.0438527822494508 0.9644229412078857 1.1428537368774414 3.3625276565551756 3.2107410430908203 3.6033716201782227 3.728696870803833 3.5679171085357666 3.9907307624816895
40 0.38930952548980713 0.3700368404388428 0.45253610610961914 3.111409068107605 2.886110782623291 3.3550562858581543 9.978955483436584 9.274816989898682 10.207841396331787 10.797028613090514 10.099455833435059 11.040236711502075
48 0.6888649702072144 0.6733896732330322 0.7305710315704346 7.914584946632385 7.5442328453063965 8.531921148300171 24.534779024124145 23.834250926971436 25.451282262802124 26.729133105278017 26.042906522750854 27.700409650802612
56 1.1546093225479126 1.0931897163391113 1.1971371173858643 17.404848170280456 16.181886672973633 18.71316933631897 53.5791748046875 51.94565010070801 54.80066800117493 57.98932385444641 56.71221208572388 60.56390690803528
64 1.760768461227417 1.7051868438720703 1.8390121459960938 36.09398736953735 33.236960649490356 37.66968512535095 107.57090601921081 104.86399435997009 111.47021508216858 115.07106499671936 111.71270108222961 117.68997979164124
72 2.6994744777679442 2.5699000358581543 2.917736768722534 68.52888834476471 61.2776939868927 71.69985318183899 200.5330825805664 190.31355500221252 207.9263789653778 214.30033280849457 206.44268083572388 223.29197883605957
80 4.160609579086303 3.965764045715332 4.413243055343628 124.0204598903656 114.79535818099976 128.60104393959045 353.135338973999 335.9334342479706 361.0396628379822 373.2520549297333 359.4194130897522 381.8565979003906
88 5.935128831863404 5.75123405456543 6.033010959625244 208.20697302818297 205.40908002853394 213.07338094711304 595.3647183895112 585.3020310401917 603.6201448440552 625.3752370834351 613.3455142974854 639.0390939712524
96 7.783886241912842 7.574444055557251 7.911603212356567 326.63650169372556 323.0268247127533 333.52667117118835 910.5237064361572 901.2565882205963 919.3483798503876 953.4719588756561 946.11363530159 959.100034236908
}\datatable
\addplot+ [
  color=two,
fill opacity=.3,
bar shift=6pt
]
table [
  x=num_qubits,
  y=detached_mean,
  y error plus expr=\thisrow{detached_max}-\thisrow{detached_mean},
  y error minus expr=\thisrow{detached_mean}-\thisrow{detached_min}
] {\datatable};

\addlegendentry{Pytket ESD}

\end{axis}
\end{tikzpicture}
    \end{subfigure}
    \caption{Circuit with 90\% two-qubit gates.}
    \label{fig:CP_90}
  \end{figure*}

  \begin{figure*}[!htbp]
    \centering
    \begin{subfigure}{0.48\textwidth}
      \centering
      \begin{tikzpicture}
\begin{axis}[
    width=\columnwidth,
    height=0.75\columnwidth,
    ybar,                     
    bar width=2.5pt,         
    symbolic x coords={128,160,192,224,256}, 
    unbounded coords=discard, 
    xtick={128,160,192,224,256},
    enlarge x limits=0.1,    
    ymin=0, ymax=10000,        
    xlabel={Number of qubits},
    ylabel={Entanglement cost},
    grid=minor,
    scaled y ticks = base 10:-4,
    minor grid style={line width=0.5pt, draw=gray!60, densely dotted},
    major grid style={dotted, draw=gray!70},
    legend style={
       at={(0.23,0.62)},
       anchor=south,
       legend columns=1,
    },
    error bars/y dir=both,
    error bars/y explicit,
    error bars/error mark options={
        rotate=90,
        mark size=1pt
    }
]

\pgfplotstableread[col sep=space,header=has colnames]{
num_qubits fgp_mean fgp_min fgp_max
128 1431.6 1404.0 1458.0
160 2212.0 2174.0 2240.0
192 3147.2 3134.0 3168.0
224 4302.4 4222.0 4356.0
256 5586.4 5538.0 5626.0
}\dataFGPtwo
\addplot+ [
  color=one,
  fill opacity=0.3,
  bar shift=-9pt,
] table [
  x=num_qubits,
  y=fgp_mean,
  y error plus expr=\thisrow{fgp_max}-\thisrow{fgp_mean},
  y error minus expr=\thisrow{fgp_mean}-\thisrow{fgp_min},
] {\dataFGPtwo};


\pgfplotstableread[col sep=space,header=has colnames]{
num_qubits fgp_mean fgp_min fgp_max
128 2421.6 2394.0 2450.0
160 3754.0 3718.0 3766.0
192 5346.4 5282.0 5414.0
224 7258.0 7184.0 7326.0
256 9420.8 9372.0 9484.0
}\dataFGPfour
\addplot+ [
  color=one,
  fill opacity=0.3,
  pattern=north east lines,   
  bar shift=3pt,
  forget plot,    
] table [
  x=num_qubits,
  y=fgp_mean,
  y error plus expr=\thisrow{fgp_max}-\thisrow{fgp_mean},
  y error minus expr=\thisrow{fgp_mean}-\thisrow{fgp_min},
] {\dataFGPfour};

\addlegendentry{FGP-rOEE}

\pgfplotstableread[col sep=space,header=has colnames]{
num_qubits r_mean r_min r_max
128 851.6 829.0 875.0
160 1341.9 1315.0 1370.0
192 1936.2 1892.0 1961.0
224 2658.0 2637.0 2677.0
256 3487.6 3446.0 3532.0
}\dataFMtwo
\addplot+ [
  color=teal,
  fill opacity=0.3,
  bar shift=-6pt,
] table [
  x=num_qubits,
  y=r_mean,
  y error plus expr=\thisrow{r_max}-\thisrow{r_mean},
  y error minus expr=\thisrow{r_mean}-\thisrow{r_min},
] {\dataFMtwo};

\pgfplotstableread[col sep=space,header=has colnames]{
num_qubits r_mean r_min r_max
128 1592.6 1577.0 1602.0
160 2523.0 2506.0 2541.0
192 3660.7 3613.0 3708.0
224 4991.7 4958.0 5027.0
256 6527.2 6485.0 6573.0
}\dataFMfour
\addplot+ [
  color=teal,
  fill opacity=0.3,
  pattern=north east lines,
  bar shift=6pt,
  forget plot,
] table [
  x=num_qubits,
  y=r_mean,
  y error plus expr=\thisrow{r_max}-\thisrow{r_mean},
  y error minus expr=\thisrow{r_mean}-\thisrow{r_min},
] {\dataFMfour};

\addlegendentry{MLFM-R}

\pgfplotstableread[col sep=space,header=has colnames]{
num_qubits part_mean part_min part_max
128 1028.2 1011.0 1043.0 
160 1626.0 1616.0 1644.0 
192 2382.6 2362.0 2415.0 
224 3224.4 3194.0 3239.0 
256 4250.6 4231.0 4280.0
}\dataTKtwo
\addplot+ [
  color=three,
  fill opacity=0.3,
  bar shift=-3pt,
] table [
  x=num_qubits,
  y=part_mean,
  y error plus expr=\thisrow{part_max}-\thisrow{part_mean},
  y error minus expr=\thisrow{part_mean}-\thisrow{part_min},
] {\dataTKtwo};

\addlegendentry{Pytket P}

\pgfplotstableread[
  col sep=space,
  header=has colnames
]{
num_qubits part_mean part_min part_max
128 1832.6 1799.0 1849.0 
160 2914.6 2888.0 2946.0 
192 4199.0 4169.0 4255.0 
224 5790.0 5766.0 5834.0 
256 7594.4 7521.0 7667.0
}\datatable
\addplot+ [
  color=gray,
fill opacity=.3,
bar shift = -10000pt,
]
table [
  x=num_qubits,
  y=,
] {\datatable};

\pgfplotstableread[col sep=space,header=has colnames]{
num_qubits part_mean part_min part_max
128 1832.6 1799.0 1849.0 
160 2914.6 2888.0 2946.0 
192 4199.0 4169.0 4255.0 
224 5790.0 5766.0 5834.0 
256 7594.4 7521.0 7667.0
}\dataTKfour
\addplot+ [
  color=three,
  fill opacity=0.3,
  pattern=north east lines,
  bar shift=9pt,
  forget plot,
] table [
  x=num_qubits,
  y=part_mean,
  y error plus expr=\thisrow{part_max}-\thisrow{part_mean},
  y error minus expr=\thisrow{part_mean}-\thisrow{part_min},
] {\dataTKfour};




\end{axis}

\node at (4.25,3.8) {%
  \begin{tikzpicture}[font=\footnotesize]
    \draw[fill=gray] (0,0) rectangle (0.3,0.3);
    \node[anchor=west] at (0.4,0.15) {2 QPUs};

    \draw[fill=gray, pattern=north east lines] (0,-0.5) rectangle (0.3,-0.2);
    \node[anchor=west] at (0.4,-0.35) {4 QPUs};

  \end{tikzpicture}%
};
\end{tikzpicture}
    \end{subfigure}
    ~
    \begin{subfigure}{0.48\textwidth}
      \centering
      \usetikzlibrary{patterns}
\usetikzlibrary{calc}
\begin{tikzpicture}
\begin{axis}[
    width=\columnwidth,
    height=0.75\columnwidth,
    ybar,                     
    bar width=2.5pt,         
    symbolic x coords={128,160,192,224,256},
    unbounded coords=discard, 
    xtick={128,160,192,224,256},
    enlarge x limits=0.1,    
    ymin=0, ymax=500,        
    xlabel={Number of qubits},
    ylabel={Time taken (s)},
    grid=minor,
    scaled y ticks = base 10:-3,
    minor grid style={line width=0.5pt, draw=gray!60, densely dotted},
    major grid style={dotted, draw=gray!70},
    legend style={
       at={(0.23,0.62)},
       anchor=south,
       legend columns=1,
    },
    error bars/y dir=both,
    error bars/y explicit,
    error bars/error mark options={
        rotate=90,
        mark size=1pt
    }
]

\pgfplotstableread[col sep=space,header=has colnames]{
num_qubits fgp_mean fgp_min fgp_max
128 3.8004552841186525 3.741442918777466 3.923525810241699
160 8.538347530364991 8.361013889312744 8.64104175567627
192 16.62597522735596 16.517704010009766 16.77318286895752
224 30.0760591506958 29.502116918563843 30.433773040771484
256 56.739788341522214 50.90563702583313 74.85412096977234
}\dataFGPtwo
\addplot+ [
  color=one,
  fill opacity=0.3,
  bar shift=-9pt,
] table [
  x=num_qubits,
  y=fgp_mean,
  y error plus expr=\thisrow{fgp_max}-\thisrow{fgp_mean},
  y error minus expr=\thisrow{fgp_mean}-\thisrow{fgp_min},
] {\dataFGPtwo};


\pgfplotstableread[col sep=space,header=has colnames]{
num_qubits fgp_mean fgp_min fgp_max
128 10.017921018600465 7.002586126327515 12.965767860412598
160 25.890225839614867 15.89801287651062 30.601945161819458
192 36.95833525657654 33.27070903778076 50.459192991256714
224 58.47969460487366 57.65534210205078 59.76092576980591
256 98.7918505191803 98.18218970298767 99.99581694602966
}\dataFGPfour
\addplot+ [
  color=one,
  fill opacity=0.3,
  pattern=north east lines,   
  bar shift=3pt,
  forget plot,    
] table [
  x=num_qubits,
  y=fgp_mean,
  y error plus expr=\thisrow{fgp_max}-\thisrow{fgp_mean},
  y error minus expr=\thisrow{fgp_mean}-\thisrow{fgp_min},
] {\dataFGPfour};

\addlegendentry{FGP-rOEE}

\pgfplotstableread[col sep=space,header=has colnames]{
num_qubits r_mean r_min r_max
128 5.3885 5.3583 5.4165
160 9.8185 9.6814 9.9164
192 15.6019 15.4284 15.7478
224 23.3926 23.2882 23.5637
256 33.5810 33.3353 33.8975
}\dataFMtwo
\addplot+ [
  color=teal,
  fill opacity=0.3,
  bar shift=-6pt,
] table [
  x=num_qubits,
  y=r_mean,
  y error plus expr=\thisrow{r_max}-\thisrow{r_mean},
  y error minus expr=\thisrow{r_mean}-\thisrow{r_min},
] {\dataFMtwo};

\pgfplotstableread[col sep=space,header=has colnames]{
num_qubits r_mean r_min r_max
128 11.9917 11.9159 12.0959
160 21.7568 21.6209 21.8382
192 34.6197 34.4640 34.7256
224 51.8080 51.5160 52.0828
256 74.2673 74.0732 74.6073
}\dataFMfour
\addplot+ [
  color=teal,
  fill opacity=0.3,
  pattern=north east lines,
  bar shift=6pt,
  forget plot,
] table [
  x=num_qubits,
  y=r_mean,
  y error plus expr=\thisrow{r_max}-\thisrow{r_mean},
  y error minus expr=\thisrow{r_mean}-\thisrow{r_min},
] {\dataFMfour};

\addlegendentry{MLFM-R}

\pgfplotstableread[col sep=space,header=has colnames]{
num_qubits part_mean part_min part_max
128 22.326348400115968 22.222893953323364 22.42086100578308 
160 42.995372486114505 22.870821237564087 48.47602605819702 
192 97.35568923950196 92.59483218193054 104.76143789291382 
224 146.04137325286865 81.05919218063354 167.11892199516296 
256 278.1659022331238 271.4041929244995 283.41053199768066
}\dataTKtwo
\addplot+ [
  color=three,
  fill opacity=0.3,
  bar shift=-3pt,
] table [
  x=num_qubits,
  y=part_mean,
  y error plus expr=\thisrow{part_max}-\thisrow{part_mean},
  y error minus expr=\thisrow{part_mean}-\thisrow{part_min},
] {\dataTKtwo};

\addlegendentry{Pytket P}

\pgfplotstableread[
  col sep=space,
  header=has colnames
]{
num_qubits part_mean part_min part_max
128 23.15481014251709 20.115867137908936 31.109879732131958 
160 55.159952545166014 45.290987968444824 69.25660490989685 
192 104.34946928024291 50.66809105873108 129.75362467765808 
224 216.6396463871002 209.49704909324646 224.5148208141327 
256 423.5220265865326 334.87830209732056 461.9474470615387
}\datatable
\addplot+ [
  color=gray,
fill opacity=.3,
bar shift = -10000pt,
]
table [
  x=num_qubits,
  y=,
] {\datatable};

\pgfplotstableread[col sep=space,header=has colnames]{
num_qubits part_mean part_min part_max
128 23.15481014251709 20.115867137908936 31.109879732131958 
160 55.159952545166014 45.290987968444824 69.25660490989685 
192 104.34946928024291 50.66809105873108 129.75362467765808 
224 216.6396463871002 209.49704909324646 224.5148208141327 
256 423.5220265865326 334.87830209732056 461.9474470615387
}\dataTKfour
\addplot+ [
  color=three,
  fill opacity=0.3,
  pattern=north east lines,
  bar shift=9pt,
  forget plot,
] table [
  x=num_qubits,
  y=part_mean,
  y error plus expr=\thisrow{part_max}-\thisrow{part_mean},
  y error minus expr=\thisrow{part_mean}-\thisrow{part_min},
] {\dataTKfour};




\end{axis}

\node at (4.25,3.8) {%
  \begin{tikzpicture}[font=\footnotesize]
    \draw[fill=gray] (0,0) rectangle (0.3,0.3);
    \node[anchor=west] at (0.4,0.15) {2 QPUs};

    \draw[fill=gray, pattern=north east lines] (0,-0.5) rectangle (0.3,-0.2);
    \node[anchor=west] at (0.4,-0.35) {4 QPUs};

  \end{tikzpicture}%
};
\end{tikzpicture}
    \end{subfigure}
    \caption{Large $CP$-fraction circuits with 50\% two-qubit gates, partitioned over 2 and 4 QPUs.}
    \label{fig:CP_large_24}
  \end{figure*}

  \begin{figure*}[!htbp]
    \centering
    \begin{subfigure}{0.48\textwidth}
      \centering
      \begin{tikzpicture}
\begin{axis}[
  width=\columnwidth,
  height=0.75\columnwidth,
    ybar,                     
    bar width=2.5pt,         
    symbolic x coords={16,32,48,64,80,96},
    unbounded coords=discard, 
    xtick={16,32,48,64,80,96},
    enlarge x limits=0.1,    
    ymin=0, ymax=550,        
    xlabel={Number of qubits},
    ylabel={Entanglement cost},
    grid=minor,
    minor grid style={line width=0.5pt, draw=gray!60, densely dotted},
    major grid style={dotted, draw=gray!70},
    scaled y ticks = base 10:-2,
    legend style={
      at={(0.23,0.62)},
      anchor=south,
      legend columns=1,
   },
    error bars/y dir=both,
    error bars/y explicit,
    error bars/error mark options={
        rotate=90,
        mark size=1pt
    }
]

\pgfplotstableread[col sep=space,header=has colnames]{
num_qubits fgp_mean fgp_min fgp_max
16 26.0 26.0 26.0
32 50.0 50.0 50.0
48 74.0 74.0 74.0
64 98.0 98.0 98.0
80 122.0 122.0 122.0
96 150.0 150.0 150.0
}\dataFGPtwo
\addplot+ [
  color=one,
  fill opacity=0.3,
  bar shift=-9pt,
] table [
  x=num_qubits,
  y=fgp_mean,
  y error plus expr=\thisrow{fgp_max}-\thisrow{fgp_mean},
  y error minus expr=\thisrow{fgp_mean}-\thisrow{fgp_min},
] {\dataFGPtwo};


\pgfplotstableread[col sep=space,header=has colnames]{
num_qubits fgp_mean fgp_min fgp_max
16 71.0 71.0 71.0
32 174.0 174.0 174.0
48 264.0 264.0 264.0
64 345.0 345.0 345.0
80 431.0 431.0 431.0
96 512.0 512.0 512.0
}\dataFGPfour
\addplot+ [
  color=one,
  fill opacity=0.3,
  pattern=north east lines,   
  bar shift=3pt,
  forget plot,    
] table [
  x=num_qubits,
  y=fgp_mean,
  y error plus expr=\thisrow{fgp_max}-\thisrow{fgp_mean},
  y error minus expr=\thisrow{fgp_mean}-\thisrow{fgp_min},
] {\dataFGPfour};

\addlegendentry{FGP-rOEE}

\pgfplotstableread[col sep=space,header=has colnames]{
num_qubits r_mean r_min r_max
16 7.0 7.0 7.0
32 15.0 15.0 15.0
48 23.0 23.0 23.0
64 31.0 31.0 31.0
80 39.0 39.0 39.0
96 47.0 47.0 47.0
}\dataFMtwo
\addplot+ [
  color=teal,
  fill opacity=0.3,
  bar shift=-6pt,
] table [
  x=num_qubits,
  y=r_mean,
  y error plus expr=\thisrow{r_max}-\thisrow{r_mean},
  y error minus expr=\thisrow{r_mean}-\thisrow{r_min},
] {\dataFMtwo};

\pgfplotstableread[col sep=space,header=has colnames]{
num_qubits r_mean r_min r_max
16 18.0 18.0 18.0
32 42.0 42.0 42.0
48 66.0 66.0 66.0
64 90.0 90.0 90.0
80 114.0 114.0 114.0
96 138.0 138.0 138.0
}\dataFMfour
\addplot+ [
  color=teal,
  fill opacity=0.3,
  pattern=north east lines,
  bar shift=6pt,
  forget plot,
] table [
  x=num_qubits,
  y=r_mean,
  y error plus expr=\thisrow{r_max}-\thisrow{r_mean},
  y error minus expr=\thisrow{r_mean}-\thisrow{r_min},
] {\dataFMfour};

\addlegendentry{MLFM-R}

\pgfplotstableread[col sep=space,header=has colnames]{
num_qubits part_mean part_min part_max embed_mean embed_min embed_max
16 9.0 9.0 9.0 9.0 9.0 9.0
32 17.4 17.0 18.0 17.4 17.0 18.0
48 30.0 30.0 30.0 30.0 30.0 30.0
64 36.0 36.0 36.0 36.0 36.0 36.0
80 41.0 41.0 41.0 41.0 41.0 41.0
96 36.0 36.0 36.0 36.0 36.0 36.0
}\dataTKtwo
\addplot+ [
  color=three,
  fill opacity=0.3,
  bar shift=-3pt,
] table [
  x=num_qubits,
  y=embed_mean,
  y error plus expr=\thisrow{embed_max}-\thisrow{embed_mean},
  y error minus expr=\thisrow{embed_mean}-\thisrow{embed_min},
] {\dataTKtwo};

\addlegendentry{Pytket PE}

\pgfplotstableread[
  col sep=space,
  header=has colnames
]{
num_qubits part_mean part_min part_max embed_mean embed_min embed_max
16 22.0 22.0 22.0 22.0 22.0 22.0
32 45.0 44.0 46.0 45.0 44.0 46.0
48 59.6 59.0 60.0 59.6 59.0 60.0
64 81.2 81.0 82.0 81.2 81.0 82.0
80 96.0 96.0 96.0 96.0 96.0 96.0
96 99.0 99.0 99.0 99.0 99.0 99.0
}\datatable
\addplot+ [
  color=gray,
fill opacity=.3,
bar shift = -10000pt,
]
table [
  x=num_qubits,
  y=,
] {\datatable};

\pgfplotstableread[col sep=space,header=has colnames]{
num_qubits part_mean part_min part_max embed_mean embed_min embed_max
16 22.0 22.0 22.0 22.0 22.0 22.0
32 45.0 44.0 46.0 45.0 44.0 46.0
48 59.6 59.0 60.0 59.6 59.0 60.0
64 81.2 81.0 82.0 81.2 81.0 82.0
80 96.0 96.0 96.0 96.0 96.0 96.0
96 99.0 99.0 99.0 99.0 99.0 99.0
}\dataTKfour
\addplot+ [
  color=three,
  fill opacity=0.3,
  pattern=north east lines,
  bar shift=9pt,
  forget plot,
] table [
  x=num_qubits,
  y=embed_mean,
  y error plus expr=\thisrow{embed_max}-\thisrow{embed_mean},
  y error minus expr=\thisrow{embed_mean}-\thisrow{embed_min},
] {\dataTKfour};




\end{axis}

\node at (4.25,3.8) {%
  \begin{tikzpicture}[font=\footnotesize]
    \draw[fill=gray] (0,0) rectangle (0.3,0.3);
    \node[anchor=west] at (0.4,0.15) {2 QPUs};

    \draw[fill=gray, pattern=north east lines] (0,-0.5) rectangle (0.3,-0.2);
    \node[anchor=west] at (0.4,-0.35) {4 QPUs};

  \end{tikzpicture}%
};
\end{tikzpicture}
    \end{subfigure}
    ~
    \begin{subfigure}{0.48\textwidth}
      \centering
      \usetikzlibrary{patterns}
\usetikzlibrary{calc}
\begin{tikzpicture}
\begin{axis}[
  width=\columnwidth,
  height=0.75\columnwidth,
    ybar,                     
    bar width=2.5pt,         
    symbolic x coords={16,32,48,64,80,96},
    unbounded coords=discard, 
    xtick={16,32,48,64,80,96},
    enlarge x limits=0.1,    
    ymin=0, ymax=40,        
    xlabel={Number of qubits},
    ylabel={Time taken (s)},
    grid=minor,
    scaled y ticks = base 10:-2,
    minor grid style={line width=0.5pt, draw=gray!60, densely dotted},
    major grid style={dotted, draw=gray!70},
    legend style={
      at={(0.23,0.62)},
      anchor=south,
      legend columns=1,
   },
    error bars/y dir=both,
    error bars/y explicit,
    error bars/error mark options={
        rotate=90,
        mark size=1pt
    }
]

\pgfplotstableread[col sep=space,header=has colnames]{
num_qubits fgp_mean fgp_min fgp_max
16 0.006645822525024414 0.006288766860961914 0.0074613094329833984
32 0.04702401161193848 0.04019522666931152 0.07022571563720703
48 0.13583245277404785 0.12503623962402344 0.17320871353149414
64 0.29343762397766116 0.28241896629333496 0.32590603828430176
80 0.5514698505401612 0.5317130088806152 0.5834732055664062
96 0.9670914649963379 0.9449999332427979 0.988947868347168
}\dataFGPtwo
\addplot+ [
  color=one,
  fill opacity=0.3,
  bar shift=-9pt,
] table [
  x=num_qubits,
  y=fgp_mean,
  y error plus expr=\thisrow{fgp_max}-\thisrow{fgp_mean},
  y error minus expr=\thisrow{fgp_mean}-\thisrow{fgp_min},
] {\dataFGPtwo};


\pgfplotstableread[col sep=space,header=has colnames]{
num_qubits fgp_mean fgp_min fgp_max
16 0.01830739974975586 0.017997026443481445 0.019161224365234375
32 0.11880936622619628 0.11471700668334961 0.12015366554260254
48 0.4140575408935547 0.39851903915405273 0.45054197311401367
64 0.9049702167510987 0.8822259902954102 0.9347920417785645
80 1.9113749504089355 1.8718841075897217 1.9360008239746094
96 3.220908498764038 3.1799471378326416 3.2797722816467285
}\dataFGPfour
\addplot+ [
  color=one,
  fill opacity=0.3,
  pattern=north east lines,   
  bar shift=3pt,
  forget plot,    
] table [
  x=num_qubits,
  y=fgp_mean,
  y error plus expr=\thisrow{fgp_max}-\thisrow{fgp_mean},
  y error minus expr=\thisrow{fgp_mean}-\thisrow{fgp_min},
] {\dataFGPfour};

\addlegendentry{FGP-rOEE}

\pgfplotstableread[col sep=space,header=has colnames]{
num_qubits r_mean r_min r_max
16 0.1317 0.1288 0.1424
32 0.7369 0.7243 0.7488
48 2.2127 2.1744 2.2668
64 4.7763 4.6757 4.8282
80 9.0424 8.9286 9.1686
96 15.1728 15.0338 15.2794
}\dataFMtwo
\addplot+ [
  color=teal,
  fill opacity=0.3,
  bar shift=-6pt,
] table [
  x=num_qubits,
  y=r_mean,
  y error plus expr=\thisrow{r_max}-\thisrow{r_mean},
  y error minus expr=\thisrow{r_mean}-\thisrow{r_min},
] {\dataFMtwo};

\pgfplotstableread[col sep=space,header=has colnames]{
num_qubits r_mean r_min r_max
16 0.2883 0.2833 0.3040
32 1.7211 1.6912 1.7805
48 5.3950 5.3239 5.4787
64 11.5584 11.3832 11.6606
80 22.4126 22.2504 22.5547
96 38.3680 38.1103 38.5574
}\dataFMfour
\addplot+ [
  color=teal,
  fill opacity=0.3,
  pattern=north east lines,
  bar shift=6pt,
  forget plot,
] table [
  x=num_qubits,
  y=r_mean,
  y error plus expr=\thisrow{r_max}-\thisrow{r_mean},
  y error minus expr=\thisrow{r_mean}-\thisrow{r_min},
] {\dataFMfour};

\addlegendentry{MLFM-R}

\pgfplotstableread[col sep=space,header=has colnames]{
num_qubits part_mean part_min part_max embed_mean embed_min embed_max
16 0.026367759704589842 0.021120071411132812 0.04179096221923828 0.015477514266967774 0.014903783798217773 0.0160980224609375
32 0.14334959983825685 0.13998889923095703 0.15278410911560059 0.21857185363769532 0.20715618133544922 0.2232191562652588
48 0.8064266204833984 0.7840709686279297 0.8161561489105225 0.873804235458374 0.8391540050506592 0.9009928703308105
64 1.5898351192474365 1.5514469146728516 1.637697696685791 1.9461799144744873 1.919619083404541 1.9827048778533936
80 2.3776942253112794 2.3127729892730713 2.510040044784546 3.5012595653533936 3.471269130706787 3.5628268718719482
96 2.694705533981323 2.5499911308288574 3.0059001445770264 5.453113174438476 5.3868348598480225 5.640100002288818
}\dataTKtwo
\addplot+ [
  color=three,
  fill opacity=0.3,
  bar shift=-3pt,
] table [
  x=num_qubits,
  y=embed_mean,
  y error plus expr=\thisrow{embed_max}-\thisrow{embed_mean},
  y error minus expr=\thisrow{embed_mean}-\thisrow{embed_min},
] {\dataTKtwo};

\addlegendentry{Pytket PE}

\pgfplotstableread[
  col sep=space,
  header=has colnames
]{
num_qubits part_mean part_min part_max embed_mean embed_min embed_max
16 0.03620457649230957 0.034049034118652344 0.03946375846862793 0.01790590286254883 0.017822980880737305 0.01800394058227539
32 0.5072051525115967 0.4921000003814697 0.518643856048584 0.21779999732971192 0.21409392356872559 0.22075510025024414
48 0.8601541519165039 0.5476970672607422 1.2944707870483398 0.8856353282928466 0.8681669235229492 0.8948049545288086
64 2.174165201187134 2.1227002143859863 2.2266929149627686 2.0053113460540772 1.9812030792236328 2.031240940093994
80 3.235993528366089 3.148844003677368 3.332810878753662 3.570457410812378 3.507014036178589 3.6366848945617676
96 2.7296295166015625 2.6631319522857666 2.8013317584991455 5.695710039138794 5.529366970062256 5.88153600692749
}\datatable
\addplot+ [
  color=gray,
fill opacity=.3,
bar shift = -10000pt,
]
table [
  x=num_qubits,
  y=,
] {\datatable};

\pgfplotstableread[col sep=space,header=has colnames]{
num_qubits part_mean part_min part_max embed_mean embed_min embed_max
16 0.03620457649230957 0.034049034118652344 0.03946375846862793 0.01790590286254883 0.017822980880737305 0.01800394058227539
32 0.5072051525115967 0.4921000003814697 0.518643856048584 0.21779999732971192 0.21409392356872559 0.22075510025024414
48 0.8601541519165039 0.5476970672607422 1.2944707870483398 0.8856353282928466 0.8681669235229492 0.8948049545288086
64 2.174165201187134 2.1227002143859863 2.2266929149627686 2.0053113460540772 1.9812030792236328 2.031240940093994
80 3.235993528366089 3.148844003677368 3.332810878753662 3.570457410812378 3.507014036178589 3.6366848945617676
96 2.7296295166015625 2.6631319522857666 2.8013317584991455 5.695710039138794 5.529366970062256 5.88153600692749
}\dataTKfour
\addplot+ [
  color=three,
  fill opacity=0.3,
  pattern=north east lines,
  bar shift=9pt,
  forget plot,
] table [
  x=num_qubits,
  y=embed_mean,
  y error plus expr=\thisrow{embed_max}-\thisrow{embed_mean},
  y error minus expr=\thisrow{embed_mean}-\thisrow{embed_min},
] {\dataTKfour};




\end{axis}

\node at (4.25,3.8) {%
  \begin{tikzpicture}[font=\footnotesize]
    \draw[fill=gray] (0,0) rectangle (0.3,0.3);
    \node[anchor=west] at (0.4,0.15) {2 QPUs};

    \draw[fill=gray, pattern=north east lines] (0,-0.5) rectangle (0.3,-0.2);
    \node[anchor=west] at (0.4,-0.35) {4 QPUs};

  \end{tikzpicture}%
};
\end{tikzpicture}
    \end{subfigure}
  \caption{QFT circuits partitioned over 2 and 4 QPUs.}
  \label{fig:QFT_24}
  \end{figure*}

  \begin{figure*}[ht]
    \centering
    \begin{subfigure}{0.48\textwidth}
      \centering
      \usetikzlibrary{patterns}
\usetikzlibrary{calc}
\begin{tikzpicture}
\begin{axis}[
  width=\columnwidth,
  height=0.75\columnwidth,
    ybar,                     
    bar width=2.5pt,         
    symbolic x coords={16,32,48,64,80,96},
    unbounded coords=discard, 
    xtick={16,32,48,64,80,96},
    enlarge x limits=0.1,    
    ymin=0, ymax=500,        
    xlabel={Number of qubits},
    ylabel={Entanglement cost},
    grid=minor,
    scaled y ticks = base 10:-2,
    minor grid style={line width=0.5pt, draw=gray!60, densely dotted},
    major grid style={dotted, draw=gray!70},
    legend style={
      at={(0.23,0.62)},
      anchor=south,
      legend columns=1,
   },
    error bars/y dir=both,
    error bars/y explicit,
    error bars/error mark options={
        rotate=90,
        mark size=1pt
    }
]

\pgfplotstableread[col sep=space,header=has colnames]{
num_qubits fgp_mean fgp_min fgp_max
16 23.2 18.0 28.0
32 59.2 42.0 84.0
48 72.4 62.0 84.0
64 92.8 82.0 104.0
80 136.8 102.0 206.0
96 151.6 140.0 172.0
}\dataFGPtwo
\addplot+ [
  color=one,
  fill opacity=0.3,
  bar shift=-9pt,
] table [
  x=num_qubits,
  y=fgp_mean,
  y error plus expr=\thisrow{fgp_max}-\thisrow{fgp_mean},
  y error minus expr=\thisrow{fgp_mean}-\thisrow{fgp_min},
] {\dataFGPtwo};


\pgfplotstableread[col sep=space,header=has colnames]{
num_qubits fgp_mean fgp_min fgp_max
16 56.4 50.0 64.0
32 147.2 142.0 152.0
48 249.6 230.0 274.0
64 325.6 290.0 396.0
80 371.2 346.0 444.0
96 447.6 414.0 486.0
}\dataFGPfour
\addplot+ [
  color=one,
  fill opacity=0.3,
  pattern=north east lines,   
  bar shift=3pt,
  forget plot,    
] table [
  x=num_qubits,
  y=fgp_mean,
  y error plus expr=\thisrow{fgp_max}-\thisrow{fgp_mean},
  y error minus expr=\thisrow{fgp_mean}-\thisrow{fgp_min},
] {\dataFGPfour};

\addlegendentry{FGP-rOEE}

\pgfplotstableread[col sep=space,header=has colnames]{
num_qubits r_mean r_min r_max
16 6.9 6.0 7.0
32 15.3 15.0 16.0
48 23.3 23.0 24.0
64 31.6 31.0 32.0
80 39.3 39.0 41.0
96 47.8 47.0 49.0
}\dataFMtwo
\addplot+ [
  color=teal,
  fill opacity=0.3,
  bar shift=-6pt,
] table [
  x=num_qubits,
  y=r_mean,
  y error plus expr=\thisrow{r_max}-\thisrow{r_mean},
  y error minus expr=\thisrow{r_mean}-\thisrow{r_min},
] {\dataFMtwo};

\pgfplotstableread[col sep=space,header=has colnames]{
num_qubits r_mean r_min r_max
16 16.4 15.0 18.0
32 43.2 42.0 44.0
48 68.5 67.0 70.0
64 92.4 90.0 96.0
80 117.6 116.0 121.0
96 141.2 138.0 144.0
}\dataFMfour
\addplot+ [
  color=teal,
  fill opacity=0.3,
  pattern=north east lines,
  bar shift=6pt,
  forget plot,
] table [
  x=num_qubits,
  y=r_mean,
  y error plus expr=\thisrow{r_max}-\thisrow{r_mean},
  y error minus expr=\thisrow{r_mean}-\thisrow{r_min},
] {\dataFMfour};

\addlegendentry{MLFM-R}

\pgfplotstableread[col sep=space,header=has colnames]{
num_qubits part_mean part_min part_max embed_mean embed_min embed_max
16 8.8 7.0 10.0 8.8 7.0 10.0
32 21.0 20.0 22.0 21.0 20.0 22.0
48 27.6 26.0 29.0 27.6 26.0 29.0
64 40.8 36.0 50.0 40.8 36.0 50.0
80 57.6 52.0 63.0 57.6 52.0 63.0
96 67.6 62.0 75.0 67.6 62.0 75.0
}\dataTKtwo
\addplot+ [
  color=three,
  fill opacity=0.3,
  bar shift=-3pt,
] table [
  x=num_qubits,
  y=embed_mean,
  y error plus expr=\thisrow{embed_max}-\thisrow{embed_mean},
  y error minus expr=\thisrow{embed_mean}-\thisrow{embed_min},
] {\dataTKtwo};

\addlegendentry{Pytket PE}

\pgfplotstableread[
  col sep=space,
  header=has colnames
]{
num_qubits part_mean part_min part_max embed_mean embed_min embed_max
16 22.6 20.0 25.0 22.6 20.0 25.0
32 54.6 51.0 57.0 54.6 51.0 57.0
48 86.8 85.0 90.0 86.8 85.0 90.0
64 121.0 117.0 126.0 121.0 117.0 126.0
80 156.8 148.0 161.0 156.8 148.0 161.0
96 188.2 185.0 193.0 188.2 185.0 193.0
}\datatable
\addplot+ [
  color=gray,
fill opacity=.3,
bar shift = -10000pt,
]
table [
  x=num_qubits,
  y=,
] {\datatable};

\pgfplotstableread[col sep=space,header=has colnames]{
num_qubits part_mean part_min part_max embed_mean embed_min embed_max
16 22.6 20.0 25.0 22.6 20.0 25.0
32 54.6 51.0 57.0 54.6 51.0 57.0
48 86.8 85.0 90.0 86.8 85.0 90.0
64 121.0 117.0 126.0 121.0 117.0 126.0
80 156.8 148.0 161.0 156.8 148.0 161.0
96 188.2 185.0 193.0 188.2 185.0 193.0
}\dataTKfour
\addplot+ [
  color=three,
  fill opacity=0.3,
  pattern=north east lines,
  bar shift=9pt,
  forget plot,
] table [
  x=num_qubits,
  y=embed_mean,
  y error plus expr=\thisrow{embed_max}-\thisrow{embed_mean},
  y error minus expr=\thisrow{embed_mean}-\thisrow{embed_min},
] {\dataTKfour};




\end{axis}

\node at (4.25,3.8) {%
  \begin{tikzpicture}[font=\footnotesize]
    \draw[fill=gray] (0,0) rectangle (0.3,0.3);
    \node[anchor=west] at (0.4,0.15) {2 QPUs};

    \draw[fill=gray, pattern=north east lines] (0,-0.5) rectangle (0.3,-0.2);
    \node[anchor=west] at (0.4,-0.35) {4 QPUs};

  \end{tikzpicture}%
};
\end{tikzpicture}
    \end{subfigure}
    ~
    \begin{subfigure}{0.48\textwidth}
      \centering
      \usetikzlibrary{calc}
\begin{tikzpicture}
\begin{axis}[
  width=\columnwidth,
  height=0.75\columnwidth,
    ybar,                     
    bar width=2.5pt,         
    symbolic x coords={16,32,48,64,80,96},
    unbounded coords=discard, 
    xtick={16,32,48,64,80,96},
    enlarge x limits=0.1,    
    ymin=0, ymax=160,        
    xlabel={Number of qubits},
    ylabel={Time taken (s)},
    grid=minor,
    scaled y ticks = base 10:-1,
    minor grid style={line width=0.5pt, draw=gray!60, densely dotted},
    major grid style={dotted, draw=gray!70},
    legend style={
      at={(0.23,0.62)},
      anchor=south,
      legend columns=1,
   },
    error bars/y dir=both,
    error bars/y explicit,
    error bars/error mark options={
        rotate=90,
        mark size=1pt
    }
]

\pgfplotstableread[col sep=space,header=has colnames]{
num_qubits fgp_mean fgp_min fgp_max
16 0.008784151077270508 0.006438016891479492 0.013995885848999023
32 0.055881929397583005 0.04309511184692383 0.09050297737121582
48 0.1588900089263916 0.12928199768066406 0.1936488151550293
64 0.3611154079437256 0.299483060836792 0.40735292434692383
80 0.7110793590545654 0.6599299907684326 0.8153328895568848
96 1.1849777698516846 1.1211650371551514 1.2776031494140625
}\dataFGPtwo
\addplot+ [
  color=one,
  fill opacity=0.3,
  bar shift=-9pt,
] table [
  x=num_qubits,
  y=fgp_mean,
  y error plus expr=\thisrow{fgp_max}-\thisrow{fgp_mean},
  y error minus expr=\thisrow{fgp_mean}-\thisrow{fgp_min},
] {\dataFGPtwo};


\pgfplotstableread[col sep=space,header=has colnames]{
num_qubits fgp_mean fgp_min fgp_max
16 0.010909605026245116 0.00946807861328125 0.012053966522216797
32 0.10203948020935058 0.07952260971069336 0.1701359748840332
48 0.3015313625335693 0.27100300788879395 0.3398118019104004
64 0.6668544292449952 0.5994830131530762 0.7221601009368896
80 1.2120769500732422 1.1617209911346436 1.3501818180084229
96 2.0422925472259523 1.9798939228057861 2.159558057785034
}\dataFGPfour
\addplot+ [
  color=one,
  fill opacity=0.3,
  pattern=north east lines,   
  bar shift=3pt,
  forget plot,    
] table [
  x=num_qubits,
  y=fgp_mean,
  y error plus expr=\thisrow{fgp_max}-\thisrow{fgp_mean},
  y error minus expr=\thisrow{fgp_mean}-\thisrow{fgp_min},
] {\dataFGPfour};

\addlegendentry{FGP-rOEE}

\pgfplotstableread[col sep=space,header=has colnames]{
num_qubits r_mean r_min r_max
16 0.2493 0.2170 0.2789
32 1.3351 1.2096 1.4656
48 3.9285 3.6469 4.1273
64 9.0067 8.4751 9.4095
80 16.3892 15.7200 17.3961
96 28.6323 28.0839 29.5493
}\dataFMtwo
\addplot+ [
  color=teal,
  fill opacity=0.3,
  bar shift=-6pt,
] table [
  x=num_qubits,
  y=r_mean,
  y error plus expr=\thisrow{r_max}-\thisrow{r_mean},
  y error minus expr=\thisrow{r_mean}-\thisrow{r_min},
] {\dataFMtwo};

\pgfplotstableread[col sep=space,header=has colnames]{
num_qubits r_mean r_min r_max
16 0.5184 0.3976 0.5890
32 3.2077 3.0005 3.3931
48 9.4950 9.1756 9.7514
64 21.6047 19.6895 22.8441
80 41.3747 38.9149 43.5827
96 71.9265 68.9753 74.3873
}\dataFMfour
\addplot+ [
  color=teal,
  fill opacity=0.3,
  pattern=north east lines,
  bar shift=6pt,
  forget plot,
] table [
  x=num_qubits,
  y=r_mean,
  y error plus expr=\thisrow{r_max}-\thisrow{r_mean},
  y error minus expr=\thisrow{r_mean}-\thisrow{r_min},
] {\dataFMfour};

\addlegendentry{MLFM-R}

\pgfplotstableread[col sep=space,header=has colnames]{
num_qubits part_mean part_min part_max embed_mean embed_min embed_max
16 0.1511821746826172 0.02539801597595215 0.6342129707336426 0.04178252220153809 0.020314931869506836 0.06342101097106934
32 0.20900917053222656 0.20119285583496094 0.2179100513458252 0.7253439903259278 0.5427711009979248 0.8551478385925293
48 0.7416351318359375 0.710352897644043 0.7878520488739014 5.126935243606567 4.103744983673096 5.515689134597778
64 2.077921676635742 1.962183952331543 2.202141046524048 20.255965089797975 16.917986154556274 24.210405349731445
80 4.625658750534058 4.337806940078735 4.782105922698975 52.784610748291016 46.72816276550293 66.91457605361938
96 9.086445522308349 8.673392057418823 9.432115077972412 140.3734495639801 123.39261603355408 166.48298692703247
}\dataTKtwo
\addplot+ [
  color=three,
  fill opacity=0.3,
  bar shift=-3pt,
] table [
  x=num_qubits,
  y=embed_mean,
  y error plus expr=\thisrow{embed_max}-\thisrow{embed_mean},
  y error minus expr=\thisrow{embed_mean}-\thisrow{embed_min},
] {\dataTKtwo};

\addlegendentry{Pytket PE}

\pgfplotstableread[
  col sep=space,
  header=has colnames
]{
num_qubits part_mean part_min part_max embed_mean embed_min embed_max
16 0.05009360313415527 0.046270132064819336 0.06241583824157715 0.0319124698638916 0.024946928024291992 0.040441036224365234
32 0.2391727924346924 0.22344589233398438 0.2620091438293457 0.5506494045257568 0.4734830856323242 0.6069128513336182
48 0.7975747585296631 0.734799861907959 0.8354649543762207 3.2192381858825683 2.7026259899139404 3.443917751312256
64 2.229425621032715 2.1592209339141846 2.34184193611145 13.132903671264648 11.801145792007446 14.787235021591187
80 4.97478756904602 4.815284013748169 5.1556360721588135 40.87770948410034 39.10500192642212 44.033445835113525
96 9.44553599357605 9.113658905029297 9.66704797744751 89.78050022125244 83.99272608757019 94.38630986213684
}\datatable
\addplot+ [
  color=gray,
fill opacity=.3,
bar shift = -10000pt,
]
table [
  x=num_qubits,
  y=,
] {\datatable};

\pgfplotstableread[col sep=space,header=has colnames]{
num_qubits part_mean part_min part_max embed_mean embed_min embed_max
16 0.05009360313415527 0.046270132064819336 0.06241583824157715 0.0319124698638916 0.024946928024291992 0.040441036224365234
32 0.2391727924346924 0.22344589233398438 0.2620091438293457 0.5506494045257568 0.4734830856323242 0.6069128513336182
48 0.7975747585296631 0.734799861907959 0.8354649543762207 3.2192381858825683 2.7026259899139404 3.443917751312256
64 2.229425621032715 2.1592209339141846 2.34184193611145 13.132903671264648 11.801145792007446 14.787235021591187
80 4.97478756904602 4.815284013748169 5.1556360721588135 40.87770948410034 39.10500192642212 44.033445835113525
96 9.44553599357605 9.113658905029297 9.66704797744751 89.78050022125244 83.99272608757019 94.38630986213684
}\dataTKfour
\addplot+ [
  color=three,
  fill opacity=0.3,
  pattern=north east lines,
  bar shift=9pt,
  forget plot,
] table [
  x=num_qubits,
  y=embed_mean,
  y error plus expr=\thisrow{embed_max}-\thisrow{embed_mean},
  y error minus expr=\thisrow{embed_mean}-\thisrow{embed_min},
] {\dataTKfour};




\end{axis}

\node at (4.25,3.8) {%
  \begin{tikzpicture}[font=\footnotesize]
    \draw[fill=gray] (0,0) rectangle (0.3,0.3);
    \node[anchor=west] at (0.4,0.15) {2 QPUs};

    \draw[fill=gray, pattern=north east lines] (0,-0.5) rectangle (0.3,-0.2);
    \node[anchor=west] at (0.4,-0.35) {4 QPUs};

  \end{tikzpicture}%
};
\end{tikzpicture}
    \end{subfigure}
  \caption{QAOA circuits partitioned over 2 and 4 QPUs.}
  \label{fig:QAOA_24}
  \end{figure*}
  
  \begin{figure*}[ht]
    \centering
    \begin{subfigure}{0.48\textwidth}
      \centering
      \begin{tikzpicture}
\begin{axis}[
  width=\columnwidth,
  height=0.75\columnwidth,
    ybar,                     
    bar width=2.5pt,         
    symbolic x coords={16,32,48,64,80,96},
    unbounded coords=discard, 
    xtick={16,32,48,64,80,96},
    enlarge x limits=0.1,    
    ymin=0, ymax=10000,        
    xlabel={Number of qubits},
    ylabel={Entanglement cost},
    grid=minor,
    scaled y ticks = base 10:-4,
    minor grid style={line width=0.5pt, draw=gray!60, densely dotted},
    major grid style={dotted, draw=gray!70},
    legend style={
      at={(0.23,0.62)},
      anchor=south,
      legend columns=1,
   },
    error bars/y dir=both,
    error bars/y explicit,
    error bars/error mark options={
        rotate=90,
        mark size=1pt
    }
]

\pgfplotstableread[col sep=space,header=has colnames]{
num_qubits fgp_mean fgp_min fgp_max
16 47.6 42.0 52.0
32 176.4 172.0 184.0
48 401.6 392.0 410.0
64 697.6 684.0 708.0
80 1088.8 1080.0 1096.0
96 1570.4 1550.0 1588.0
}\dataFGPtwo
\addplot+ [
  color=one,
  fill opacity=0.3,
  bar shift=-9pt,
] table [
  x=num_qubits,
  y=fgp_mean,
  y error plus expr=\thisrow{fgp_max}-\thisrow{fgp_mean},
  y error minus expr=\thisrow{fgp_mean}-\thisrow{fgp_min},
] {\dataFGPtwo};


\pgfplotstableread[col sep=space,header=has colnames]{
num_qubits fgp_mean fgp_min fgp_max
16 89.8 85.0 100.0
32 348.4 343.0 355.0
48 740.2 710.0 798.0
64 1291.2 1253.0 1326.0
80 1998.0 1954.0 2031.0
96 2828.4 2749.0 2928.0
}\dataFGPfour
\addplot+ [
  color=one,
  fill opacity=0.3,
  pattern=north east lines,   
  bar shift=3pt,
  forget plot,    
] table [
  x=num_qubits,
  y=fgp_mean,
  y error plus expr=\thisrow{fgp_max}-\thisrow{fgp_mean},
  y error minus expr=\thisrow{fgp_mean}-\thisrow{fgp_min},
] {\dataFGPfour};

\addlegendentry{FGP-rOEE}

\pgfplotstableread[col sep=space,header=has colnames]{
num_qubits r_mean r_min r_max
16 36.6 34.0 40.0
32 145.7 144.0 150.0
48 326.9 320.0 339.0
64 588.6 578.0 601.0
80 908.5 891.0 936.0
96 1327.5 1283.0 1348.0
}\dataFMtwo
\addplot+ [
  color=teal,
  fill opacity=0.3,
  bar shift=-6pt,
] table [
  x=num_qubits,
  y=r_mean,
  y error plus expr=\thisrow{r_max}-\thisrow{r_mean},
  y error minus expr=\thisrow{r_mean}-\thisrow{r_min},
] {\dataFMtwo};

\pgfplotstableread[col sep=space,header=has colnames]{
num_qubits r_mean r_min r_max
16 66.7 64.0 75.0
32 264.0 256.0 271.0
48 585.7 575.0 595.0
64 1051.0 1029.0 1093.0
80 1626.2 1600.0 1655.0
96 2344.6 2320.0 2392.0
}\dataFMfour
\addplot+ [
  color=teal,
  fill opacity=0.3,
  pattern=north east lines,
  bar shift=6pt,
  forget plot,
] table [
  x=num_qubits,
  y=r_mean,
  y error plus expr=\thisrow{r_max}-\thisrow{r_mean},
  y error minus expr=\thisrow{r_mean}-\thisrow{r_min},
] {\dataFMfour};

\addlegendentry{MLFM-R}

\pgfplotstableread[col sep=space,header=has colnames]{
num_qubits part_mean part_min part_max
16 148.5 138.0 156.0
32 615.0 606.0 624.0
48 1435.5 1398.0 1476.0
64 2623.5 2586.0 2652.0
80 4276.5 4230.0 4332.0
96 6217.5 6168.0 6300.0
}\dataTKtwo
\addplot+ [
  color=three,
  fill opacity=0.3,
  bar shift=-3pt,
] table [
  x=num_qubits,
  y=part_mean,
  y error plus expr=\thisrow{part_max}-\thisrow{part_mean},
  y error minus expr=\thisrow{part_mean}-\thisrow{part_min},
] {\dataTKtwo};

\addlegendentry{Pytket P}

\pgfplotstableread[
  col sep=space,
  header=has colnames
]{
num_qubits part_mean part_min part_max
16 246.0 240.0 252.0
32 966.6 948.0 981.0
48 2269.8 2241.0 2283.0
64 4101.4 4052.0 4152.0
80 6502.0 6435.0 6588.0
96 9504.2 9396.0 9575.0
}\datatable
\addplot+ [
  color=gray,
fill opacity=.3,
bar shift = -10000pt,
]
table [
  x=num_qubits,
  y=,
] {\datatable};

\pgfplotstableread[col sep=space,header=has colnames]{
num_qubits part_mean part_min part_max
16 246.0 240.0 252.0
32 966.6 948.0 981.0
48 2269.8 2241.0 2283.0
64 4101.4 4052.0 4152.0
80 6502.0 6435.0 6588.0
96 9504.2 9396.0 9575.0
}\dataTKfour
\addplot+ [
  color=three,
  fill opacity=0.3,
  pattern=north east lines,
  bar shift=9pt,
  forget plot,
] table [
  x=num_qubits,
  y=part_mean,
  y error plus expr=\thisrow{part_max}-\thisrow{part_mean},
  y error minus expr=\thisrow{part_mean}-\thisrow{part_min},
] {\dataTKfour};




\end{axis}

\node at (4.25,3.8) {%
  \begin{tikzpicture}[font=\footnotesize]
    \draw[fill=gray] (0,0) rectangle (0.3,0.3);
    \node[anchor=west] at (0.4,0.15) {2 QPUs};

    \draw[fill=gray, pattern=north east lines] (0,-0.5) rectangle (0.3,-0.2);
    \node[anchor=west] at (0.4,-0.35) {4 QPUs};

  \end{tikzpicture}%
};
\end{tikzpicture}
    \end{subfigure}
    ~
    \begin{subfigure}{0.48\textwidth}
      \centering
      \usetikzlibrary{patterns}
\usetikzlibrary{calc}
\begin{tikzpicture}
\begin{axis}[
  width=\columnwidth,
  height=0.75\columnwidth,
    ybar,                     
    bar width=2.5pt,         
    symbolic x coords={16,32,48,64,80,96},
    unbounded coords=discard, 
    xtick={16,32,48,64,80,96},
    enlarge x limits=0.1,    
    ymin=0, ymax=110,        
    xlabel={Number of qubits},
    ylabel={Time taken (s)},
    grid=minor,
    scaled y ticks = base 10:-2,
    minor grid style={line width=0.5pt, draw=gray!60, densely dotted},
    major grid style={dotted, draw=gray!70},
    legend style={
      at={(0.23,0.62)},
      anchor=south,
      legend columns=1,
   },
    error bars/y dir=both,
    error bars/y explicit,
    error bars/error mark options={
        rotate=90,
        mark size=1pt
    }
]

\pgfplotstableread[col sep=space,header=has colnames]{
num_qubits fgp_mean fgp_min fgp_max
16 0.01867203712463379 0.018115997314453125 0.019710063934326172
32 0.15607666969299316 0.12565183639526367 0.17751073837280273
48 0.4991586685180664 0.48156309127807617 0.52799391746521
64 1.2037981510162354 1.1437981128692627 1.247290849685669
80 2.378938150405884 2.3071348667144775 2.507223129272461
96 4.211400890350342 4.183417081832886 4.233396053314209
}\dataFGPtwo
\addplot+ [
  color=one,
  fill opacity=0.3,
  bar shift=-9pt,
] table [
  x=num_qubits,
  y=fgp_mean,
  y error plus expr=\thisrow{fgp_max}-\thisrow{fgp_mean},
  y error minus expr=\thisrow{fgp_mean}-\thisrow{fgp_min},
] {\dataFGPtwo};


\pgfplotstableread[col sep=space,header=has colnames]{
num_qubits fgp_mean fgp_min fgp_max
16 0.03192281723022461 0.025510072708129883 0.03864407539367676
32 0.2666463375091553 0.22939682006835938 0.31854796409606934
48 0.8728063106536865 0.8149552345275879 0.9896221160888672
64 2.404084825515747 2.132862091064453 2.8411977291107178
80 5.378231239318848 5.212825059890747 5.636458396911621
96 9.973511028289796 8.833284854888916 11.022814750671387
}\dataFGPfour
\addplot+ [
  color=one,
  fill opacity=0.3,
  pattern=north east lines,   
  bar shift=3pt,
  forget plot,    
] table [
  x=num_qubits,
  y=fgp_mean,
  y error plus expr=\thisrow{fgp_max}-\thisrow{fgp_mean},
  y error minus expr=\thisrow{fgp_mean}-\thisrow{fgp_min},
] {\dataFGPfour};

\addlegendentry{FGP-rOEE}

\pgfplotstableread[col sep=space,header=has colnames]{
num_qubits r_mean r_min r_max
16 0.3709 0.3597 0.4009
32 1.6673 1.6082 1.6955
48 4.3370 4.2880 4.3980
64 8.8299 8.7008 8.9130
80 15.7605 15.5547 15.9179
96 26.2638 26.0991 26.3918
}\dataFMtwo
\addplot+ [
  color=teal,
  fill opacity=0.3,
  bar shift=-6pt,
] table [
  x=num_qubits,
  y=r_mean,
  y error plus expr=\thisrow{r_max}-\thisrow{r_mean},
  y error minus expr=\thisrow{r_mean}-\thisrow{r_min},
] {\dataFMtwo};

\pgfplotstableread[col sep=space,header=has colnames]{
num_qubits r_mean r_min r_max
16 0.7577 0.7416 0.7977
32 3.4926 3.4530 3.5435
48 9.1630 9.0631 9.2559
64 18.7783 18.6633 18.8825
80 33.5481 33.3107 33.9386
96 55.8969 55.7295 56.1077
}\dataFMfour
\addplot+ [
  color=teal,
  fill opacity=0.3,
  pattern=north east lines,
  bar shift=6pt,
  forget plot,
] table [
  x=num_qubits,
  y=r_mean,
  y error plus expr=\thisrow{r_max}-\thisrow{r_mean},
  y error minus expr=\thisrow{r_mean}-\thisrow{r_min},
] {\dataFMfour};

\addlegendentry{MLFM-R}

\pgfplotstableread[col sep=space,header=has colnames]{
num_qubits part_mean part_min part_max
16 0.15658199787139893 0.15382599830627441 0.15834879875183105
32 2.1311588883399963 2.0949089527130127 2.1529018878936768
48 7.334447741508484 7.158915996551514 7.593524932861328
64 17.9982847571373 17.79779577255249 18.126810789108276
80 36.19772803783417 35.93020701408386 36.57015609741211
96 65.77314245700836 64.79618191719055 66.92008018493652
}\dataTKtwo
\addplot+ [
  color=three,
  fill opacity=0.3,
  bar shift=-3pt,
] table [
  x=num_qubits,
  y=part_mean,
  y error plus expr=\thisrow{part_max}-\thisrow{part_mean},
  y error minus expr=\thisrow{part_mean}-\thisrow{part_min},
] {\dataTKtwo};

\addlegendentry{Pytket P}

\pgfplotstableread[
  col sep=space,
  header=has colnames
]{
num_qubits part_mean part_min part_max
16 0.33707499504089355 0.2696359157562256 0.5709941387176514
32 3.269359588623047 2.5363962650299072 3.7859508991241455
48 11.434198427200318 8.096641063690186 12.348763227462769
64 28.153446006774903 19.260003089904785 30.956566095352173
80 59.59758701324463 58.58549213409424 60.4005389213562
96 104.17176899909973 102.78261971473694 105.90074825286865
}\datatable
\addplot+ [
  color=gray,
fill opacity=.3,
bar shift = -10000pt,
]
table [
  x=num_qubits,
  y=,
] {\datatable};

\pgfplotstableread[col sep=space,header=has colnames]{
num_qubits part_mean part_min part_max
16 0.33707499504089355 0.2696359157562256 0.5709941387176514
32 3.269359588623047 2.5363962650299072 3.7859508991241455
48 11.434198427200318 8.096641063690186 12.348763227462769
64 28.153446006774903 19.260003089904785 30.956566095352173
80 59.59758701324463 58.58549213409424 60.4005389213562
96 104.17176899909973 102.78261971473694 105.90074825286865
}\dataTKfour
\addplot+ [
  color=three,
  fill opacity=0.3,
  pattern=north east lines,
  bar shift=9pt,
  forget plot,
] table [
  x=num_qubits,
  y=part_mean,
  y error plus expr=\thisrow{part_max}-\thisrow{part_mean},
  y error minus expr=\thisrow{part_mean}-\thisrow{part_min},
] {\dataTKfour};




\end{axis}

\node at (4.25,3.8) {%
  \begin{tikzpicture}[font=\footnotesize]
    \draw[fill=gray] (0,0) rectangle (0.3,0.3);
    \node[anchor=west] at (0.4,0.15) {2 QPUs};

    \draw[fill=gray, pattern=north east lines] (0,-0.5) rectangle (0.3,-0.2);
    \node[anchor=west] at (0.4,-0.35) {4 QPUs};

  \end{tikzpicture}%
};
\end{tikzpicture}
    \end{subfigure}
  \caption{QV circuits, partitioned over 2 and 4 QPUs.}
  \label{fig:QV_24}
  \end{figure*}

\subsubsection{Fine-grained partitioning}\label{sec:fgp}

A pioneering technique in the quantum circuit partitioning literature was introduced by Baker et al. \cite{bakerTimeslicedQuantumCircuit2020}. The paper provides a method for covering all non-local gates using state teleportation, referred to as fine-grained partitioning. The essence of the method is to split a quantum circuit into a sequence of static graphs indicating the qubit interactions at each time step, with time decaying contributions for future time steps. A refinement heuristic, called \text{overall extreme exchange}, is employed to achieve a fully local partitioning for each time step, using the assignment for each time step as the starting point for the next. In similar fashion to our approach, the end result is a sequence of assignments of qubits to QPUs, one for each time step of the circuit. The difference is that the assignment guarantees that all two-qubit gates occur locally, using state teleportations to transition between each contiguous assignment. The algorithm used is referred to as FGP-rOEE. FGP-rOEE does not allow for gate teleportation or multi-gate teleportation, so it typically underperforms in circuits that permit large groups of teleportation compatible groups.

\section{Discussion}

The results indicate that the proposed framework can achieve lower entanglement costs than existing, state-of-the-art techniques, provided that an appropriate algorithm is used to partition the resulting hypergraph. The results in \cref{fig:group_vs} show the effectiveness of employing a gate grouping routine, allowing entanglement costs to be reduced via multi-gate teleportation. However, \cref{fig:group_vs_comm} highlights that this is traded off with the communication qubit capacity required to achieve these entanglement costs. While this means some systems may be limited in the extent to which they can benefit from multi-gate teleportation, we believe that this should serve as an indication that large numbers of communication qubits are desirable for DQC architectures. \cref{fig:explore_result} shows the effectiveness of using an exploratory variant of the FM algorithm which is more effective at escaping local minima. This is particularly important given the structure of the hypergraph, where using a static initial assignment of qubits to partitions initialises the algorithm in a local minimum, as seen in \cref{fig:loc_min}. The exploratory variant allows larger sequences of moves to be probed, which may move the same node more than once, ultimately breaking into lower cost solution regions. 

Investigation of the multilevel framework shows that all coarsening strategies result in both lower entanglement costs and shorter runtime than the basic FM approach, shown in \cref{fig:coarsening_performance}. The recursive coarsening strategy proves to be both the most effective and the most time efficient, so is used as the main method for comparison with other techniques. \cref{fig:CP_30,fig:CP_50,fig:CP_70,fig:CP_90} compare the recursive multilevel FM (MLFM-R) with two methods from Pytket-DQC (Partition and PartitionEmbed) and the FGP-rOEE algorithm from Ref. \cite{bakerTimeslicedQuantumCircuit2020}. MLFM-R achieves the lowest entanglement costs for two-qubit gate proportions of $0.3$ and $0.5$, roughly matches the best performing methods for $0.7$, though underperforms for $0.9$. This is likely due to the fact that the Pytket-DQC methods optimise for detached gates, those which are executed in a partition which is not local to either the root or the receiver qubit. This is an effective technique for extremely high proportions of two-qubit gates. Since we do not consider this possibility, we limit the scope of multi-gate teleportation. We aim to extend the generality of the framework in this way in future. We also note that Pytket-DQC requires its own transpiler pass before distribution, which may merge together adjacent two-qubit gates when the proportion of two-qubit gates is very high.

In the time results, MLFM-R is significantly faster than Pytket ESD, moderately faster than Pytket PE, but slower than FGP-rOEE. Notably, the core FM algorithm also scales with the number of partitions. In these tests, the number of QPUs is increased by one each time the number of qubits is increased, such that the time is scaling with both the circuit size and the number of partitions. This is a limitation that we aim to mitigate in the future. However, in \cref{fig:CP_large_24} we show that, for a constant number of partitions (2 and 4), the MLFM-R outperforms even the fastest Pytket-DQC method, Partition, in both entanglement costs and time, as well as the FGP-rOEE algorithm. We were unable to complete these tests for Pytket PE or ESD, since the runtime was too long. This demonstrates that MLFM-R is able to scale to circuits with much larger numbers of qubits when there is an appropriate limit on the number of partitions. 

\begin{table*}[!htbp]
    \centering
    \caption{Overall comparison of MLFM-R and Pytket PE across all tests completed. We calculate a size independent metric for each circuit, where we divide the mean best e-bit cost achieved by the total number of two-qubit gates in each circuit, averaging over all circuits for each benchmark. A * indicates where we were unable to use Pytket PE due to runtime, so instead used P. }
    \label{tab:frac}
        \pgfplotstabletypeset[
          col sep=space,
          header=true,
          every head row/.style={
            output empty row,
            before row={
              \toprule
              & & & \multicolumn{2}{c}{E-bit Fraction} \\
              Circuit & Partitions & Ref. & MLFM-R & PE\\
              \midrule
            },
            after row={},
        },
        every last row/.style={
            after row={\bottomrule},
        },
          columns={circuit, partitions, ref, rcost, pecost},
                   columns/circuit/.style={
                    column name={Circuit},
                    column type=l,
                    string type,
                    string replace={_}{\_}, 
                },
        columns/partitions/.style={
            column name={Partitions},
            column type=l,
            string type,
            string replace={_}{\_}, 
        },
        columns/ref/.style={
            column name={Ref.},
            column type=l,
            string type,
            string replace={_}{\_}, 
        },
        columns/rcost/.style={
            column name={MLFM-R E-bit},
            column type=l,
            fixed,
            precision = 3, 
        },
        columns/pecost/.style={
            column name={PE E-bit},
            column type=l,
            fixed,
            precision = 3, 
        }
        ]{
            circuit partitions ref rcost pecost
            CP 2-12 [\ref{fig:CP_30},\ref{fig:CP_50},\ref{fig:CP_70},\ref{fig:CP_90}] 0.421040 0.41397374372701823
            CP-large* 2-4 [\ref{fig:CP_large_24},\ref{fig:CP_large_24}] 0.304219 0.35767585671421837
            QAOA 2-4 [\ref{fig:QAOA_24},\ref{fig:QAOA_24}] 0.043815 0.05929962374817798
            QASM 2-4 [\ref{tab:QASM_res}] 0.038028 0.09797962967099245
            QASM-large 2-4 [\ref{tab:QASM_res_large}] 0.022527 0.05445700568452912
            QFT 2-4 [\ref{fig:QFT_24},\ref{fig:QFT_24}] 0.043601 0.04616182557624888
            QV* 2-4 [\ref{fig:QV_24},\ref{fig:QV_24}] 0.132574 0.5529436451496635
            Mean N/A N/A 0.143686 0.22238863376717294 
        } 

  \end{table*}

Furthermore, it is possible to reduce the time even further by terminating before the finest levels of refinement, though we do not investigate this here. In \cref{fig:QFT_24} we investigate the performance of the algorithm on QFT circuits. Interestingly, as mentioned in \cref{sec:QFT}, we note that no improvement is achieved for any QFT circuits using any FM methods, though we achieve a very low entanglement cost. Previously we speculated that the initial placement may be a global optimum after the greedy gate grouping. In practice, it may not be necessary to perform any optimisation, and the initial placement can be used alongside the greedy gate grouping routine. However, for larger QFT circuits, we noted that Pytket PE was able to achieve lower entanglement costs than our initial placement. However, the QFT cost after $80$ qubits seems to plateau, indicating that some approximation in the QFT circuit may be leading to lower costs. Pytket-DQC requires its own transpiler pass before distribution, which may include some truncation of small angle rotations, though we have not verified this. While large hyper-edges in QFT circuits leads to slower runtimes for MLFM-R, in practice we would not need to use the FM algorithm for QFT circuits, since the greedy gate grouping pass is sufficient to achieve low entanglement costs. 

\cref{fig:QAOA_24} displays the results generated from QAOA circuits, in which MLFM-R achieves the lowest entanglement costs. The time taken here is much longer than for FGP-rOEE, but shorter than Pytket PE, which displays erratic performance on these tests. 

\cref{fig:QV_24} shows the results for QV circuits. Previously, QV circuits were identified as difficult circuits to partition using only gate teleportation \cite{andres-martinezDistributingCircuitsHeterogeneous2024, burtGeneralisedCircuitPartitioning2024b}, since they contain no potential for gate grouping. Consequently, methods such as FGP-rOEE, which use only state teleportation, are very effective at distributing QV circuits. Despite this, MLFM-R is still able to outperform FGP-rOEE in entanglement costs, highlighting the versatility of the temporal graph construction. From Pytket-DQC, we were only able to use Partition, since PartitionEmbed and EmbedSteinerDetached were too slow for QV circuits and showed no benefit over Partition at small scales. Partition is the slowest method in this case. 

Finally, Tab. \ref{tab:QASM_res} and Tab. \ref{tab:QASM_res_large} show results from the QASM benchmark suite circuits. For each circuit, we optimise the partitioning over $2,3$ and $4$ QPUs. In the majority of cases, MLFM-R achieves the lowest entanglement cost, though is marginally slower than its competitors in many of these cases. We summarise the comparison between MLFM-R and the next best performing method in Pytket-DQC in Tab. \ref{tab:frac}. Pytket EmbedSteinerDetached is too slow to run on many of the benchmarks, so we primarily compare with PartitionEmbed. However, in some cases, the embedding refinement was too slow and we had to use just Partition, which usually gives similar results and is significantly faster. These cases are indicated on the table. We use a size-independent metric, in which the optimised entanglement cost in terms of e-bit count is divided by the total number of qubits, giving the \textit{e-bit fraction}. For the e-bit fraction, over all tests, we outperform the next best method by over 35\%, which is significant. MLFM-R is often faster than PartitionEmbed and is also faster than Partition for the large benchmarks (though Partition is faster in most other cases). FGP-rOEE is almost always the fastest method, but the limitation of only allowing state teleportation means that the entanglement costs are usually significantly higher, so we do not include the results in Tab. \ref{tab:frac}.

\section{Conclusions}

In this work, we introduced and investigated a multilevel framework for partitioning quantum circuits over distributed quantum processing units. We formulated quantum circuit partitioning as a problem of partitioning temporally-extended hypergraphs with an objective that counts entanglement costs from quantum state teleportation, multi-gate teleportation and a new procedure termed \textit{nested state teleportation}. The partitioning is based on the Fiduccia-Mattheyses algorithm -- a well-known, efficient heuristic for hypergraph partitioning. We adapted the algorithm to fit the constraints of the problem and designed a unique objective corresponding to entanglement requirements. We illustrated how to incorporate this objective into the algorithm efficiently. We extended the framework to include temporal coarsening of problem hypergraphs, allowing us to partition at multiple resolutions of time. We demonstrated that multilevel partitioning can both reduce runtime and improve solution quality. We were able to significantly outperform state-of-the-art methods in terms of entanglement requirements, for similar or lower runtime in most cases. The results indicate that multilevel partitioning, coupled with a sufficiently general problem formulation, forms a highly effective framework for entanglement minimisation that should be integrated into future distributed quantum computing compiler stacks.

\section{Future work}\label{sec:future_work}

While the multilevel partitioning framework has proved to be very effective, a number of improvements still need to be made. A limitation of the framework is that communication qubits are only accounted for in the final circuit extraction phase. This means that the algorithm may produce a partitioning that is unfeasible due to a lack of communication qubits. We plan to introduce problem variants where we explicitly account for communication qubits available in the system, while also providing post-processing methods for mitigating communication qubit limitations. Additionally, techniques such as detached gates and embedding of non-local unitaries, identified and used in Refs. \cite{gsundaramEfficientDistributionQuantum2021,wuEntanglementefficientBipartitedistributedQuantum2023a,andres-martinezDistributingCircuitsHeterogeneous2024}, could be incorporated into the framework to further reduce entanglement costs. 

We note also that the present work has not explicitly considered general quantum networks, with limited connectivity and noisy hardware. This is avoided here, since the focus has been on clearly defining the core partitioning problem. However, we have already addressed many concerns in follow-up work, providing techniques for extending the multilevel framework to large-scale constrained networks \cite{burtEntanglementEfficientDistributionQuantum2025}. In addition to the current work, these extensions have been incorporated into the \texttt{disqco} library, with the hope of providing a complete framework for quantum circuit partitioning and distribution.

\subsection*{A note on fault-tolerance}

Finally, we make a brief note on the relevance of this work to the context of fault-tolerant quantum computing. It is likely that small scale implementations of NISQ-era DQC will coincide with small-scale monolithic implementations of fault-tolerant quantum algorithms. However, the future of large-scale quantum computing lies in distributed, fault-tolerant architectures \cite{rametteFaulttolerantConnectionErrorcorrected2024}. While we do not explicitly address fault-tolerant DQC in this work, we believe many of the techniques and ideas are directly transferrable. Since we make no assumptions about intra-QPU connectivity and optimise purely for the bottleneck of the system -- inter-QPU entanglement distribution -- the techniques are directly applicable to fault-tolerant architectures where QPUs host multiple code patches.

In such cases, we can either replace the state and gate teleportation primitives with their transversal, logical, equivalents \cite{ryan-andersonHighfidelityTeleportationLogical2024,stackAssessingTeleportationLogical2025}, or use non-local variations of lattice surgery \cite{horsmanSurfaceCodeQuantum2012}. It has already been demonstrated that lattice surgery, mediated by e-bits, can be used to perform both state and gate teleportation at the logical level \cite{rametteFaulttolerantConnectionErrorcorrected2024,martonLatticeSurgerybasedLogical2025a} and early-stage experimental demonstrations have been performed \cite{erhardEntanglingLogicalQubits2021,ryan-andersonHighfidelityTeleportationLogical2024}. There is also active research in optimising non-local lattice surgery to reduce resource requirements and error propagation \cite{haugLatticeSurgeryBell2025,jacintoNetworkRequirementsDistributed2025a,shalbyOptimizedNoiseresilientSurface2025c}.

Other proposals for fault-tolerant DQC consider higher rate codes spanning full modules using specialised communication links \cite{yoderTourGrossModular2025}. In such cases, logic can be performed locally and non-locally using generalisations to lattice surgery based on the extractors framework \cite{heExtractorsQLDPCArchitectures2025}. A bolder proposition is to consider individual codes distributed over multiple modules \cite{sutcliffeDistributedQuantumError2025,higgottConstructionsPerformanceHyperbolic2024a}, permitting full exploitation of the scaling properties of high-rate codes.

While the mapping is less direct in the latter two cases, inter-module logical operations are likely to be a bottleneck, regardless of whether they occur by interacting multiple code patches using surgery, or interaction of logicals within a single code.

In future work, we aim to explicitly consider fault-tolerant DQC architectures, incorporating the partitioning techniques we have developed into a complete compiler stack for fault-tolerant distributed quantum computing.

\section*{Acknowledgements}

The authors acknowledge funding from the Engineering and Physical Sciences Research Council (EPSRC) funded Distributed Quantum Computing project, grant number EP/W032643/1.

\newpage
\newpage
\bibliography{references}

\clearpage
\clearpage

\appendix

\section{FM implementation}\label{app:fm_impl}
\subsection{Auxiliary functions}\label{sec:fm_aux}

Here we include some of the auxiliary functions used in the FM algorithm. Alg. \ref{alg:InitFMGCP} contains functions used for initialising the main algorithm. This includes calculating the gains for all moves and converting them into a structure which allows for easy retrieval and updates. We omit the basic helper functions for gain calculation and cost retrieval, since these are defined in the main text. 
\begin{algorithm}[ht]
  \caption{Initial routines for FM}
  \label{alg:InitFMGCP}
  \DontPrintSemicolon
  
  \SetKwProg{FnCost}{Cost}{:}{end}
  \SetKwProg{FnGain}{Gain}{:}{end}
  \SetKwProg{FnComputeAllGains}{ComputeAllGains}{:}{end}
  \SetKwProg{FnBuildGainBuckets}{BuildGainBuckets}{:}{end}
  \SetKwProg{FnEdgeCost}{EdgeCost}{:}{end}
  
  \SetKwFunction{FnMoveNode}{MoveNode}
  \SetKwFunction{FnRetrieveEdgeCost}{RetrieveEdgeCost}
  \SetKwFunction{CallEdgeCost}{EdgeCost}
  \SetKwFunction{CallGain}{Gain}

  \KwIn{\\
  $H(V,E;\tau,\kappa)$: Temporal hypergraph extracted from a quantum circuit.\\
  $\mathcal{C}$: Precomputed cost table for hyper-edge configurations. \\
  $\Phi : V \to \{1,\ldots,K\}$: Current partition assignment. \\
  $K$: Number of partitions (QPUs).
  }

  \FnComputeAllGains{$(H,\,\Phi,\,K,\,\mathcal{C})$}{
    Construct the gain function $\Gamma : V \times \{1,\ldots,K\} \to \mathbb{R} \cup \{\infty\}$.\;
    \ForEach{$v \in V$}{
      \For{$p \gets 1$ \KwTo $K$}{
        \If{$p = \Phi(v)$}{
          $\Gamma(v,p) \gets \infty$\tcp*{Moving to current partition is disallowed}
        }
        \Else{
          $\Gamma(v,p) \gets \CallGain(v,p,\Phi,\mathcal{C})$\;
        }
      }
    }
    \Return $\Gamma$\;
  }
  
  \BlankLine
  
  \FnBuildGainBuckets{$(\Gamma)$}{
    Build bucket structure $\mathcal{B}$ grouping moves $(v,p)$ by their gain $g$.\;
    \tcp{$\mathcal{B}(g)$ denotes the set of moves with gain $g$}
    $\mathcal{B} \gets \emptyset$\;
    \ForEach{$(v,p)$ \textup{with} $\Gamma(v,p) \neq \infty$}{
      $g \gets \Gamma(v,p)$\;
      \If{$g \notin \mathrm{dom}(\mathcal{B})$}{
        $\mathcal{B}(g) \gets \emptyset$\;
      }
      $\mathcal{B}(g) \gets \mathcal{B}(g) \cup \{(v,p)\}$\;
    }
    \Return $\mathcal{B}$\;
  }
  
\end{algorithm}

Alg. \ref{alg:AuxFMGCP} contains the sub-routines for choosing the best moves at each iteration, as well as the $\texttt{UpdateGains}$ function, which is the bulk of the computation in the main FM algorithm. 

\begin{algorithm}[!htpb]
  \caption{Auxiliary routines for FM}
  \label{alg:AuxFMGCP}
  
  \SetKwProg{FnUpdateGains}{UpdateGains}{:}{end}
  \SetKwProg{FnBestMove}{BestMove}{:}{end}

  \SetKwFunction{FnMoveNode}{MoveNode}
  \SetKwFunction{FnDeltaGainContrib}{DeltaGainContrib}
  \SetKwFunction{FnRetrieveCost}{RetrieveCost}
  \SetKwFunction{FnChooseRandom}{ChooseRandom}

  \KwIn{\\
  $H(V,E;\tau,\kappa)$: Temporal hypergraph extracted from a quantum circuit.\\
  $\Gamma : V \times \{1,\ldots,K\} \to \mathbb{R} \cup \{\infty\}$: Gain function.\\
  $\mathcal{B}$: Gain bucket structure built from $\Gamma$.\\
  $\mathcal{C}$: Precomputed cost table for hyper-edge configurations. \\
  $\Phi : V \to \{1,\ldots,K\}$: Current partition assignment. \\
  $\mathcal{X} \subseteq V$: Set of locked nodes.\\
  $\mathrm{Cap} : \{1,\ldots,K\} \to \mathbb{N}$: Capacity limits for each partition.
  }
  
  \FnBestMove{$(H, \mathcal{B},\,\mathcal{X},\,\mathrm{Cap},\,\Phi)$}{
    Select the best admissible move according.\;
    \ForEach{$g$ in gains of $\mathcal{B}$ ascending}{
      \While{$\mathcal{B}(g) \neq \emptyset$}{
        $(v^\star,p^\star) \gets \FnChooseRandom(\mathcal{B}(g))$\;
        $\mathcal{B}(g) \gets \mathcal{B}(g) \setminus \{(v^\star,p^\star)\}$\;
        
        \If{$v^\star \notin \mathcal{X}$ \textup{and} $\vert \{u \in V^{(\tau(v^\star))} \vert \Phi(u) = p^\star\} \vert < \textup{Cap}[p]$}{%
          \Return $(v^\star,p^\star)$\;
        }
      }
    }
    \Return $\bot$\tcp*{No valid moves}
  }
  
  
  
  \BlankLine
  
  \FnUpdateGains{$(H,\,\Gamma,\,\mathcal{B},\,v^\star,\,p^\star,\,\Phi,\,\mathcal{C})$}{
    Update the gain function after moving $v^\star$ to $p^\star$.\;
    
    \ForEach{$e \in E$ with $v^\star \in e$}{
      \ForEach{$u \in e$ with $u \neq v^\star$}{
        \For{$p \gets 1$ \KwTo $K$}{
          
          $\delta \gets \FnDeltaGainContrib(u, e, p, \Phi, \mathcal{C})$\;
          $g_{\text{old}} \gets \Gamma(u,p)$\;
          \If{$g_{\text{old}} \neq \infty$}{
            $\mathcal{B}(g_{\text{old}}) \gets \mathcal{B}(g_{\text{old}}) \setminus \{(u,p)\}$\;
          }
          
          $\Gamma(u,p) \gets \Gamma(u,p) + \delta$\;
          $g_{\text{new}} \gets \Gamma(u,p)$\;
          
          \If{$g_{\text{new}} \neq \infty$}{
          $\mathcal{B}(g_{\text{new}}) \gets \mathcal{B}(g_{\text{new}}) \cup \{(u,p)\}$\;}
          
        }
      }
    }
  }
  
\end{algorithm}

There is some freedom involved in choosing the best move given that it is common to have multiple moves with the same gain. Here, we choose randomly among the best moves, which helps to add some stochasticity to the algorithm, potentially avoiding poor local minima. The gain update function is where the structure of our algorithm differs most from standard FM implementations. Notably, since gains are calculated from differences in edge costs, which depend on hyper-edge configurations, we update the gains on a per-edge basis. This means, after a given move, we first identify which edges are affected by the move. We then iterate over the nodes in each edge, calculating the delta gain contribution from that edge to each node's gain for each possible move. This is slightly different to normal FM implementations, which often find the neighbours of the moved node first, then calculate the delta gains for each neighbour. Since we iterate over hyper-edges first, we avoid redundant calculations from edges which are not affected by the move. 

\begin{algorithm}[!htpb]
  \caption{Delta-gain calculation for a single edge $e$}
  \label{alg:DeltaGainConfig}

  \SetKwProg{FnDeltaGainContrib}{DeltaGainContrib}{:}{end}
  \SetKwProg{FnGetConfig}{GetConfig}{:}{end}
  \SetKwProg{FnUpdateConfig}{UpdateConfig}{:}{end}
  \SetKwFunction{CallUpdateConfig}{UpdateCfg}
  \SetKwFunction{FnRetrieveCounts}{RetrieveCounts}
  \SetKwFunction{FnRetrieveCfg}{RetrieveCfg}
  
  \FnDeltaGainContrib{$(v,\,e,\,p,\,\Phi,\,\mathcal{C})$}{%
    Compute $\delta_e(\Phi,\Phi') = c_e(\Phi') - c_e(\Phi)$ for the move $v : \Phi(v) \to p$.
    \tcp{1. Retrieve root/receiver multiplicities and total edge configuration}
    $(\mathbf{r}_e,\,\mathbf{s}_e) \gets \FnRetrieveCounts(e)$\;    
    $\mathrm{cfg}_{\text{old}} \gets \FnRetrieveCfg(e)$\;
    $c_{\text{old}} \gets \mathcal{C}(\mathrm{cfg}_{\text{old}})$\;
    $\mathbf{r}_e[\Phi(v)] \gets \mathbf{r}_e[\Phi(v)] - 1$\;
    $\mathbf{r}_e[p] \gets \mathbf{r}_e[p] + 1$\;
    \tcp{(Receiver counts $\mathbf{s}_e$ updated analogously if $v$ is a receiver node)}
    $\mathrm{cfg}_{\text{new}} \gets \CallUpdateConfig(\mathbf{r}_e,\mathbf{s}_e, \mathrm{cfg_{old}})$\;
    $c_{\text{new}} \gets \mathcal{C}(\mathrm{cfg}_{\text{new}})$\;
    \Return $c_{\text{new}} - c_{\text{old}}$\;
  }
  
  \BlankLine
  

  \FnUpdateConfig{$(\mathbf{r}_e',\,\mathbf{s}_e',\,\mathrm{src},\,\text{p})$}{%
    \tcp{Update $cfg^{(e)}(\Phi)$ after moving node from src to p}

    \If{$\mathbf{s}_e'[\mathrm{src}] = 0$}{
      $\mathrm{Cfg}[\mathrm{src}] \gets 0$\;
    }
    \Else{
      \If{$\mathbf{r}_e'[\mathrm{src}] = 0$}{
        $\mathrm{Cfg}[\mathrm{src}] \gets 1$\;
      }
      \Else{
        $\mathrm{Cfg}[\mathrm{src}] \gets 0$\;
      }
    }
    \If{$\mathbf{s}_e'[\mathrm{p}] = 0$}{
        $\mathrm{Cfg}[\mathrm{p}] \gets 0$\;
      }
    \Else{
      \If{
        $\mathbf{r}_e'[\mathrm{p}] = 0$}{
        $\mathrm{Cfg}[\text{p}] \gets 1$\;
      }
      \Else{
        $\mathrm{Cfg}[\mathrm{p}] \gets 0$\;
      }
    }
    \Return $\mathrm{Cfg}$\;
  }
  
\end{algorithm}

At first glance, it may seem that we are double counting some contribution, since each neighbour may be involved in multiple edges with the moved node. However, since we are calculating the delta gain contribution from each edge separately, we recover the correct total delta gain for each neighbour after summing over all edges, as seen in Eq. \ref{eq:delta_gain}. After this, we proceed to re-sort the gain buckets according to these changes. Since the calculation of the delta gain contributions is somewhat involved, we dedicate Alg. \ref{alg:DeltaGainConfig} to describing this process. 

The bulk of these routines consists in retrieving and updating the hyper-edge configurations for each possible move. Asymptotically, this is efficient, since we can update the configurations and retrieve the costs in constant time, but the implementation involves retrieving and updating a number of different objects. As alluded to in the main text, in addition to storing and updating the root and receiver configurations, we also store the root and receiver \textit{counts} objects, which track the number of nodes in each partition for the root and receiver sets, respectively. For each move, we increment the counts at the source and the destination of the move, then check whether the counts have changed from zero to non-zero or vice versa, indicating that the configuration must change too. Importantly, for each move, since nodes are either in the root or receiver sets, we need only update the root or receiver counts. Furthermore, since only the entries for the source and destination partition change, we only update these entries. For the full process, we need the edge costs for $\Phi$, $\Phi'$, $\tilde{\Phi}$ and $\tilde{\Phi}'$. In each case, $'$ corresponds to the node which has actually been moved, whereas $\tilde{}$ corresponds to the prospective move for which we are updating the delta gain. In this way, $\tilde{\Phi}$ corresponds to the prospective move being made \textit{before} the actual node was moved. This means that, for each prospective move, we need to update the configuration for two assignment functions, corresponding to $\tilde{\Phi}$ and $\tilde{\Phi}'$. We already have the costs for $\Phi$ and $\Phi'$, since $\Phi$ was already stored, and $\Phi'$ is updated and stored after the completed move was chosen.

\subsection{Pre-computation costs} \label{sec:precomp}

Here we briefly address the issue of pre-computation overhead. The efficiency guarantee of FM for our custom cost function is reliant on the fact that we can pre-compute the cost of all edge configurations in pre-processing. However, the number of configurations is exponential in the number of partitions, and so this is only feasible for less than $20$ partitions. This will not be an issue for most early stage quantum networks, though in the future we will want algorithms that can handle this. 

There are a number of things that could be done to mitigate this. We may choose to omit the pre-computation, and rather store the cost of each configuration as it appears in computation, such that we need only perform the calculation once per configuration. In fact, it is often better to do this in practice, since many configurations will never appear in the computation, and so we can save on both time and memory by only storing the configurations that are actually used. Alternatively, we may choose to perform the partitioning \textit{recursively}. In contrast to the recursive coarsening spoken about in the main text, recursive partitioning refers to multi-partitioning techniques where we partition a graph for a small number of partitions, then continue to partition the resulting subgraphs until we reach the desired number of partitions. This is also a common technique in many large-scale graph partitioning algorithms. Such techniques become necessary not only when the number of partitions is large, but also when the network has constrained connectivity, such that not all partitions are directly connected to each other. While we do not address this case here, it is clear that this case requires additional calculation to accurately model costs, since edges which span multiple partitions may need to be routed through intermediate partitions. We have addressed these issues in follow-up work, showing that recursive methods permit scaling to constrained networks of over $50$ QPUs \cite{burtEntanglementEfficientDistributionQuantum2025}.

\section{Coarsening}

\subsection{Temporal coarsening strategies}\label{app:tc_strats}

Here we detail the algorithms employed for the coarsening process in the FM algorithm. Since we perform only temporal contraction, we simply need to track which time steps have been contracted in order to transform a coarse solution to a finer solution. After partitioning nodes in a coarsened graph, we read from the contraction path which nodes will be uncoarsened at the next level, and assign them the same partition as the parent node from which they have been uncoarsened. Alg. \ref{alg:coarsen-window} details the window-based coarsening strategy, which coarsens nodes in a window of size $w$ at each time step. The window is moved along the time axis, coarsening nodes in the window at each step. We set the input to the algorithm as the number of levels, from which we calculate the window size. This allows us to restrict the number of levels to a fixed number. As mentioned briefly in Sec. \ref{sec:MLCP}, after nodes have been uncoarsened and refined once, we lock them in place to reduce the number of nodes which are moved at each level. This number is then proportional to $n_qw$, such that the uncoarsening acts like a moving window scanning and partitioning the nodes over time. Alg. \ref{alg:coarsen-blocks} details the block-based coarsening strategy, which instead identifies blocks of contiguous nodes to coarsen into a single node. Similarly to the window method, we would like to be able to set the number of levels. This determines the number of blocks. We then coarsen the blocks one time step per-level, such that the resulting number of levels is the number of time steps in each block. We also choose to add an extra level, which performs a full coarsening of the final blocks to a single time step. Alg. \ref{alg:coarsen-recursive} details the recursive coarsening strategy, which coarsens pairs of neighbouring nodes at each time step, such that the total number of nodes is halved at each time step. This turns out to be the most effective strategy and is also leads to the fastest results when we don't cap the number of nodes moved per pass. We conjecture that this is because, at each level, all nodes are roughly as coarse as each other, i.e. each node ``contains'' roughly the same number of nodes. This means that the gains will not be skewed towards large super-nodes, thus controlling the partitioning at an appropriate level of granularity.

  
  
    
  
  

  \begin{algorithm}[ht]
  \caption{Window coarsening}
  \label{alg:coarsen-window}
  
  \SetKwProg{FnCoarsenWindow}{CoarsenWindow}{:}{end}
  \SetKwFunction{FnContractTime}{ContractTime}
  
  \KwIn{\\
    \textbf{$H_0 = (V_0,E_0,\tau,\kappa)$}: temporal hypergraph with time-layers $1,\dots,d$.\\
    $L$: number of coarsening levels.
  }
  \KwOut{\\
    Coarsening hierarchy $\mathcal{H} = (H_0,H_1,\dots,H_M)$ for some $M \leq L$.
  }
  
  \BlankLine
  
  \FnCoarsenWindow{$(H_0,\,d,\,L)$}{
    $\mathcal{H} \gets (H_0)$\;
    $H \gets H_0$\;
    $w \gets \lfloor d / L \rfloor$\tcp*{Window size}
    $\ell \gets d$\;
  
    \While{$\ell > 1$}{
      $\ell_{\text{start}} \gets \max(1,\,\ell - w)$\;
      \For{$\ell' \gets \ell$ \textbf{\textup{down to}} $\ell_{\text{start}} + 1$}
      {
        $H' \gets$ \FnContractTime{$H, \ell', \ell'-1$}\;
        $H \gets H'$\;
      }
      $\mathcal{H} \gets \mathcal{H} \,\Vert\, H$\;
      $\ell \gets \ell - w$\;
    }
  
    \Return $\mathcal{H}$\;
  }
\end{algorithm}
  
  
  

\begin{algorithm}[ht]
  \caption{Block coarsening}
  \label{alg:coarsen-blocks}
  \DontPrintSemicolon
  
  \SetKwProg{FnCoarsenBlocks}{CoarsenBlocks}{:}{end}
  \SetKwFunction{FnContractTime}{ContractTime}
  
  \KwIn{\\
    $H_0 = (V_0,E_0,\tau,\kappa)$: temporal hypergraph with time-layers $1,\dots,d$.\\
    $L$: number of coarsening levels (blocks).
  }
  \KwOut{\\
    Coarsening hierarchy $\mathcal{H} = (H_0,H_1,\dots,H_M)$ for some $M \leq L$.
  }
  
  \BlankLine
  
  \FnCoarsenBlocks{$(H_0,\,d,\,L)$}{
    $\mathcal{H} \gets (H_0)$\;
    $H \gets H_0$\;
    
    $b \gets \left\lfloor \dfrac{d}{L} \right\rfloor$\tcp*{Block size}
    $\ell \gets d$\;
  
    \For{$\ell -1 > d - b$}{
      $t \gets \ell$\;
      \While{$t > 1$}{
        $H' \gets$ \FnContractTime{$H, t, t-1$}\;
        $H \gets H'$\;
        $t \gets t - b$\;
      }
      $\mathcal{H} \gets \mathcal{H} \,\Vert\, H$\;
      $\ell \gets \ell - 1$\tcp*{Slide block window by one layer}
    }
    \tcp{Optionally merge all blocks into one layer}
    \Return $\mathcal{H}$\;
  }
\end{algorithm}

\subsection{Spatial coarsening} \label{sec:spatial_coarsening}

While the paper explores only ``temporal'' coarsening, i.e., coarsening along the time axis, we briefly discuss the idea of spatial coarsening and how it could be incorporated in future work. Spatial coarsening corresponds to coarsening gate-like edges, merging the nodes involved together. Consider first a basic graph, for which no gate grouping has occurred. Contracting a gate-like edge merges nodes from two qubits at the same time step together, such that partitioning of this node determines exactly where this gate will be performed, locally. In this way, if all gate edges are contracted, any partitioning of the nodes results in all gates are covered by state teleportation, since there are no gate-edges to cut. For circuits where state teleportation methods are most effective, such as quantum volume, this would be a useful strategy. However, in many other cases, this would be a limitation, as we are removing the possibility of gate grouping. When we consider hypergraphs, there is more flexibility involved in spatial coarsening. Since hyper-edges can extend over multiple time steps, we could choose to contract nodes which are not at the same time step. However, this makes the process of enforcing capacity constraints more complicated, since these constraints are defined at each time step. 

Alternatively, we may want to consider spatial coarsening after a full coarsening along the time axis. For example, if we have a very large circuit, with over $1000$ qubits, then we still have a $1000$ nodes after coarsening along the time axis. We can reduce the problem size by performing further coarsening along the spatial axis, merging qubit nodes together. This could be a useful strategy for dealing with very large circuits, though we do not investigate this here. If we are including gate grouping as well, there are certain considerations to make, since hyper-edges will extend over a number of time steps. 

We leave a full investigation of integrating spatial and temporal coarsening to future work, though we note that many compiler problems in quantum computing, such as qubit routing, may benefit from both spatial and temporal coarsening strategies.

\onecolumn
\newpage
\section{QASM results tables}

In Tab. \ref{tab:QASM_res} and Tab. \ref{tab:QASM_res_large}, we show the results of the QASM benchmark suite circuits \cite{liQASMBenchLowlevelQASM2022}, partitioned over 2, 3 and 4 QPUs. The results are generated using the exploratory MLFM-R algorithm with a cap of $n_{q}$ on the number of nodes moved per pass, for 10 passes, as for other results. Note that we transpile circuits into the $U(\theta, \phi, \lambda), CP(\theta)$ gate set before partitioning, using the \textit{Qiskit} transpiler \cite{javadi-abhariQuantumComputingQiskit2024}, though Pytket-DQC will run additional transpilation for compatibility with its distributors. We test circuits from the \textit{larger} section of the benchmark suite, though omit circuits of depth greater than $1000$ after transpilation. Additionally, we omit QFT, QAOA and QV circuits, since they have already been tested.

\begin{table*}[htbp]
  \centering
\caption{Comparison of entanglement cost and time taken for QASM benchmark suite circuits.}
  \label{tab:QASM_res}
  \resizebox{\textwidth}{!}{%
      \begin{filecontents*}{plots/QASM/FM_tket_table_QASM.dat}
circuit_name   num_partitions  part_cost  embed_cost  esd_cost  MLFM_rec_cost  part_time  embed_time  esd_time  MLFM_rec_time 
adder28        2               4.000      3.600       3.000     2.200          0.079      0.205       0.699     1.257
adder28        3               2.000      2.000       5.000     3.300          0.070      0.118       0.712     2.039
adder28        4               12.000     10.800      9.000     7.300          0.080      0.226       0.694     3.102
bv30           2               1.000      1.000       1.000     1.000          0.008      0.010       0.020     0.203
bv30           3               2.000      2.000       2.000     2.000          0.010      0.011       0.020     0.299
bv30           4               2.600      2.600       3.000     2.800          0.011      0.013       0.023     0.433
knn31          2               2.400      2.400       4.000     1.000          0.049      0.075       0.719     0.956
knn31          3               4.400      4.400       5.000     2.000          0.066      0.093       0.429     1.754
knn31          4               10.800     10.800      21.000    3.500          0.098      0.127       0.643     2.345
cc32           2               1.000      1.000       1.000     1.000          0.008      0.012       0.034     0.758
cc32           3               2.200      2.200       2.000     2.000          0.018      0.022       0.031     1.255
cc32           4               3.000      3.000       3.000     3.000          0.013      0.016       0.033     1.734
dnn33          2               4.000      4.000       4.000     4.000          0.095      0.179       1.115     1.165
dnn33          3               11.000     11.000      11.000    7.000          0.107      0.201       0.987     1.908
dnn33          4               28.800     28.800      36.000    9.300          0.221      0.318       1.111     2.757
ising34        2               1.000      1.000       1.000     1.000          0.023      0.034       0.111     0.073
ising34        3               6.600      6.600       9.000     2.600          0.030      0.042       0.116     0.109
ising34        4               13.400     13.400      17.000    3.500          0.035      0.049       0.124     0.144
cat35          2               1.000      1.000       1.000     1.000          0.010      0.014       0.025     0.354
cat35          3               9.000      9.000       25.000    2.000          0.028      0.032       0.045     0.573
cat35          4               9.000      9.000       23.000    3.000          0.027      0.032       0.045     0.751
wstate36       2               2.000      2.000       2.000     1.000          0.028      0.039       0.119     0.414
wstate36       3               4.000      4.000       4.000     3.000          0.033      0.044       0.124     0.574
wstate36       4               6.000      6.000       6.000     3.000          0.036      0.047       0.129     0.880
qugan39        2               5.000      5.000       5.000     4.000          0.121      0.233       2.092     1.411
qugan39        3               12.000     12.000      12.000    8.400          0.130      0.255       1.901     2.152
qugan39        4               22.600     21.600      21.000    14.100         0.169      0.448       1.764     3.290
ghz40          2               1.000      1.000       1.000     1.000          0.012      0.017       0.032     0.462
ghz40          3               4.400      4.400       14.000    2.000          0.023      0.028       0.045     0.743
ghz40          4               3.000      3.000       3.000     3.000          0.017      0.022       0.037     0.981
knn41          2               1.000      1.000       1.000     1.000          0.073      0.118       2.044     2.028
knn41          3               5.000      5.000       5.000     2.100          0.092      0.140       1.252     3.441
knn41          4               10.800     10.800      15.000    3.000          0.159      0.209       0.940     5.069
swaptest41     2               1.000      1.000       1.000     1.000          0.070      0.113       2.021     2.028
swaptest41     3               5.000      5.000       5.000     2.100          0.089      0.135       1.233     3.441
swaptest41     4               9.200      9.200       15.000    3.000          0.125      0.173       0.904     5.069
ising42        2               1.000      1.000       1.000     1.000          0.031      0.048       0.176     0.101
ising42        3               2.000      2.000       2.000     2.200          0.035      0.053       0.182     0.166
ising42        4               14.400     14.400      19.000    3.600          0.049      0.068       0.184     0.179
\end{filecontents*}
\pgfplotstabletypeset[
        col sep=space,
        header=true,
        every head row/.style={
            before row={\toprule},
            after row={\midrule},
        },
        every last row/.style={
            after row={\bottomrule},
        },
        columns={circuit_name, num_partitions, part_cost, embed_cost, esd_cost, MLFM_rec_cost, part_time,
                 embed_time, esd_time, MLFM_rec_time},
                 columns/circuit_name/.style={
                  column name={Circuit},
                  column type=l,
                  string type,
                  string replace={_}{\_}, 
              },
              columns/circuit_name/.style={
          column name={Circuit},
          column type=l,
          string type,
          string replace={_}{ }, 
      },
      columns/num_partitions/.style={
          column name={Partitions},
          column type=l,
          fixed,
          precision = 0, 
      },
      columns/part_cost/.style={
          column name={P Cost},
          column type=l,
          fixed,
          precision = 1, 
      },
      columns/part_time/.style={
          column name={P Time},
          column type=l,
          fixed,
          precision = 3, 
      },
      columns/embed_cost/.style={
          column name={PE Cost},
          column type=l,
          fixed,
          precision = 1, 
      },
      columns/embed_time/.style={
          column name={PE Time},
          column type=l,
          fixed,
          precision = 3, 
      },
      columns/esd_cost/.style={
          column name={ESD Cost},
          column type=l,
          fixed,
          precision = 1, 
      },
      columns/esd_time/.style={
          column name={ESD Time},
          column type=l,
          fixed,
          precision = 3, 
      },
      columns/MLFM_rec_cost/.style={
          column name={MLFM-R Cost},
          column type=l,
          fixed,
          precision = 1, 
      },
      columns/MLFM_rec_time/.style={
          column name={MLFM-R Time},
          column type=l,
          fixed,
          precision = 3, 
      }
      ]{plots/QASM/FM_tket_table_QASM.dat}
  }

\end{table*}

\begin{table*}[htbp]
  \centering
\caption{Comparison of entanglement cost and time taken for larger QASM benchmark suite circuits.}
  \label{tab:QASM_res_large}
  \resizebox{\textwidth}{!}{%
      \begin{filecontents*}{plots/QASM/FM_tket_table_QASM_larger.dat}
circuit_name   num_partitions  part_cost  embed_cost  esd_cost  MLFM_rec_cost  part_time  embed_time  esd_time  MLFM_rec_time 
dnn51          2               4.000      4.000       4.000     3.200          0.199      0.395       5.024     2.660
dnn51          3               11.000     10.200      10.000    7.100          0.211      0.652       3.103     4.897
dnn51          4               15.000     15.000      15.000    14.100         0.229      0.456       3.278     6.530
adder64        2               4.000      3.200       3.000     4.600          0.276      0.858       5.765     7.269
adder64        3               9.000      8.400       8.000     9.800          0.284      0.883       5.780     13.660
adder64        4               12.000     10.800      9.000     13.100         0.295      0.919       5.730     18.543
cc64           2               1.000      1.000       1.000     1.000          0.025      0.039       0.163     3.600
cc64           3               2.000      2.000       2.000     2.000          0.032      0.046       0.150     6.424
cc64           4               3.000      3.000       3.000     3.000          0.033      0.047       0.133     8.608
cat65          2               1.000      1.000       1.000     1.000          0.028      0.041       0.075     1.400
cat65          3               3.800      3.800       4.000     2.000          0.040      0.053       0.084     2.453
cat65          4               8.000      8.000       21.000    3.000          0.051      0.064       0.123     3.100
ising66        2               1.000      1.000       1.000     1.200          0.071      0.113       0.505     0.153
ising66        3               2.000      2.000       2.000     2.400          0.076      0.118       0.502     0.232
ising66        4               3.000      3.000       3.000     4.000          0.078      0.120       0.506     0.297
knn67          2               1.000      1.000       1.000     1.000          0.187      0.309       11.154    6.218
knn67          3               2.000      2.000       2.000     2.000          0.199      0.321       8.785     12.465
knn67          4               6.000      6.000       6.000     5.000          0.215      0.339       4.692     16.765
bv70           2               1.000      1.000       1.000     1.000          0.017      0.025       0.072     0.957
bv70           3               1.000      1.000       1.000     2.000          0.019      0.027       0.079     1.669
bv70           4               2.000      2.000       3.000     3.000          0.022      0.030       0.061     2.221
qugan71        2               5.000      5.000       5.000     4.000          0.346      0.725       11.392    5.530
qugan71        3               10.000     10.000      10.000    11.900         0.375      0.769       8.146     9.694
qugan71        4               15.000     15.000      15.000    14.500         0.403      0.794       9.485     13.137
wstate76       2               2.000      2.000       2.000     1.000          0.105      0.156       0.666     2.114
wstate76       3               4.000      4.000       4.000     3.000          0.115      0.165       0.687     3.133
wstate76       4               6.000      6.000       6.000     3.000          0.117      0.167       0.703     4.777
ghz78          2               1.000      1.000       1.000     1.000          0.039      0.057       0.103     2.138
ghz78          3               2.000      2.000       2.000     2.000          0.047      0.069       0.126     3.785
ghz78          4               3.000      3.000       3.000     3.000          0.047      0.066       0.114     4.898
swaptest83     2               1.000      1.000       1.000     1.000          0.258      0.444       26.224    11.010
swaptest83     3               2.000      2.000       2.000     2.000          0.277      0.466       20.204    19.046
swaptest83     4               6.000      6.000       6.000     5.000          0.294      0.482       17.082    29.794
ising98        2               1.000      1.000       1.000     1.500          0.147      0.241       1.354     0.236
ising98        3               2.000      2.000       2.000     2.800          0.156      0.250       1.369     0.383
ising98        4               3.000      3.000       3.000     4.000          0.164      0.258       1.378     0.466
\end{filecontents*}
\pgfplotstabletypeset[
        col sep=space,
        header=true,
        every head row/.style={
            before row={\toprule},
            after row={\midrule},
        },
        every last row/.style={
            after row={\bottomrule},
        },
        columns={circuit_name, num_partitions, part_cost, embed_cost, esd_cost, MLFM_rec_cost, part_time,
                 embed_time, esd_time, MLFM_rec_time},
                 columns/circuit_name/.style={
                  column name={Circuit},
                  column type=l,
                  string type,
                  string replace={_}{\_}, 
              },
              columns/circuit_name/.style={
          column name={Circuit},
          column type=l,
          string type,
          string replace={_}{ }, 
      },
      columns/num_partitions/.style={
          column name={Partitions},
          column type=l,
          fixed,
          precision = 0, 
      },
      columns/part_cost/.style={
          column name={P Cost},
          column type=l,
          fixed,
          precision = 1, 
      },
      columns/part_time/.style={
          column name={P Time},
          column type=l,
          fixed,
          precision = 3, 
      },
      columns/embed_cost/.style={
          column name={PE Cost},
          column type=l,
          fixed,
          precision = 1, 
      },
      columns/embed_time/.style={
          column name={PE Time},
          column type=l,
          fixed,
          precision = 3, 
      },
      columns/esd_cost/.style={
          column name={ESD Cost},
          column type=l,
          fixed,
          precision = 1, 
      },
      columns/esd_time/.style={
          column name={ESD Time},
          column type=l,
          fixed,
          precision = 3, 
      },
      columns/MLFM_rec_cost/.style={
          column name={MLFM-R Cost},
          column type=l,
          fixed,
          precision = 1, 
      },
      columns/MLFM_rec_time/.style={
          column name={MLFM-R Time},
          column type=l,
          fixed,
          precision = 3, 
      }
      ]{plots/QASM/FM_tket_table_QASM_larger.dat}
  }
\end{table*}

\end{document}